\documentclass[conf]{new-aiaa}
\usepackage[utf8]{inputenc}

\usepackage{multirow}
\usepackage{graphicx}
\usepackage{amsmath}
\usepackage[version=4]{mhchem}
\usepackage{siunitx}
\usepackage{longtable,tabularx}
\title{Open-Source Parametric Airfoils to Study Geometric Effects on Buffet}

\author{Zauner M.\footnote{Researcher, Aviation Technology Directorate, Aircraft Lifecycle Innovation Hub, zauner.markus@jaxa.jp.}, 
Lusher D. J.\footnote{Invited Researcher, Aviation Technology Directorate, Aircraft Lifecycle Innovation Hub, lusher.david@jaxa.jp.},
Moise P.\footnote{Visitor, University of Southampton, p.moise@soton.ac.uk},
Sansica A.\footnote{Researcher, Aviation Technology Directorate, Aircraft Lifecycle Innovation Hub, sansica.andrea@jaxa.jp.},
Hashimoto A.\footnote{Manager, Aviation Technology Directorate, Aircraft Lifecycle Innovation Hub, hashimoto.atsushi@jaxa.jp.},
and 
Sandham N. D.\footnote{Professor, Aerodynamics and Flight Mechanics Group, University of Southampton, n.sandham@soton.ac.uk}.}

\affil{Japan Aerospace Exploration Agency, Chofu, Tokyo, 182-8522 Japan}
\affil{University of Southampton, Southampton, Hampshire, SO17 1BJ, UK}
\begin{document}

\maketitle

\begin{abstract}

% attempted a rewrite of the abstract (feel free to change)
Recent research into buffet in the transonic flow regime has been focused on a limited number of proprietary airfoil geometries and has mainly considered parametric variations in Mach number and angle of attack. In contrast, relatively little is known about the sensitivity of buffet frequencies and amplitudes to geometric properties of airfoils. In the present contribution, an airfoil geometry construction method based on a small number of parameters is developed. The resulting airfoils and computational grids are high-order continuous everywhere except at the trailing edge corners. 
The effects of four key geometric parameters, defined by local extrema of coordinates on the airfoil at the design condition and denoted as `crest' points, are studied using large-eddy simulation, considering both free-transitional and tripped boundary layers. For both states of the boundary layer, buffet amplitude and frequency are found to be highly sensitive to the axial and vertical position of the suction-side crest point, while the vertical crest position on the pressure side affects the mean lift. The present work confirms that buffet can appear for free-transitional (laminar buffet) and tripped conditions (turbulent buffet) with similar sensitivities. 
For test cases near onset conditions intermediate-frequency phenomena have been observed, which were linked to unsteadiness of separation bubbles. Frequencies scaling based on mean-flow properties of the separation bubble were shown to be in good agreement with previous findings on different airfoil geometries in Zauner \textit{et al.} (Flow Turb. \& Comb., Vol. 110, 2023, pp. 1023-1057).
Airfoil geometries are provided as open source: \url{https://doi.org/10.5281/zenodo.12204411}.

\end{abstract}

\section{Introduction}

The safe flight envelope of aircraft is typically limited by an unsteady flow phenomenon called ``buffet'', leading to strong vibrations of wings and control surfaces. 
Buffet can be categorised according to the speed regime (i.e. low- and high-speed buffet), spatial properties (i.e. extruded two-dimensional geometry invariant in the spanwise direction, swept profiles, and fully three-dimensional wing configurations), and airfoil properties (i.e. symmetric and asymmetric).
Buffet research in literature has a strong focus on high-speed applications, where compressibility effects cannot be ignored, and is often referred to as `transonic buffet'. A possible connection between low- and high-speed buffet is still under investigation and discussion \cite{PDBLS2021, Moise2024}, but the focus of the current work lies on transonic buffet.
Transonic buffet has been linked to linear instabilities observed within a narrow band before stall for two-dimensional airfoils, swept wings \cite{Crouch2018}, as well as fully three-dimensional aircraft configurations \cite{Timme2016, Sansica2023}. Even though two-dimensional buffet phenomena can co-exist for swept wings \citep{Paladini2019a} as well as half-wing-body configurations \citep{Ohmichi2018, DAguanno2021}, three-dimensional phenomena like buffet cells \cite{Iovnovich2012}, which have been shown to be essentially stall cells \cite{PDBLS2021}, are also likely to be present and often dominate the flow dynamics. Even for unswept infinite wings, \citet{lusher2024highfidelity} identified both 2D shock-oscillation and 3D buffet-cell modes at different frequencies, applying high-fidelity methods and aspect ratios of up to $AR=3$. 
Here, we focus on buffet over unswept wing sections on a narrow domain, exhibiting so-called `two-dimensional buffet'. Furthermore, we consider only supercritical airfoil profiles exhibiting Type II buffet \citep{Giannelis2017}, (for which shock waves are present and oscillate only on the suction side). 
We investigate laminar-flow profiles for free-transitional boundary layers, which are characterised by relatively flat suction-side surfaces and aim to delay the transition point as long as possible. For tripped boundary layers, we consider conventional supercritical airfoil shapes, which are similar to the popular OAT15A profile developed by ONERA.
%considering free-transitional boundary layers over laminar-flow profiles. Such profiles aim to maintain laminar boundary layers up to the shock foot to reduce skin-friction drag. In the future we will extend our studies to more conventional airfoil geometries considering tripped boundary layers.

\begin{figure}[hbt!]
\centering
	\includegraphics[width=\textwidth, trim=10 850 10 100, clip]{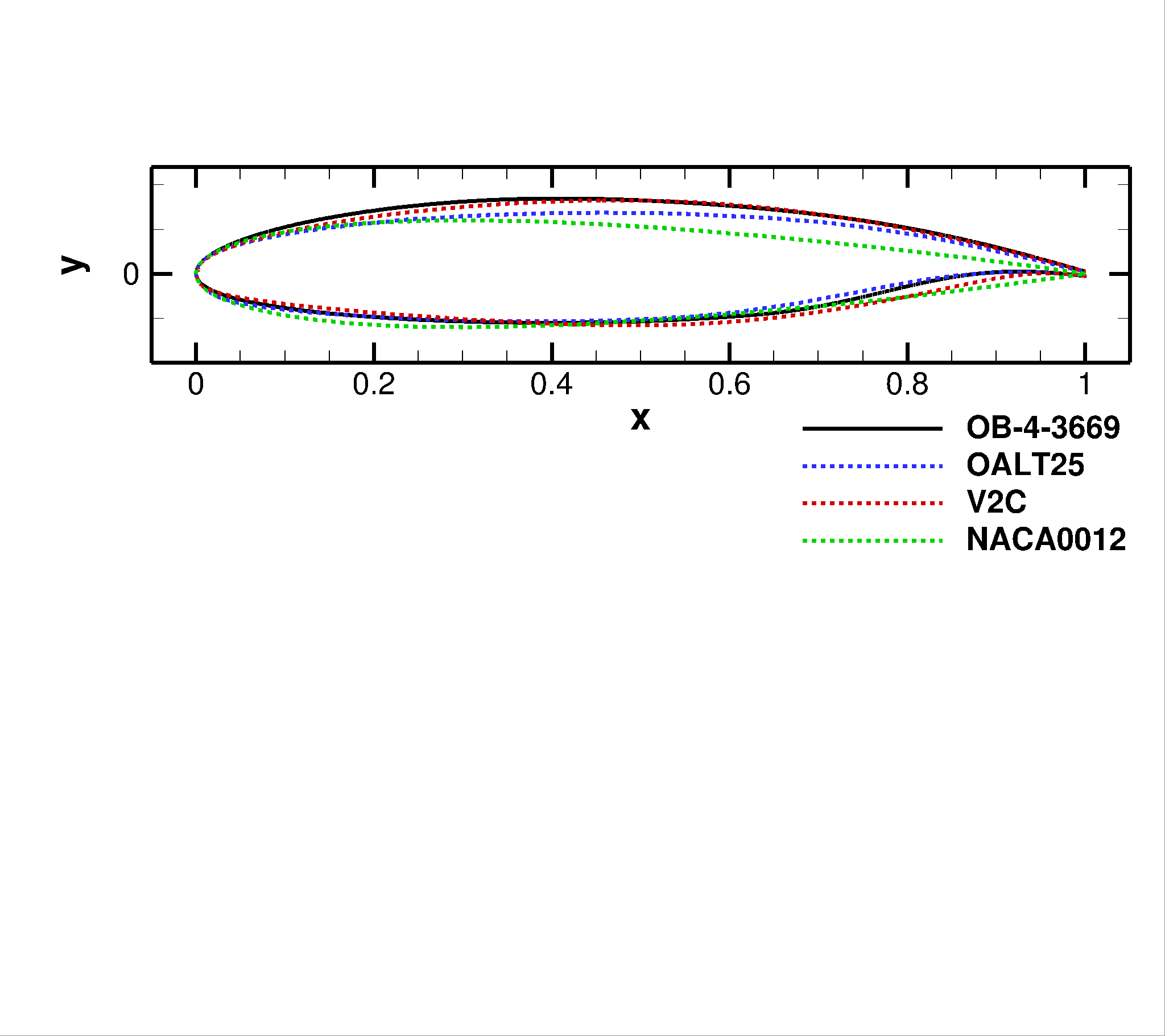}
\caption{Comparison of different laminar-flow airfoil geometries. The OB-4-3669 has its upper and lower crest positions at $X_U/Y_U=0.3/0.06$ and $X_L/Y_L=0.6/-0.09$, respectively.\label{fig:airfoil_comp}}
\end{figure}

% Despite significant progress in transonic buffet research, we still do not understand the precise underlying mechanism. 
%It is particularly challenging to predict frequencies and onset conditions associated with transonic buffet in a universal manner.
\citet{Crouch2007} showed that we can predict frequencies and onset conditions associated with transonic buffet using global stability analysis. However, we still do not fully understand the driving mechanism and how this mechanism depends on the shape of the airfoil.
For different geometries, we often observe large differences in buffet frequencies, $f_B$, which are commonly normalised by the chord length, $c$ and free-stream velocity, $U_{\infty}$ (i.e., Strouhal number, here denoted as $St_B=f_B U_{\infty} /c$). Also, onset and offset conditions can vary significantly across different airfoil geometries. 
To set the scene for the present study,  simulation results of Dassault Aviation's V2C profile, ONERA's OALT25 profile (both supercritical laminar-flow profiles), and the symmetric NACA0012 profile are shown in table \ref{tab:airfoils} for a constant angle of attack of $\alpha=4^{\circ}$ and free-transitional boundary layers. Besides Reynolds number ($Re$) and Mach number ($M$), also the angle of attack ($\alpha$) and boundary-layer transition mode are listed. In addition, we provide the chord-based Strouhal numbers as well as approximate amplitudes of lift oscillations associated with buffet, denoted as $St_B$ and $\Delta C_{L,max}$, respectively.
We also list the location of so-called upper ($X_U,Y_U$) and lower ($X_L,Y_L$) crest points after rotating the profile by $\alpha$, which will allow comparison with present parametric airfoil geometries and will be detailed later. 
Corresponding airfoil geometries at $\alpha=0$ are illustrated in figure \ref{fig:airfoil_comp}, including also a present OpenBuffet geometry with similar geometric properties as the V2C profile. At the design angle of $\alpha_D=4^{\circ}$, the crest points of this profile will be located at $(X_U,Y_U)=(0.3,0.06)$ and $(X_L,Y_L)=(0.6,0.09)$, which will be discussed in more detail later.
Considering moderate Reynolds numbers of $Re=500,\!000$ at $M=0.7$, \citet{Zauner2018a} and \citet{Moise2021} reported Strouhal numbers of $St\approx0.1$ for direct numerical and large-eddy simulations of the V2C profile, respectively, which agree well with wind-tunnel experiments by \citet{Placek2016} and {Davidson2016a}. Considering the OALT25 profile at the same $\alpha$, $M$ and $Re$, large-eddy simulations (LES) of \citet{Zauner2023b} show very weak low-frequency oscillations associated with incipient transonic buffet at much lower frequencies corresponding to $St_B\approx0.035$. At an increase Mach number of $M=0.735$, buffet frequencies for V2C and OALT25 profiles respectively increase to $St_B=0.15$ and $0.082$, but remain very different \cite{Zauner2023b}. 
%In order to compare with experimental results for the OALT25 profile, simulation Reynolds numbers were increased to $Re=3,\!000,\!000$. Even though the buffet amplitude decreased, frequencies remained almost constant and agreed reasonably well with experiments \cite{Brion2019} and LES \cite{Dandois2018}. 
Despite significant geometric differences, LES of the symmetric NACA0012 by \citet{Moise2024} show similar buffet frequencies compared to the V2C profile, while onset conditions differ. 
\begin{table}    
    \begin{center}
    \def~{\hphantom{0}}
        \begin{tabular}{l|ccc|c|cc|cccc}
        Airfoil  & Re        & M     & $\alpha$ & Transition & $St_B$   & approx. $\Delta C_{L,max}$ & $X_U$ & $Y_U$ & $X_L$ & $Y_L$                            \\
        % V2C      & 1,500,000 & 0.7   & 4     & free       & 0.11  &          &                             \\
        \hline
        V2C      & 500,000   & 0.7   & $4^{\circ}$     & free       & 0.12  & 0.20  & 0.33 & 0.055 & 0.59 & -0.095                                \\
        V2C      & 500,000   & 0.735 & $4^{\circ}$     & free       & 0.15  & 0.30 & & & &                               \\
        \hline
        OALT25   & 500,000   & 0.7   & $4^{\circ}$     & free       & 0.035 & $< 0.05$ & 0.26 & 0.044 & 0.54 & -0.087           \\
        OALT25   & 500,000   & 0.735 & $4^{\circ}$     & free       & 0.082 & 0.45  & & & &    \\
        OALT25   & 3,000,000 & 0.735 & $4^{\circ}$     & free       & 0.082 & 0.10  & & & &  \\                
        \hline
        % CRM      & 500,000   & 0.7   & $4^\circ$     & free       & 0.05  & 0.24     & very preliminary simulation \\
        NACA0012 & 500,000   & 0.7    & $4^\circ$     & free      & --     & --        & 0.18 & 0.043 & 0.53  & -0.088       \\
        NACA0012 & 500,000   & 0.75   & $4^\circ$     & free      & 0.12     &  0.4    & & & &\\
        % OAT15A   & 500,000   & 0.7   & 4     & free       & ?     & ?        & 
        \hline
        \end{tabular}
  \end{center}
\caption{Comparison of buffet characteristics of V2C, OALT25, and NACA0012 airfoils from simulation data reported in \cite{Moise2021}, \cite{Zauner2023b}, and \cite{Moise2024}, respectively. Crest-point coordinates of airfoil geometries at an angle of attack of $\alpha=4^{\circ}$ are denoted by $X_U$ and $Y_U$ (suction side), as well as $X_L$ and $Y_L$ (pressure side).  \label{tab:airfoils}
}
\end{table}

\begin{figure}[hbt!]
\centering
    \includegraphics[width=\textwidth, trim=10 800 10 100, clip]{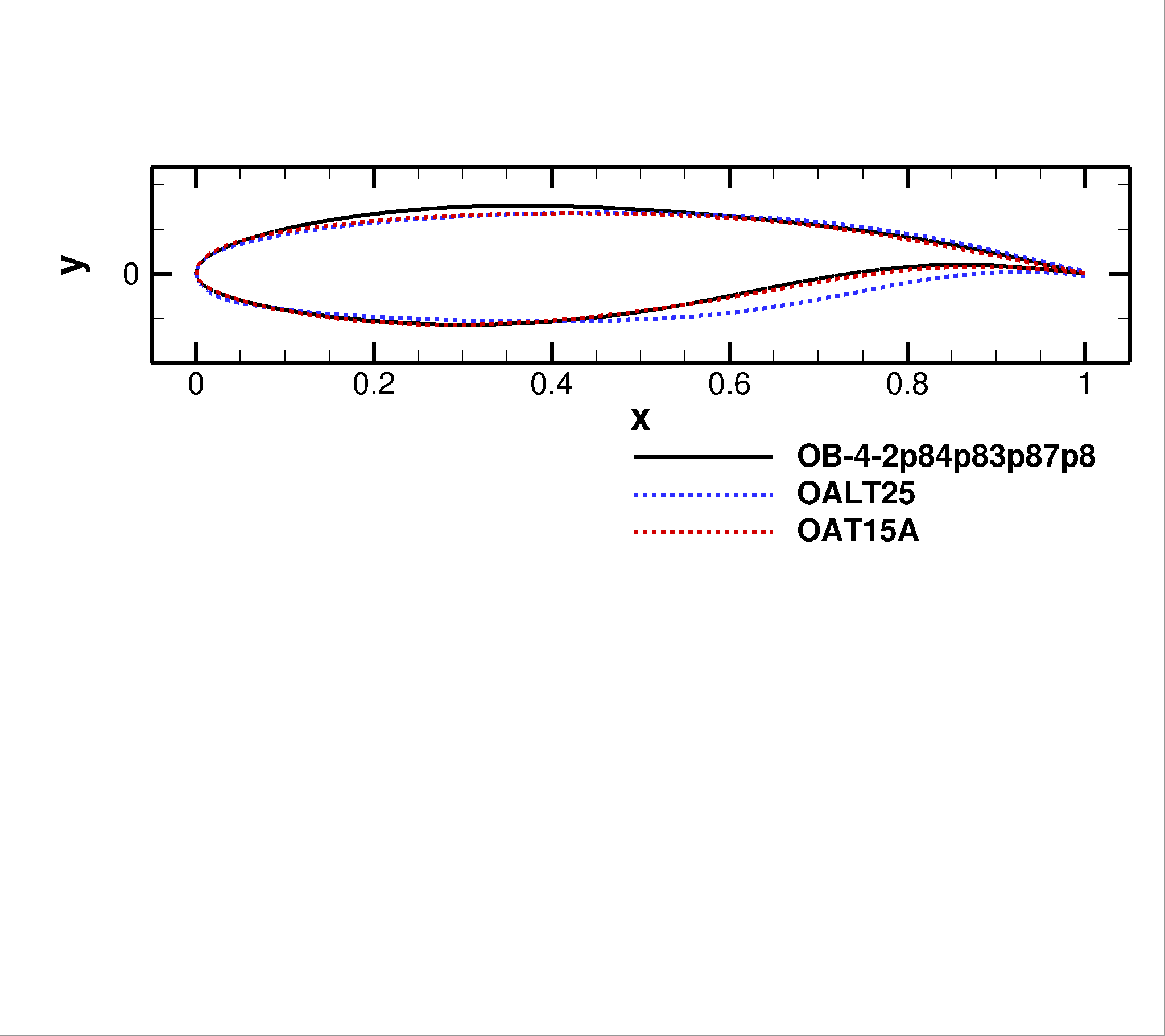}
\caption{Comparison of different conventional supercritical airfoil geometries. The OB-3p5-2p8'4p8'3p8'7p8 geometry is used for case OB-T-D2-4p8 and specified in table \ref{tab:parameters_tripped}. \label{fig:airfoil_comp2}}
\end{figure}

% tripped vs free-transitional
Even though the mode of boundary-layer transition has significant impact on buffet onset conditions and amplitudes, frequencies seem to be less affected \cite{Moise2023}. In their wind-tunnel experiments on ONERA's OALT25 profile, \citet{Brion2019} showed that low-frequency oscillations appear for tripped and free-transitional boundary layers at $St_B\approx0.07$ and $0.05$, respectively. For the latter test case, however, a co-existing intermediate-frequency phenomenon at $St_I \approx 1.1$ dominates the flow dynamics and has been associated with unsteadiness of the shock-induced separation bubble \cite{Dandois2018, Zauner2023b}. Even though the buffet frequencies measured in experiments for the tripped OALT25 profile agree reasonably well with those of \citet{JMDMS2009} reported for ONERA's OAT15A ($St_B \approx 0.07$), we should bear in mind that flow conditions differ, as shown in table \ref{tab:airfoils_tripped}. Corresponding airfoil geometries are illustrated in figure \ref{fig:airfoil_comp2}, including also a present OpenBuffet geometry with similar geometric properties as the OAT15A profile (geometric details of the OB-3p5-2p8'4p8'3p8'7p8 geometry are specified in table \ref{tab:parameters_tripped} for test case OB-T-D2-4p9 and will be explained later).
\begin{table}[htb]    
    \begin{center}
    \def~{\hphantom{0}}
        \begin{tabular}{l|ccc|c|cc|cccc}
        Airfoil  & Re        & M     & $\alpha$ & Transition & $St_B$   & approx. PSD & $X_U$ & $Y_U$ & $X_L$ & $Y_L$                            \\
        % V2C      & 1,500,000 & 0.7   & 4     & free       & 0.11  &          &                             \\
        \hline
        OALT25\cite{Brion2019}   & 3,000,000 & 0.735 & $4.0^{\circ}$     & free       & 0.05  & $4\cdot 10^{-6} [p^2/q_0^2]$  & 0.26 & 0.044 & 0.54 & -0.087 \\ 
        %crest-to-crest thickness 0.131 at 4deg / 0.122 at 0deg
        OALT25\cite{Brion2019}   & 3,000,000 & 0.735 & $4.0^{\circ}$     & tripped       & 0.07  & $4\cdot 10^{-4} [p^2/q_0^2]$  & & & &   \\
        \hline
        OAT15A\cite{JMDMS2009}   & 3,000,000 & 0.73 & $3.5^{\circ}$     & tripped       & 0.07  & $3\cdot 10^{11} [p^2]$  & 0.25 & 0.048 & 0.37 & -0.077  \\
        %crest-to-crest thickness 0.125 at 3.5deg / 0.125 at 0deg
        \hline
        \end{tabular}
  \end{center}
\caption{Comparison of buffet characteristics of OALT25 and OAT15A airfoils. Cases at $Re=3,\!000,\!000$ correspond to experimental data of \cite{Brion2019} and \cite{JMDMS2009}. The power-spectral density (PSD) measurements for OALT25 and OAT15A profiles are respectively based on pressure-sensor data at $x/c=0.73$ and $0.45$.   \label{tab:airfoils_tripped}}
\end{table}

In general, one could try to explain differences in buffet frequencies by acoustic feedback models similar to \citet{Roos1980} and \citet{LEE1990}. However, even though such models may match results for certain profiles under certain conditions for the OAT15A profile \citep{Deck2005, Xiao2006}, predictions of other studies did not agree with observations \citep{Jacquin2009,Garnier2010,Moise2021}. 
For the aforementioned free-transitional test case, where freestream conditions are similar, buffet occurs for the OALT25 profile at almost four times lower frequencies compared to the V2C or NACA0012 profiles. Such significant variations are hard to explain by acoustic feedback models. Also for the example comparing tripped and free-transitional OALT25 experiments \citep{Brion2019}, significant variations of the shock position have been observed, while buffet frequencies remained similar.

On the other hand, geometric features of the airfoils are very likely to influence gradients in the flowfield and global stability characteristics.
In tables \ref{tab:airfoils} and \ref{tab:airfoils_tripped}, the coordinates of the crest points (the points denoting local extrema of the $y$-coordinate after rotation by $\alpha$) are listed for the considered airfoil geometries. We can see that the upper crest is located further downstream for the V2C profile compared to the OALT25 airfoil (see also, Fig.~\ref{fig:airfoil_comp}). It also has an approximately $15\%$ larger crest-to-crest thickness (the difference between the $y$-coordinates of upper and lower crest points after rotation by $\alpha$). However, it is hard to interpret the significance of differences in crest-point locations with respect to transonic buffet.
It is also interesting to note that main geometric differences between ONERA's OALT25 and OAT15A profiles appear on the pressure side. However, based on results in literature it is impossible to conclude whether the good agreement of buffet frequencies for both airfoils is due to geometric similarities of the suction side profiles or mixed (competing) effects of variations of geometry and flow conditions.
Despite the availability of extensive literature on experimental as well as numerical studies on transonic buffet, they are very often limited to single airfoil geometries. Given the fact that transonic buffet is highly sensitive to $M$, $Re$, $\alpha$, and the mode of boundary-layer transition (free or tripped), results are often difficult to compare and it is hard to understand the isolated influence of geometric features on the dynamical behaviour of transonic buffet (i.e. onset, frequencies, amplitude, etc.).

In the present contribution, we want to analyse the sensitivity of buffet to geometric features such as crest-point locations for tripped and free-transitional boundary layers. Results can help us to learn more about the underlying buffet mechanism and provide insights into airfoil design, which poses significant challenges due to the high dimensionality of the geometry. Various parameterisation approaches have been adopted to simplify this design process to reduce the dimensionality of this problem. The set of parameters needs to be carefully chosen such that they allow for sufficient variation in the airfoil geometry, cause physically intuitive variations in the geometry, produce smooth profiles and have minimal redundancy. 
We can find a wide range of airfoil construction methods in the literature, which can be used for such parametric studies. Most of these methods aim to facilitate the conceptual design of new airfoils. Based on thin-airfoil theory, it is often beneficial to parametrise and optimise the camber and thickness separately to study a wide range of parameters, such as angles of attack and Mach numbers. The NACA airfoil series is a very popular example of this approach. Other methods, such as orthogonal basis function (OBF)\cite{Robinson2001}  or class-shape function transformation (CST)\cite{Olson2015, Karman2019} methods, use polynomial functions to describe the shape of the airfoil contour, which facilitates the use of optimisation methods to analyse the aerodynamic performance for a wide range of flight conditions. 
For the present work, we are less interested in the airfoil performance or lift-drag polars. As we want to better understand the sensitivity of transonic buffet to geometric characteristics, we require increased control over the airfoil curvature at particular flow conditions, where buffet typically occurs. On the other hand, we want to keep the number of geometric input parameters as low as possible to facilitate the physical interpretations of results and trends, while maintaining a low number of cases and, consequently, minimising computational costs. In the context of using high-order discretisation schemes for high-fidelity CFD simulations, it is beneficial to describe the airfoil surface by high-order polynomials to avoid discontinuities in the metric terms. This is implemented in the present work, where the entire grid is generated by blending polynomial functions to guarantee continuous metric terms, where possible. The method developed is similar to the `Parametrised Section' (PARSEC) method of \citet{Sobieczky1998}, where an airfoil was defined by two sixth–order polynomials describing upper and lower sides. For the PARSEC airfoils, position and curvature of upper and lower crest points, maximum and trailing-edge thickness, leading-edge radius, and trailing-edge angles at zero incidence, sum up to a total of 11 parameters, but we emphasise that their focus was not on transonic buffet. 
We should keep in mind that when we design or parametrise an airfoil at zero inclination, airfoil properties like crest locations change with the angle of attack. While effect of chord-rotation or camber on aerodynamic mean properties can be explained by the thin-airfoil theory, the effects of variations of crest points and the projected airfoil thickness (or crest-to-crest thickness) are more difficult to predict.

% Mind that when we change the angle of attack, we do not only modify the inclination of the chord, but also the location of the crest points with respect to the free stream. While effect of chord-rotation on aerodynamic properties can be explained by the thin-arifoil theory (i.e. potential flow), we could show in the present study that buffet characteristics are highly sensitive to the crest-point locations. Therefore, we need to be careful when concluding form angle-of-attack studies on the nature of transonic buffet.

In the current approach, segments between the leading edge and crest points on either side of the airfoil are described by blended ellipses, whilst sixth-order polynomials are used to connect the crest points to the trailing edge. In contrast to the PARSEC method, we define these parameters at a specified design angle $\alpha_D$, where we expect buffet to occur. Parameters (\textit{e.g.} crest-point locations) can then be varied about this point without the need to consider rotation transforms at a given angle of attack. 
The angles at the trailing-edge corners (`trailing-edge angle' and `boat-tail angle' correspond to upper and lower corners of the trailing edge, respectively) are here defined with respect to the horizontal axis at the design angle.
Considering also trailing-edge thickness leaves us with a total of 10 parameters to describe the present `OpenBuffet' (OB) airfoil geometries, which are listed in table \ref{tab:parameters} together with those of the PARSEC method. More details about the airfoil construction method will be provided in the Methodology section. Compared to the PARSEC airfoils, the present approach is more intuitive, as it does not require the specification of derivatives. This makes it also straight forward to mimic existing airfoils, as crest points are easier to measure than derivatives.
We want to re-emphasise at this point that this study does not aim to replicate established airfoil geometries, for which better methods exist \cite{Sobester2009, Masters2015}. Instead, our objective is to study geometric effects on unsteady flow phenomena at buffet conditions using a minimum set of intuitive design parameters. In order to reduce the parameter space for our current contribution, we set the semi-axes defining the leading-edge rounding to $a_{LE}=X_U$ and $b_{LE}=Y_U$, and fix the remaining parameters to obtain a four-digit airfoil series for a given design angle. The last column of table \ref{tab:parameters} contains an example of parameters of a representative parametric profile, which will be detailed in the next section.
% Add this later?
% Based on previous studies, where buffet occurs at $Re=500,\!000$ and $M=0.7$ at an angle of attack of $\alpha=4^{\circ}$, we select in the present contribution a design angle of $\alpha=4^{\circ}$ for free-transitional boundary layers. For cases considering tripped boundary layers, the design angle is set to $\alpha=3.5^{\circ}$, which corresponds to buffet experiments of \cite{JMDMS2009} for the OAT15A profile. That way, the crest points correspond for each case to their coordinates at the angle of attack of interest. 
%The last column of table \ref{tab:parameters} contains an example of parameters. % for our baseline profile for free-transitional cases.
\begin{table}[htb]
    \begin{center}
    \def~{\hphantom{0}}
        \begin{tabular}{l|c|c|r}
        Parameter               & PARSEC & OpenBuffet & OB-4-3666         \\
        \hline
        Design angle            & $0^{\circ}$ & $\alpha_{d}$ & $4^{\circ}$ \\
        \hline
        Upper crest streamwise position    & $X_U$ & $X_U=D_1 / 10$  & $0.3$   \\
        Upper crest vertical position       & $Y_U$ & $Y_U=D_2 / 100$ & $0.06$  \\
        Lower crest streamwise position    & $X_L$ & $X_L=D_3 / 10$  & $0.6$   \\
        Lower crest vertical position        & $Y_L$ & $Y_L=-D_4 / 100$& $-0.06$ \\
        \hline
        Leading-edge rounding   & Upper leading-edge radius $r_U$ & Semi-major axis of ellipse $a_{LE}$  & $X_U$ \\
                                & Lower leading-edge radius $r_L$ & Semi-minor axis of ellipse $b_{LE}$  & $Y_U$ \\
        \hline
        Trailing-edge angle     & $\alpha_{TE}$  & $\alpha_{TE}=D5/10$ & $20^{\circ}$         \\
        Boat-tail angle         & $\beta_{TE}$   & $\beta_{TE}=D6/10$  & $10^{\circ}$          \\
        Trailing-edge thickness & $\Delta_{TE}$  & $\Delta_{TE}=D7/1000$ & $0.005$ \\
        \hline
        Upper crest curvature   & $Y_u''(x=X_U)$       & N/A  & N/A     \\
        Lower crest curvature   & $Y_l''(x=X_L) $      & N/A  & N/A      \\
        \hline
        \end{tabular}
  \end{center}
  \caption{Comparison between parameters of present OpenBuffet airfoils and PARSEC airfoil geometries.}\label{tab:parameters}
\end{table}

%P: We can summarise using something like this:
In summary, this study aims to shed light on the effect of geometry on transonic buffet, by formulating parametric airfoil profiles that allow for a systematic variation of certain key geometrical features at a given angle of attack. To reduce the parameter space, we focus on the coordinates of upper and lower crest points provided by a four-digit airfoil series. Airfoil geometries have been made openly available (\url{https://doi.org/10.5281/zenodo.12204411}).

In the following sections, we will first describe our simulation methods in \S \ref{sec:methodsLES} \& \ref{sec:methodsOpenSBLI} and grid-generation \S \ref{sec:methodsGrid}, before detailing our proposed parametric airfoil geometries in \S \ref{sec:methodsGeometry}. 
We then show results for these parametric airfoil geometries, investigating the influence of the upper crest location in \S \ref{sec:OpenBuffet_Upper}, where free-transitional as well as tripped boundary layers are considered in \S \ref{sec:OpenBuffet_Upper_Free} and \S \ref{sec:OpenBuffet_Upper_Tripped}, respectively.
To extend the parameter space, additional simulations for free-transitional boundary layers are shown in \S \ref{sec:modified_profiles} for modified airfoil geometries, including also analysis of the lower crest locations.
We will conclude this contribution in \S \ref{sec:conclusion}.

\section{Methodology} \label{sec:methods}
This work has been carried out in the scope of a collaboration between the Japanese Aerospace Exploration Agency (JAXA) and the University of Southampton. Simulations considering free-transitional boundary layers have been carried out at the University of Southampton using their in-house code SBLI. For cases with tripped boundary layers, simulations were performed by JAXA using an open-source framework called OpenSBLI, which is able to leverage GPU architectures. Both numerical methods are described in the following sub sections.
\subsection{Spectral-Error based Large-Eddy Simulations (SE-LES) used for free-transitional cases} \label{sec:methodsLES}
For cases considering free-transitional boundary layers, we employ the in-house code SBLI \citep{Yao2009} for scale-resolving numerical simulations. This code, extensively validated, has been utilised in various investigations, including studies on shock-wave/boundary-layer interaction \citep{Touber2009, Sansica2014, Sansica2016}, subsonic wing sections \citep{Jones2008, DeTullio2018}, and transonic buffet \citep{Zauner2019a, Zauner2020, Moise2023}. SBLI demonstrates robust performance in massively-parallelised simulations using structured multi-block grids across diverse high-performance computer architectures.

The compressible three-dimensional Navier-Stokes equations are normalised by the airfoil chord ($c$), freestream density ($\rho_{\infty}$), streamwise velocity ($U_{\infty}$), and temperature ($T_{\infty}$). These equations are solved in non-dimensional form using fourth-order finite difference schemes (central at interior and the Carpenter scheme \citep{Carpenter1999} at boundaries) for spatial discretisation. For temporal discretisation, we use a low-storage third-order Runge-Kutta scheme, and a total variation diminishing scheme is applied to capture shock wave features \citep{Sansica2015}. A non-dimensional time step of $\Delta t = 3.5\times 10^{-5}$ Convective Time Units (CTUs) is maintained for all free-transitional simulations. For boundary conditions, zonal characteristic boundary conditions \cite{Sandberg2006_char} are enforced at the outlet, while integral characteristic boundary conditions \cite{Sandhu1994} are applied at the remaining outer boundaries.

In this contribution, we adopt a Spectral-Error based Implicit Large-Eddy Simulation (SE-ILES) scheme \cite{Zauner2019c,Zauner2020}. This approach involves evaluating spectral error indicators \citep{Jacobs2017} locally every $N_E=30$ time steps corresponding to $0.00105$ CTUs. These indicators control a sixth-order filter \citep{Visbal2002} to eliminate scales insufficiently resolved by the grid. Further details of this approach can be found in \cite{Zauner2019b}, where validation against DNS for a transonic buffet test case is also shown. Requirements for grid and domain independence, and the effect of various flow conditions ($M$, $Re$, $\alpha$, transition type, \textit{etc.}) have been examined using SBLI and reported in multiple studies \citep{Zauner2020, Moise2021, Moise2023,Zauner2023b}. For the OALT25 profile, good agreement between simulation results of \cite{Zauner2023b} and experimental measurements of \cite{Brion2019} was found.

\subsection{Implicit Large-Eddy Simulations using OpenSBLI for tripped turbulent cases} \label{sec:methodsOpenSBLI}

For simulations considering boundary-layer tripping, we use the multi-block OpenSBLI automatic code-generation framework for compressible CFD simulations \citep{LJS2018,LJS2021}, developed at the University of Southampton and JAXA. OpenSBLI utilises symbolic algebra to automatically generate a complete finite-difference CFD solver in the Oxford Parallel Structured (OPS) \cite{Reguly_2014_OPSC} Domain-Specific Language (DSL). Users can define systems of partial differential equations to solve, which are expanded and discretized symbolically to create a simulation code tailored to the problem specified. The OPS library is embedded in C/C++ code, enabling massively-parallel execution of the code on a variety of high-performance-computing architectures via source-to-source translation, including GPUs. The OpenSBLI code was recently cross-validated against six other independently-developed flow solvers using a range of different numerical methodologies in \cite{NATO2024_TGV} for a supersonic Taylor-Green vortex case involving shock-waves and transition to turbulence.

OpenSBLI applies high-order, non-dissipative central finite-differences to compute spatial derivatives, in cubic-split formulations to improve numerical stability \citep{Coppola_CubicSplit_2019}. Shock-capturing is performed by high-order Weighted Essentially Non-Oscillatory (WENO) schemes, applied in a characteristic-based filter framework \citep{Yee2018}. For this study, fourth and fifth order schemes are used for central and WENO-based derivatives respectively. Grid metrics are computed with the same order of central scheme. A fourth-order one-sided boundary-closure \cite{Carpenter1999} is applied at domain boundaries. A low-storage fourth-order explicit Runge-Kutta scheme is used for time-advancement, with a non-dimensional time-step of $\Delta t = 5\times 10^{-5}$. An 11-point Dispersion-Relation-Preserving (DRP) explicit filter \cite{BOGEY2004194} is also applied to provide further stabilisation. To minimize the numerical dissipation introduced by the filter, a targeted filter approach (as described in \cite{LZSH2023}) is applied. A pressure-gradient-based sensor is computed to detect strong oscillations in the flow. The filter is applied only at these regions, and only at a frequency of once every 25 time-steps corresponding to $0.00125$ CTUs. The airfoil surface is enforced by an isothermal no-slip boundary condition with wall temperature equal to the freestream value. Sponge zones are applied on the outer boundaries to maintain uniform freestream conditions. Data is written out from the solver every 0.05 CTUs to calculate the aerodynamic coefficients and flow visualisations. 

Both high-fidelity codes (OpenSBLI and SBLI), have recently been cross-validated for a laminar buffet test case on the Dassault Aviation's V2C profile at $Re=500,000$ \citep{LZSH2023}. Excellent agreement was found for all aerodynamic coefficients and buffet quantities. Furthermore, OpenSBLI has been used to perform large-scale simulations of transonic buffet on turbulent wing profiles such as the NASA-CRM geometry in \cite{LSH2024_narrow_buffet,lusher2024highfidelity}. In \cite{LSH2024_narrow_buffet}, OpenSBLI was cross-validated against results from URANS and global stability analysis obtained by JAXA's unstructured solver FaSTAR \cite{FaSTAR_Hashimoto_2012}, with excellent agreement found for buffet onset conditions, the main low-frequency buffet peak, and time-averaged aerodynamic quantities. In addition to sensitivity to numerical tripping strength, domain-width sensitivities were also examined for airfoils with span-widths equal to between 2.5\% and 50\% of chordwise length $(AR=0.025 - 0.5)$. In \cite{lusher2024highfidelity}, OpenSBLI was applied to wide-span turbulent transonic buffet on aspect ratios of up to $AR=3$. Three-dimensional buffet cells were identified with high-fidelity (ILES) methods for the first time. A Spectral Proper Orthogonal Decomposition (SPOD) method was applied to the data, identifying both 2D shock-oscillation and 3D buffet-cell modes. For the present work on the influence of parametric airfoil geometries on buffet, an aspect ratio of $AR=0.1$ is used as this was shown to provide domain-independent representations of trailing-edge separation for tripped cases in \cite{LSH2024_narrow_buffet}. Further details of the open-source airfoil code used to perform the tripped cases in this work are given in \cite{OpenSBLI_V3_CPC}.

To obtain turbulent conditions for this study, boundary-layer forcing is applied to trip the initially laminar boundary layer. The forcing is applied to the wall-normal velocity component near the leading edge, which is then used to set the momentum and total energy on the wall. Away from the forcing strip, the wall is a standard isothermal no-slip viscous boundary condition. The forcing is taken to be a modified form of the one given in \cite{Moise2023} as
\begin{equation}\label{eq:tripping_eqn}
\rho v_w=\sum_{i=1}^3 A \exp \left(-\frac{\left(x-x_t\right)^2}{2 \sigma^2}\right) \sin \left(\frac{k_i z}{0.05c}\right) \sin \left(\omega_i t+\Phi_i\right),
\end{equation}
for simulation time $t$, trip amplitude $A=0.075$, trip location $x_t=0.07$, and Gaussian scaling factor $\sigma=0.00833$. The three modes $\left(0, 1, 2\right)$ have spatial wavenumbers of $k_i = \left(6\pi, 8\pi, 8\pi\right)$, with phases $\Phi_i = \left(0, \pi, -\pi/2\right)$, and temporal frequencies of $\omega_i = \left(26, 88, 200\right)$. The effect of varied trip strength $A$, was investigated in \cite{LSH2024_narrow_buffet}. Laminar, transitional, and fully-turbulent buffet interactions were identified based on the state of the boundary-layer upstream of the SBLI. The $A=0.075$ level of tripping used in this study was shown in \cite{LSH2024_narrow_buffet} to produce fully-turbulent buffet conditions as required. The other flow conditions for Reynolds, freestream Mach, and Prandtl numbers are taken to be $Re=500,000$, $M_\infty = 0.73$, and $Pr=0.71$. Sutherland's law for dynamic viscosity is used with a reference freestream temperature of $T_{\infty} = 273.15K$.

\subsection{Grid generation} \label{sec:methodsGrid}
The grids used in the present study have been generated using an open-source parametric grid-generation tool for high-fidelity grids called PolyGridWizZ and has been developed at the University of Southampton \citep{polygridwizz, Zauner2018b}. We use $6^{th}$-order polynomials to define the geometry of grid lines and corresponding grid-point distributions. Such body-fitted grids are not necessarily limited to a single geometry, as grid lines are designed with respect to the position (coordinates) and local curvature with respect to the airfoil contour. Therefore, it is possible to use grids with very similar characteristics and resolutions for different airfoil geometries.
For free-transitional cases, grids previously used for simulations of Dassault Aviation's V2C profile and ONERA's OALT25 airfoil and already assessed in terms of spatial resolution \citep{Zauner2018b, Zauner2018c} were adapted to the OpenBuffet parametric airfoil geometries. Tripped cases require finer grids to resolve the tripping location and breakdown to turbulence adequately. 
Grids for free-transitional and tripped cases consist of $75,\!350,\!600$ and $346,\!720,\!500$ points in total, respectively. Both grids consist of one C-type block and two H-type blocks, with details about resolutions shown in table \ref{tab:grid_details}. The number of grid points in circumferential and radial directions of the C-type block containing the airfoil are denoted by $N_{\xi}$ and $N_{\eta}$, respectively. The two H-type blocks containing the wake region below and above the trailing edge of the airfoil are resolved by $N_{x}$ and $N_{y}$ grid points in freestream and perpendicular direction, respectively.
\begin{table}[h]
    \begin{center}
    \def~{\hphantom{0}}
    \begin{tabular}{c|c|cc|cc|cc}
        Transition type & Total grid count  & \multicolumn{2}{c|}{C-block} & \multicolumn{2}{c|}{H-block} & $L_z$ & $N_z$ \\
         & & $N_{\xi}$ & $N_{\eta}$ & $N_{x}$ & $N_{y}$\\
        \hline
        Free-transitional & $75,\!350,\!600$ & $1495$ & $480$ & $799$ & $494$ & $0.05$ & $50$\\
        \hline
        Tripped & $346,\!720,\!500$ & $2251$ & $630$ & $709$ & $630$ & $0.1$ & $150$\\
        \hline
    \end{tabular}
    \end{center}
    \caption{Grid resolutions of free-transitional and tripped LES grids.}
    \label{tab:grid_details}
\end{table}

\subsection{OpenBuffet parametric airfoil geometries (``OB")} \label{sec:methodsGeometry}
The OpenBuffet parametric airfoil geometries are composed of elliptical shapes connecting the leading edge with local extreme points on both the upper and lower sides of the airfoil. These specific locations, indicative of crests, will be henceforth referred to as `crest points'. %As demonstrated later for certain airfoils (\textit{e.g.} the OB-3663 geometry), the lower-side crest point, typically indicating a local minimum of the y-coordinate (convex crest), can transition into a local maximum (concave shape) after surpassing a critical value. 
To connect the crest points with the trailing edge, we use $6^{th}$-order polynomials. 

\begin{figure}[hbt!]
\centering
	\includegraphics[width=1.00\textwidth]{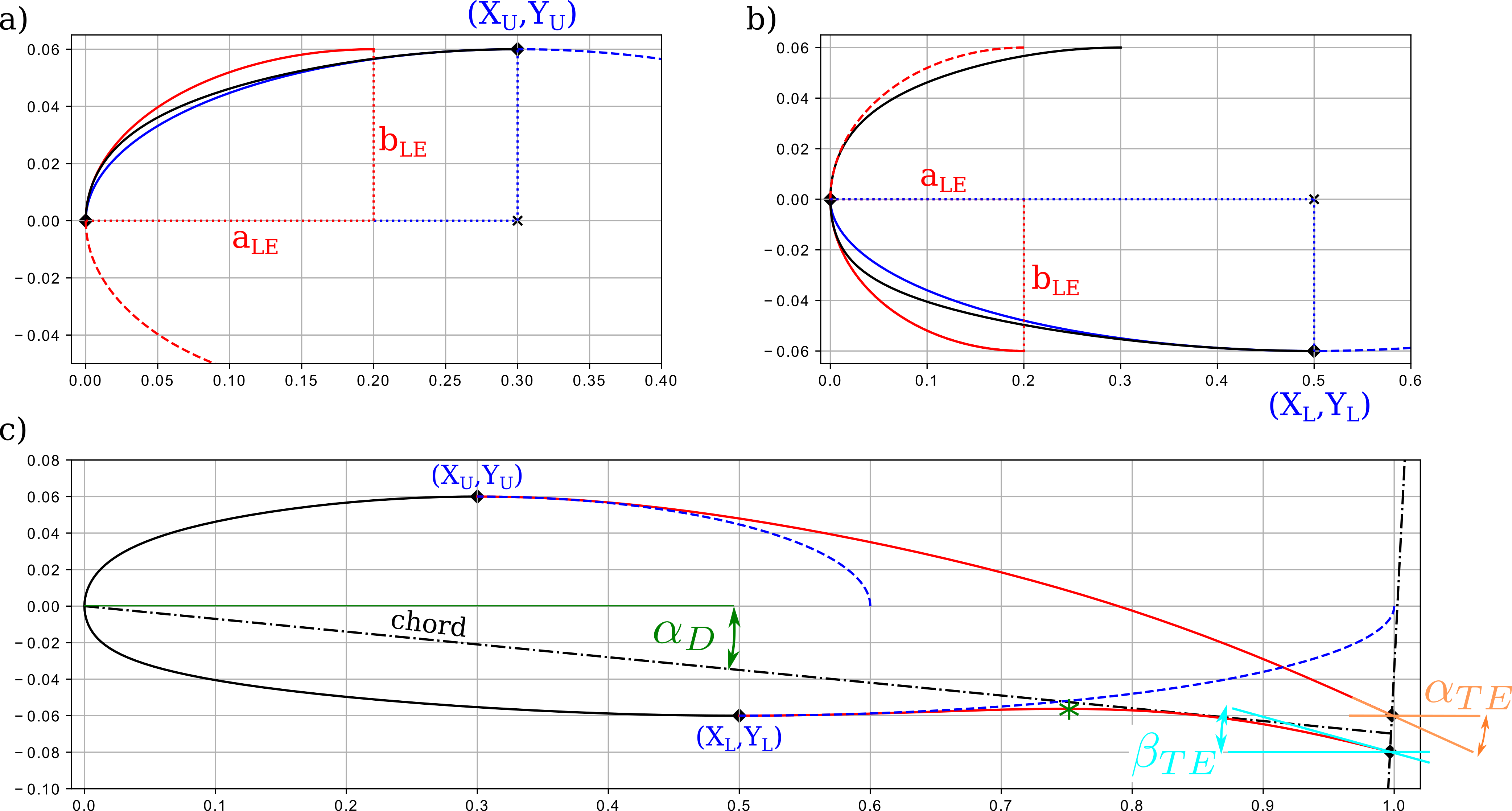}
\caption{Sketch of an OB-$4$-$3656$-$26$-$1p81$-$5$ profile: (a) and (b) show the construction of the leading edge, while (c) shows (in red) the polynomials connecting to the trailing edge and (in blue dashed lines) the semi-ellipses used in the construction. \label{fig:sketch}}
\end{figure}

The shape of an ellipse is, generally speaking, more intuitive compared to high-order polynomials and is defined by only two parameters denoted by their semi-axes according to
\begin{align}
    x_{e}&=a_{e} \cos(\psi)+x_0, \label{eqn:general_ellipse_x} \\ 
    y_{e}&=b_{e} \sin(\psi), \label{eqn:general_ellipse_y}
\end{align}
where $\psi$ is linearly discretised by $N_{\psi}$ points ranging from $\pi \geq \psi \geq\pi/2$. In the present context, ellipses are always centered around $y_0=0$, while the $x$-location of the center ($x_0$) can vary. 
The discretisation of the profile (\textit{i.e.} $N_{\psi}$) needs to be sufficiently higher than the wall resolution of the final CFD grid. For the present cases, where the entire airfoil surfaces is resolved by $N_{xi}\approx 1,\!500$ points in the circumferential direction using a minimum spacing of $\Delta s=0.0003$, we chose $100,\!000$ points to describe the raw geometry (High resolutions are favourable for accurate computation of gradients and interpolation of polynomial grid-distribution functions).
The coordinates of the crest points (\textit{i.e.} the endpoints of the ellipses) are defined at a design angle of attack $\alpha_D$. This allows us to directly parameterise the crest points at an angle of attack of interest (\textit{e.g.}, where buffet occurs), without accounting for rotation transforms.
The full name-tag of OpenBuffet parametric airfoils is composed of 10 parameters according to the second column of table \ref{tab:parameters}:
OB-$\alpha_D$-$D_1 D_2 D_3 D_4$-$a_{LE} b_{LE}$-$\alpha_{TE} \beta_{TE}$-$\Delta_{TE}$ (if applicable, the decimal point of parameters is replaced by the letter `p').We will briefly introduce the parameters of the name tag referring to figure \ref{fig:sketch}(c), while remaining details of this figure will follow below. The design angle $\alpha_D$ is marked green and denotes the rotation of the chord, while the upper and lower crest points defined by digits $D_1-D_4$ are labeled by their respective coordinates ($X_U$,$Y_U$) and ($X_L$,$Y_L$) in blue. Angles $\alpha_{TE}$ and $\beta_{TE}$, defining the angles at the corner of the blunt trailing edge are labeled orange and cyan, respectively. The trailing-edge thickness $\Delta_{TE}$ denotes the distance between upper and lower trailing-edge corners.

The construction of OpenBuffet airfoil geometries is done in four main steps, with figure \ref{fig:sketch} providing visual aids for clarity. The aspect ratio between $x$ and $y$ coordinates is intentionally distorted in the figure to emphasize geometric features.
The $(x,y)$ coordinates for the leading edge, ($0,0$), upper and lower crest, ($X_U$,$Y_U$) and ($X_L$,$Y_L$), respectively, are defined \textit{a priori} for a given airfoil. The first four digits of the airfoil, $D_1$ to $D_4$, can be used to calculate ($X_U$,$Y_U$) and ($X_L$,$Y_L$), as given in table~\ref{tab:parameters}. Strictly speaking, there are two crest points denoting local extrema on the lower side. The crest point, we control is usually the convex one (denoted by the black diamond symbol) upstream of the inflection point, where the second derivative switches sign. The second concave crest point is denoted by the green symbol. Increasing $Y_L$, the inflection point moves upstream and coincides at a certain value with the crest point we are controlling. When further increasing $Y_L$, the inflection point moves further upstream and, eventually, we control the concave crest. This is, for example, the case for the OB-4-3663 airfoil.
%The coordinate of the blunt trailing edge's mid-point, which terminates the chord line can be calculated by a rotational transformation of a straight line by the design angle of attack, \textit{i.e.}, $(\cos\alpha_D,-\sin\alpha_D)$. The upper and lower end points of the blunt trailing edge can then be computed assuming that they are equidistant from the mid-point and orthogonal to the chord line. Based on these coordinates, curves can be fitted so as to connect them and form the airfoil profile. 
The curves connecting the leading edge and the crest points (\textit{i.e.} the fore part of the airfoil) are determined by three ellipses. The first ellipse (shown in red in figure \ref{fig:sketch}(a) and (b)) is used to define the leading edge curvature and is common for both the upper and lower surfaces according to 
\begin{align}
    (x-a_{LE})^2/a_{LE}^2+y^2/b_{LE}^2=1.
\end{align}
The other two ellipses are used to match the coordinates of the upper and lower crest points (shown in blue in figure \ref{fig:sketch}(a) and (b)) and can be expressed as:
\begin{align}
    (x-X_U)^2/X_U^2+y^2/Y_U^2=1 \\
    (x-X_L)^2/X_L^2+y^2/Y_L^2=1
\end{align}

To ensure a smooth transition from the first ellipse to the other two on each side, a blending function is used. This blending function, $f(i)$ with $1 \le i \le N_{\psi}$, ranges from $f(1)=1$ to $f(N_{\psi})=0$ and can be defined by a $6^{th}$-order polynomial
\begin{align}
   f(i)=Ai^6 + Bi^5 + Ci^4 + Di^3 + Ei^2 + Fi + G.
\end{align}
Higher derivatives at the bounds are set to zero ($f'(1)=f''(1)=0$ and $f'(N_{\psi})=f''(N_{\psi})=f'''(N_{\psi})=0$). The additional third derivative at the starting point (leading edge) delays the transition from the leading-edge ellipse to the ellipse defined by the crest point. 
The smooth transition from the red ellipse at the leading edge to the blue one at the crest point is demonstrated in figures \ref{fig:sketch}(a), where the resulting blended black curve is discretised according to:
\begin{align}
    x_0(i)&=X_U-f(i)(X_U-a_{LE}) \\
    a_{e}(i)&=X_U-f(i)(X_U-a_{LE}) \\
    b_{e}(i)&=Y_U-f(i)(Y_U-b_{LE}),
\end{align}
which can then be substituted into equations \ref{eqn:general_ellipse_x} and \ref{eqn:general_ellipse_y} to obtain corresponding coordinates.
Note that semi-axes as well as origin of the blended ellipse are now functions of $\psi$ (discretised by $i$), where $i=1$ recovers the leading-edge ellipse, while $i=N_{\psi}$ corresponds to the ellipse spanned by the crest point.
The same procedure is performed on the lower side as shown in figure \ref{fig:sketch}(b). 
Before connecting the crest points with the corners of the trailing edge, we rotate the chord by the design angle $\alpha_D$, \textit{i.e.}, $(\cos\alpha_D,-\sin\alpha_D)$. The rotated chord is depicted by a black dash-dotted curve in figure \ref{fig:sketch}(c), where the corners of the trailing edge are located (orthogonal to the rotated chord) at a distance of $\pm \Delta_{TE}/2$. Having determined the starting and ending points, we can now define $6^{th}$-order polynomials $P(i)$ (with $1<i<=N_{\psi}$) to connect the crest points with the corners at the trailing edge.
The derivatives at the crest points need to match those of the ellipse (\textit{i.e.} $P'(1)=0$ and $P''(1)=-b_{e}(N_{\psi})/(a_{e}(N_{\psi})^2)$). At the trailing edge, first-order derivatives are given by trailing-edge and boat-tail angles $\alpha_{TE}$ and $\beta_{TE}$ (with respect to the $x$ axis), respectively. Both angles are shown in figure  \ref{fig:sketch}(c) and marked orange and cyan, respectively. Second- and third-order derivatives are set to zero. Polynomials for the upper and lower sides are denoted by red curves for the example in figure \ref{fig:sketch}(c).
After having obtained the full geometry at the design angle, we can rotate the coordinates back to zero degrees to obtain coordinates relative to the chord line in the usual convention.

For the 4-digit OpenBuffet airfoil series, we will limit the parameter space for a given design angle to four digits ($D_1$, $D_2$, $D_3$ and $D_4$): OB-$\alpha$-$D_1 D_2 D_3 D_4$. The semi-axes defining the leading-edge curvature are fixed as the coordinates of the upper crest (\textit{i.e.} $a_{e}=X_U$ and $b_{e}=Y_U$). Thus, the leading edge-rounding on the upper and lower sides of the airfoil will change with the location of the upper crest points. While the forepart of the upper side is now defined by a single ellipse, there is still blending of two ellipses required on the pressure side. Parameters defining the trailing edge are similar to V2C and OALT25 geometries.

The baseline geometry for free-transitional cases of the present contribution corresponds to an OB-4-3666 profile, while we consider an OB-3p5-2p8'5p7'3p8'7p8 profile for tripped cases. The full set of parameters are listed in table \ref{tab:parameters_2}. 
The latter OpenBuffet airfoil is defined by several parameters containing decimal numbers to obtain a geometry similar to ONERA's OAT15A profile, which ensures buffet at the given flow conditions. For clarity, we use the prime symbol ' to separate parameters $D_1-D_4$ when we encounter multiple decimal number. 
We will also use name-tags for each test case in the parametric study, specifying only the parameters that differ from the reference geometry, instead of the lengthy airfoil definitions.
\begin{table}
\centering
\begin{tabular}{lcccccccccc}
Boundary layer & Angle of Attack & $X_U$ & $Y_U$ & $X_L$ & $Y_L$ & $a_{LE}$ & $b_{LE}$ & $\alpha_{TE}$ & $\beta_{TE}$ & $\Delta_{TE}$\\
\hline
Free-trans. & $4^{\circ}$   & 0.30   & 0.060 & 0.60  & -0.060  & $X_U$ & $Y_U$ & $18^{\circ}$ & $10^{\circ}$ & 0.005 \\
Tripped     & $3.5^{\circ}$ & 0.28   & 0.057 & 0.38 & -0.078 & $X_U$ & $Y_U$ & $17^{\circ}$ & $10^{\circ}$ & 0.000\\
\hline \\
\end{tabular}
\caption{Parameters for baseline case with tripped and free-transitional boundary layers.}
\label{tab:parameters_2}
\end{table}
% \begin{figure}[hbt!]
% \centering
% 	\includegraphics[width=1.00\textwidth]{Figures_verification/comp.png}
% \caption{Comparison of different airfoil geometries. \label{fig:airfoil_comp}}
% \end{figure}
Regarding Dassault Aviation's V2C geometry mentioned in the introduction, the geometric features of the OB-4-3669 profile appear to be the most qualitatively similar. These geometries are shown in figure \ref{fig:airfoil_comp} together with ONERA's OALT25 and the NACA0012 airfoils.

As already mentioned before, we consider in addition to the original OpenBuffet geometries (described in \S \ref{sec:methodsGeometry}) a slight variation of the polynomial connecting crest points with the trailing edge. This leads to a flatter suction-side surface and increased robustness for extreme parameter choices as detailed in \S \ref{sec:modified_profiles}.

% Here comes now the restructured bit
\section{Buffet Sensitivities to the Upper-Side Crest Position for OpenBuffet Airfoils}\label{sec:OpenBuffet_Upper}

As discussed in the introduction, based on literature the upper-side geometry may has strong influence on buffet frequencies. %While buffet frequencies agree well for OALT25 and OAT15A profiles\cite{Brion2019} with fairly similar upper crest locations\footnote{Crest coordinates for OALT25 and OAT15A profiles are $X_U/Y_U=0.26/0.044$ at $\alpha={4^{\circ}}$ and $0.25/0.048$ at $\alpha={3.5^{\circ}}$, respectively.}, we observe significant differences in buffet and geometric properties for the V2C profile (see table \ref{tab:airfoils}). %, which differs not only in terms of crest positions, but also thickness.
Therefore, we want to investigate in the present section sensitivities of transonic buffet to the location (i.e. coordinates) of the upper crest point denoted by $X_U$ and $Y_U$.  
In a first part we will assess test cases subjected to free-transitional boundary layers, using parametric OpenBuffet profiles that are similar to laminar-flow wing profiles like Dassault Aviation's V2C profile. At such conditions, we typically observe full break-down to turbulence in vicinity of the shock location and relatively large shock-induced separation bubbles.
In the second part we will then apply boundary-layer tripping to OpenBuffet profiles, which are similar to conventional supercritical airfoils like ONERA's OAT15A geometry. %For these cases, we typically observe much smaller separation bubbles.

\subsection{Free-transitional boundary layers}\label{sec:OpenBuffet_Upper_Free}

In this sub-section, we will study trends for aerodynamic properties and buffet characteristics for different airfoil shapes considering free-transitional boundary layers. 
Flow conditions are similar to previous studies of \citet{Moise2021} and \citet{Zauner2023b} corresponding to $\alpha=4^{\circ}$, $M=0.7$, and $Re=500,\!000$, where grid resolution and domain size requirements have been well established. 
Table \ref{tab:parameters_free} provides an overview of four-digit OpenBuffet geometries considered for this study. This airfoil family is representative of supercritical laminar-flow airfoils. 
For brevity, labels for test cases are composed of the profile name (`OB' for OpenBuffet), the transition mode (`F' for free-transitional), the digit (`D1' or `D2') which differs from the reference case (OB-F-Ref), and the value of the latter.
\begin{table}
\centering
\begin{tabular}{lcccccccccc}
Label & Angle of Attack & $X_U$ & $Y_U$ & $X_L$ & $Y_L$ & $a_{LE}$ & $b_{LE}$ & $\alpha_{TE}$ & $\beta_{TE}$ & $\Delta_{TE}$\\
\hline
OB-F-Ref & $4^{\circ}$   & 0.30   & 0.060 & 0.60  & -0.060  & $X_U$ & $Y_U$ & $18^{\circ}$ & $10^{\circ}$ & 0.005 \\
\hline
OB-F-D1-3p5 & $4^{\circ}$   & \textbf{0.35}   & 0.060 & 0.60  & -0.060  & $X_U$ & $Y_U$ & $18^{\circ}$ & $10^{\circ}$ & 0.005 \\
OB-F-D1-4 & $4^{\circ}$     & \textbf{0.40}   & 0.060 & 0.60  & -0.060  & $X_U$ & $Y_U$ & $18^{\circ}$ & $10^{\circ}$ & 0.005 \\
\hline
OB-F-D2-4p5 & $4^{\circ}$   & 0.30   & \textbf{0.045} & 0.60  & -0.060  & $X_U$ & $Y_U$ & $18^{\circ}$ & $10^{\circ}$ & 0.005 \\
OB-F-D2-4 & $4^{\circ}$     & 0.30   & \textbf{0.040} & 0.60  & -0.060  & $X_U$ & $Y_U$ & $18^{\circ}$ & $10^{\circ}$ & 0.005 \\
\hline \\
\end{tabular}
\caption{Summary and labels for test cases considering free-transitional boundary layers.}
\label{tab:parameters_free}
\end{table}

\subsubsection{Reference case OB-4-3666}\label{sec:reference}

% As a reference test case we select the four-digit OB-4-3666 profile, which is representative of laminar-flow airfoils. %We will adopt flow conditions similar to previous studies\cite{Moise2021,Zauner2023b} considering free-transitional boundary layers.
\begin{figure}[hbt!]
\centering
  \begin{tabular}{ll}
    a) & b) \\
    \includegraphics[width=0.45\textwidth,trim={30mm 10mm 80mm 40mm},clip]{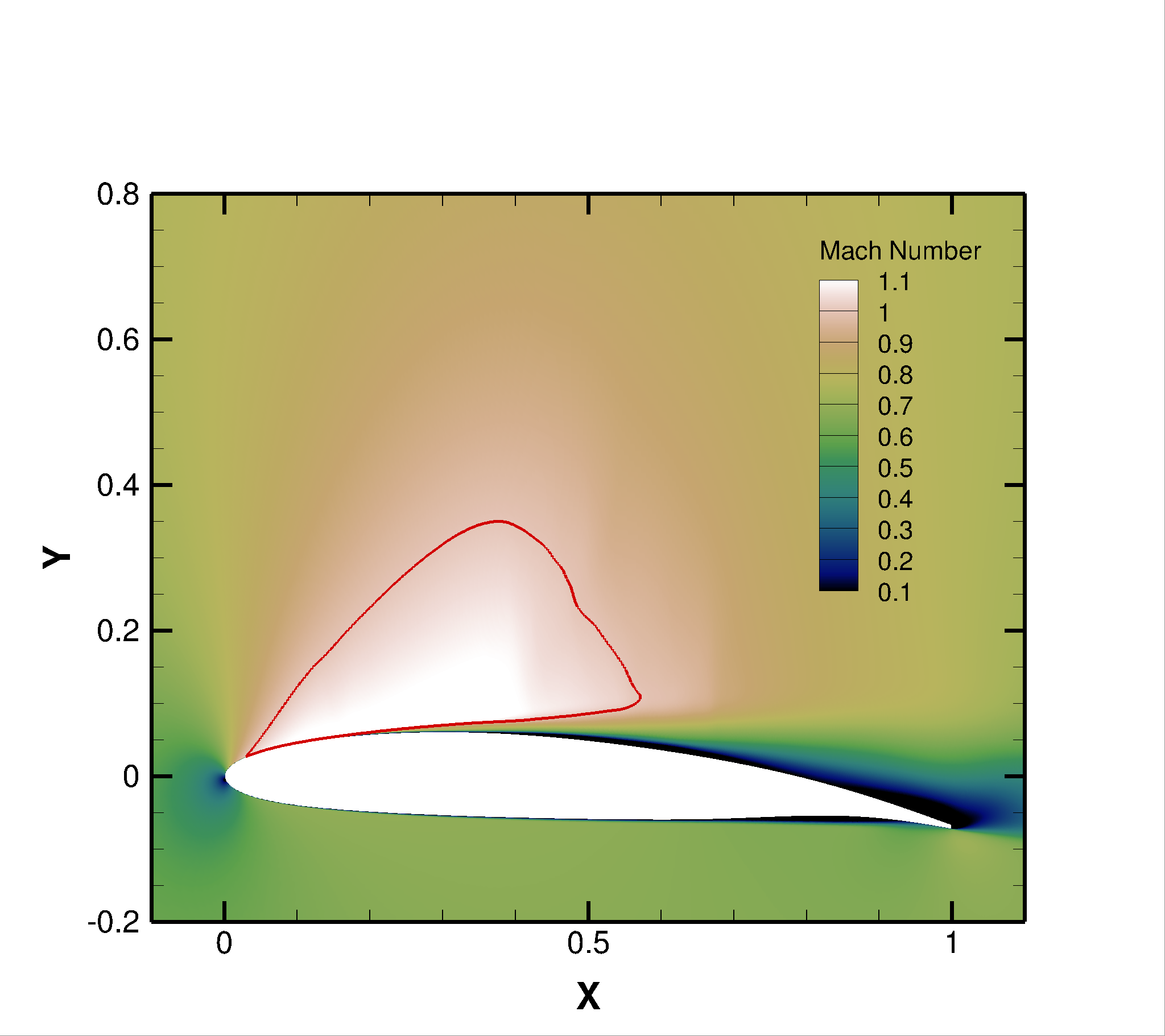} &
    \includegraphics[width=0.45\textwidth,trim={10mm 10mm 10mm 10mm},clip]{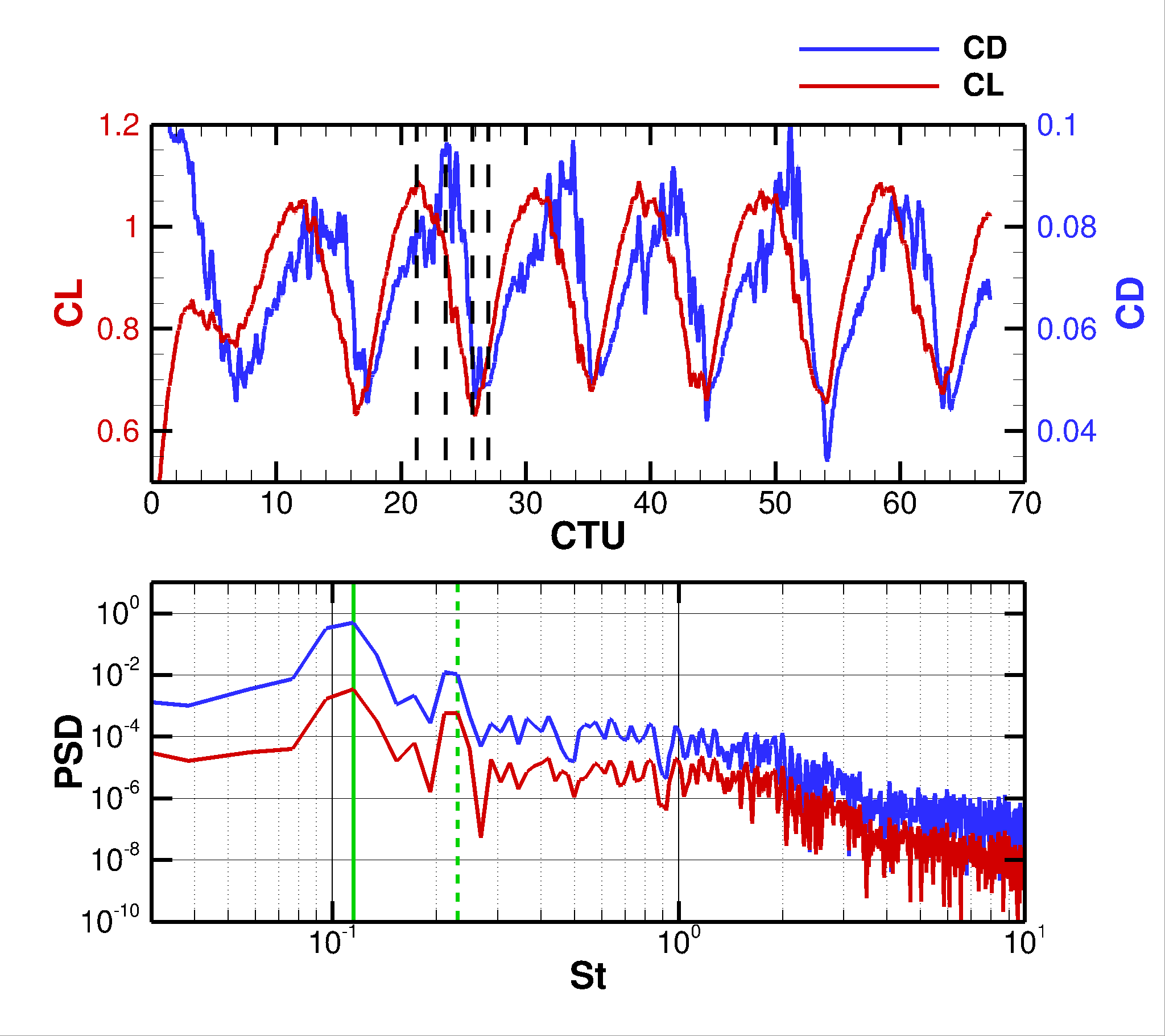} \\
  \end{tabular}
\caption{(a) Time- and span-averaged flow field showing Mach number contours for the OB-4-3666 airfoil at $\alpha=4^{\circ}$, considering free-transitional boundary layers, $M=0.7$, and $Re=500,\!000$. (b) Histories and spectra of lift (red) and drag coefficients (blue). Dashed lines mark time instances of snapshots in figure \ref{fig:snap}.\label{fig:ref}}
\end{figure}
As a reference test case we select the four-digit OB-4-3666 profile at $\alpha=4^{\circ}$, $M=0.7$ and $Re=500,\!000$. Figure \ref{fig:ref}(a) shows the time- and span-averaged mean flow around this airfoil with free-transitional boundary layers. The red sonic line delineates the supersonic region with local Mach numbers larger than unity. 
The top plot of figure \ref{fig:ref}(b) shows respectively histories of lift and drag coefficients $C_L$ and $C_D$. 
We observe large-scale oscillations associated with transonic buffet, where the lift coefficient (red curve: left-hand-side scale) leads the drag coefficient (blue curve: right-hand-side scale). Spectra of the bottom plot of figure \ref{fig:ref}(b) are computed using Fast Fourier Transform (FFT) considering a single Hanning window and ignoring the initial transient of the first 15 convective time units (CTU), unless otherwise stated. The spectral resolution for the present time series corresponds to $\Delta St \approx 0.019$.
We observe clear peaks for the buffet frequency corresponding to a Strouhal number of $St_B=0.115$ and its harmonic at $St=0.23$, which are respectively indicated by solid and dashed green lines. 
We will see later for different geometries where buffet is less dominant a secondary spectral peak arising at intermediate frequencies around $St\approx0.5$ corresponding to a period of approximately $2$ CTUs. Even though we do not observe such a clear peak in spectra of figure \ref{fig:ref}(b) for the baseline case, we can notice corresponding intermittent spikes in the $C_D$ history during phases, where the drag is increased due to flow-separation phenomena (\textit{e.g.} after 34 CTUs).
To characterise buffet, we can measure maximum and minimum vales of the lift coefficient ($C_{L,max}$ and $C_{L,min}$, respectively) as well as a maximum amplitude ($\Delta C_{L,max}=C_{L,max}-C_{L,min}$). For the reference case we obtain $\Delta C_{L,max}=0.46$, $C_{L,max}=1.09$, $C_{L,min}=0.63$. We will use these as key parameters of interest for our parametric studies to characterise transonic buffet and its intensity.

\begin{figure}[hbt!]
\centering
  \begin{tabular}{ll}
    a) & b) \\
    \includegraphics[width=0.45\textwidth,trim={10mm 10mm 10mm 10mm},clip]{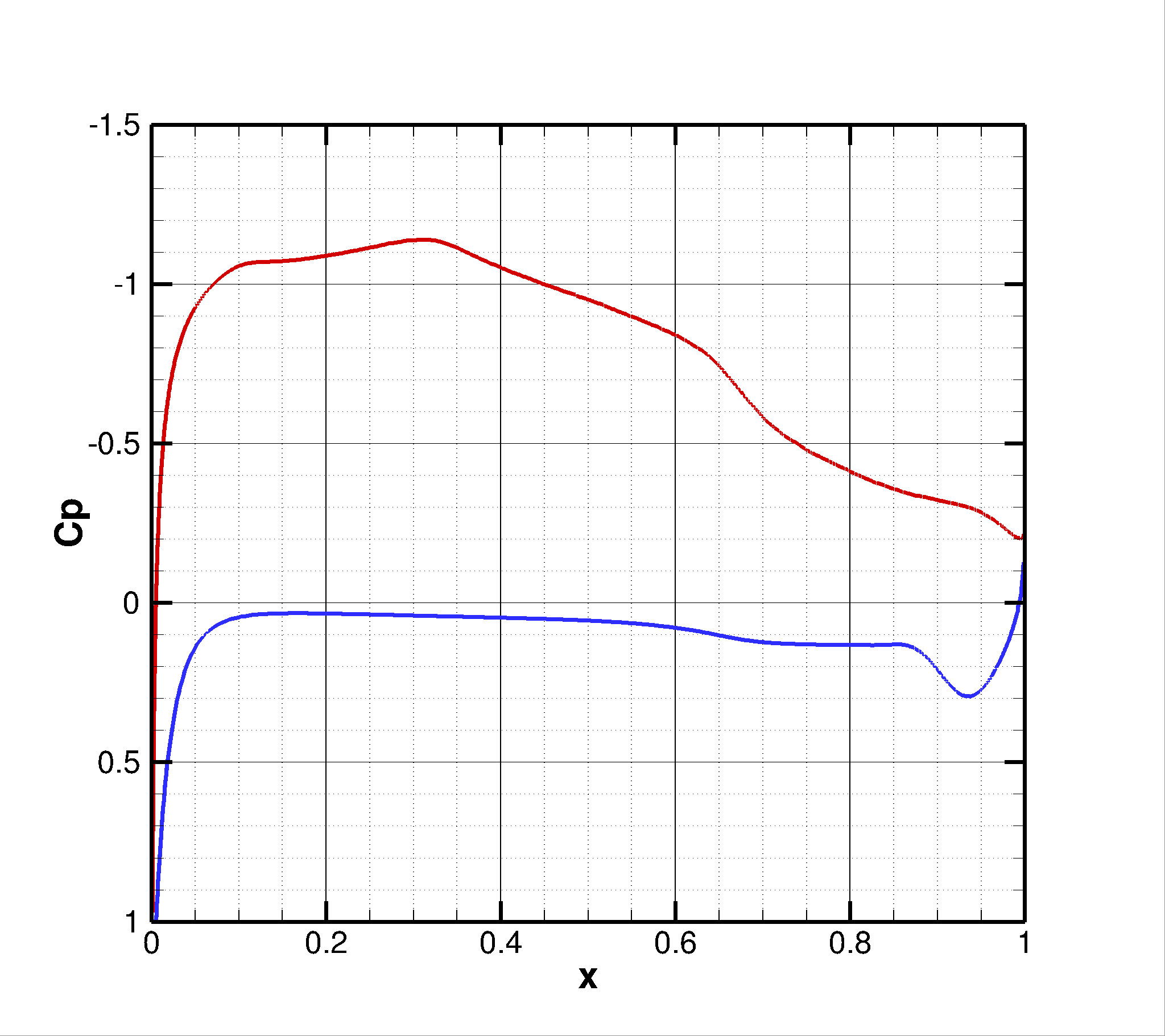} &
    \includegraphics[width=0.45\textwidth,trim={10mm 10mm 10mm 10mm},clip]{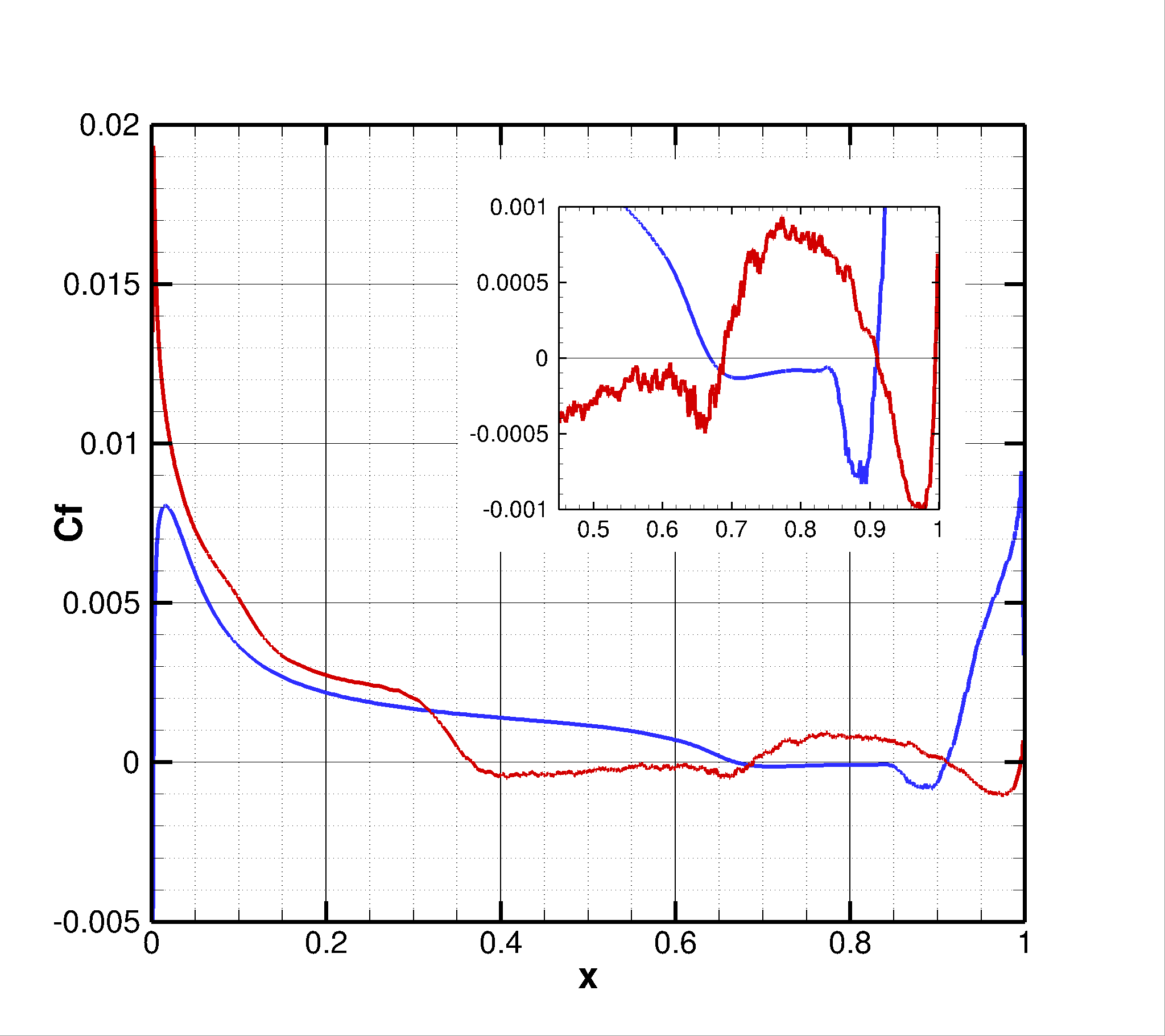} \\
  \end{tabular}
\caption{Time-averaged (a) wall-pressure coefficient $C_p$ and (b) skin-friction coefficient $C_f$ distributions for the OB-4-3666 profile with free-transitional boundary-layers. Blue and red curves correspond to pressure and suction side distributions, respectively.\label{fig:CpCf_free}}
\end{figure}
Figure \ref{fig:CpCf_free} shows time averaged (a) wall-pressure coefficient $C_p$ and (b) skin-friction coefficient $C_f$ distributions for the baseline profile for the free-transitional cases (underlying simulation data is sampled at time intervals of $0.035$ CTUs at a 2D slice located at $z=0$). The $C_p$ reaches its maximum around $x\approx0.3$. Due to the back and forth motion of the shock wave, the sharp pressure gradient across the shock wave appears as a smooth slope within the range of $0.30<x<0.65$ in the time-averaged flow field.
Regions of $C_f<0$ in figure \ref{fig:CpCf_free}(b) indicate laminar/transitional separation bubbles on the suction side in the range of $0.35<x<0.70$. While the mean flow re-attaches at $x\approx0.7$, a secondary (turbulent) separation zone is observed near the suction-side trailing edge. While the mean flow on the pressure side remains attached near the trailing edge, we can observe a transitional separation bubble in the concave part around $0.7<x<0.9$. The inset of figure \ref{fig:CpCf_free}(b) provides a close-up of the re-attachment points for both sides.

\begin{figure}[hbt!]
\centering
  \begin{tabular}{ll}
    a) & b) \\
    \includegraphics[width=0.45\textwidth,trim={0mm 0mm 0mm 0mm},clip]{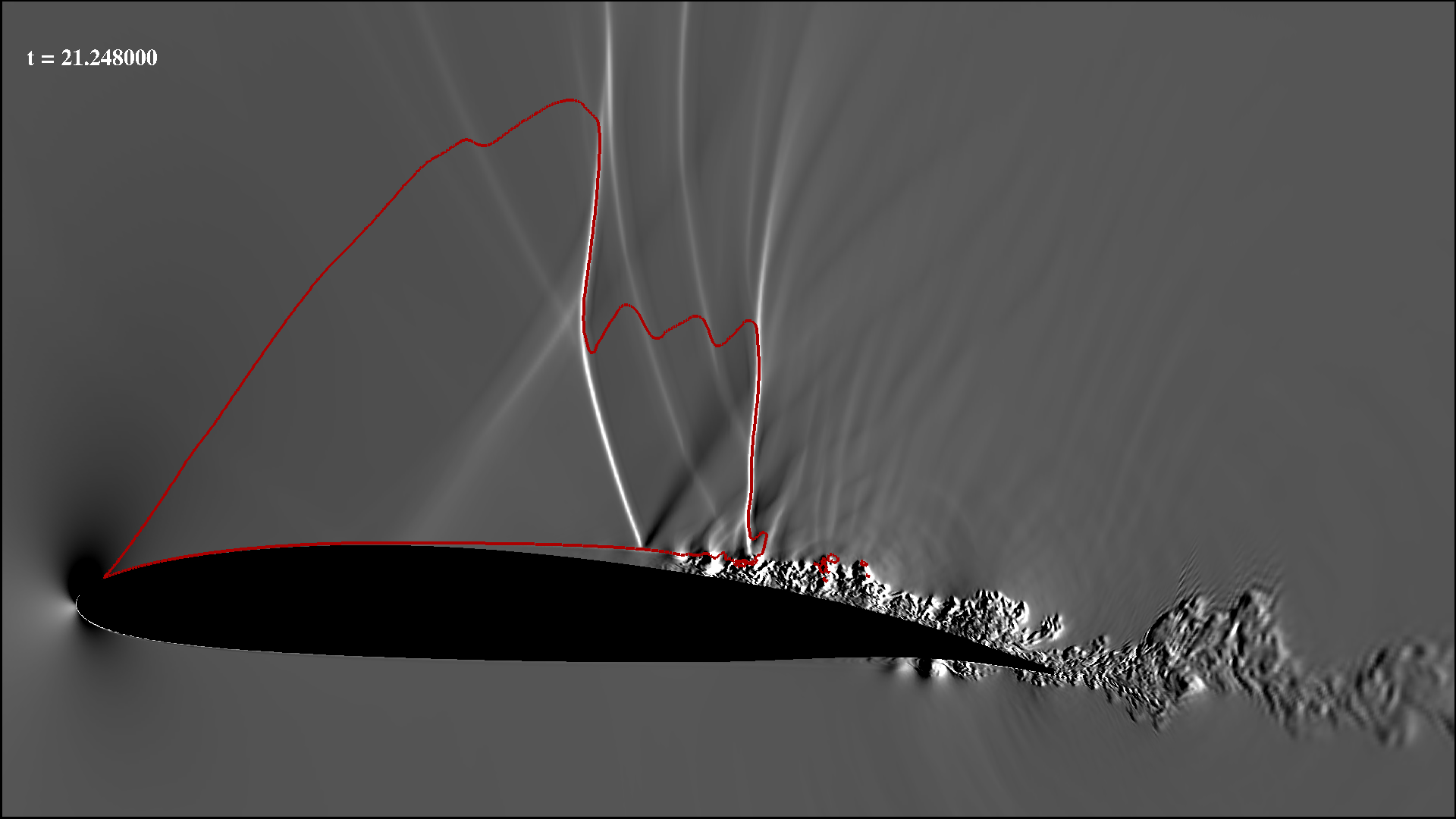} &
    \includegraphics[width=0.45\textwidth,trim={0mm 0mm 0mm 0mm},clip]{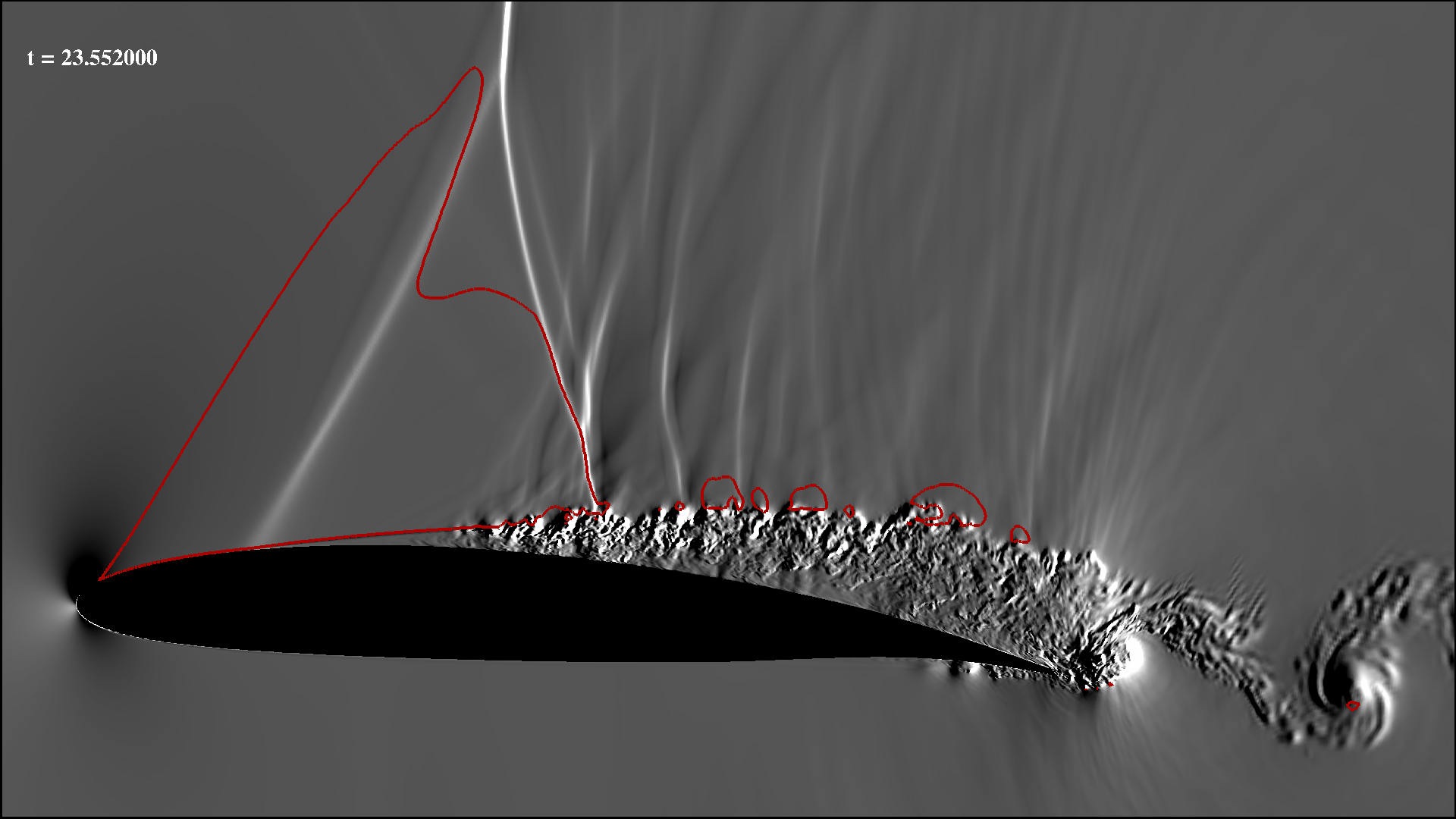} \\
    c) & d) \\
    \includegraphics[width=0.45\textwidth,trim={0mm 0mm 0mm 0mm},clip]{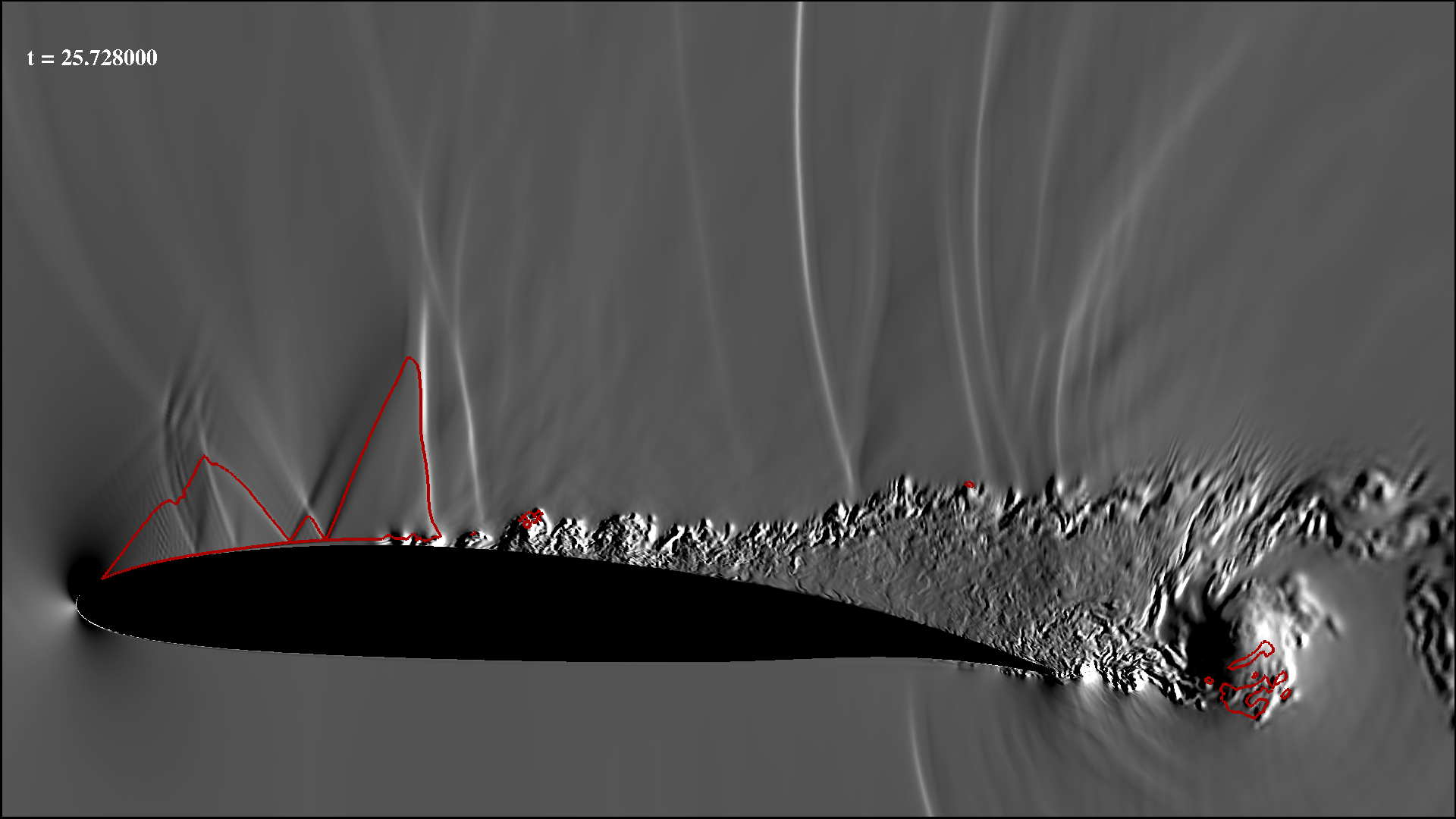} &
    \includegraphics[width=0.45\textwidth,trim={0mm 0mm 0mm 0mm},clip]{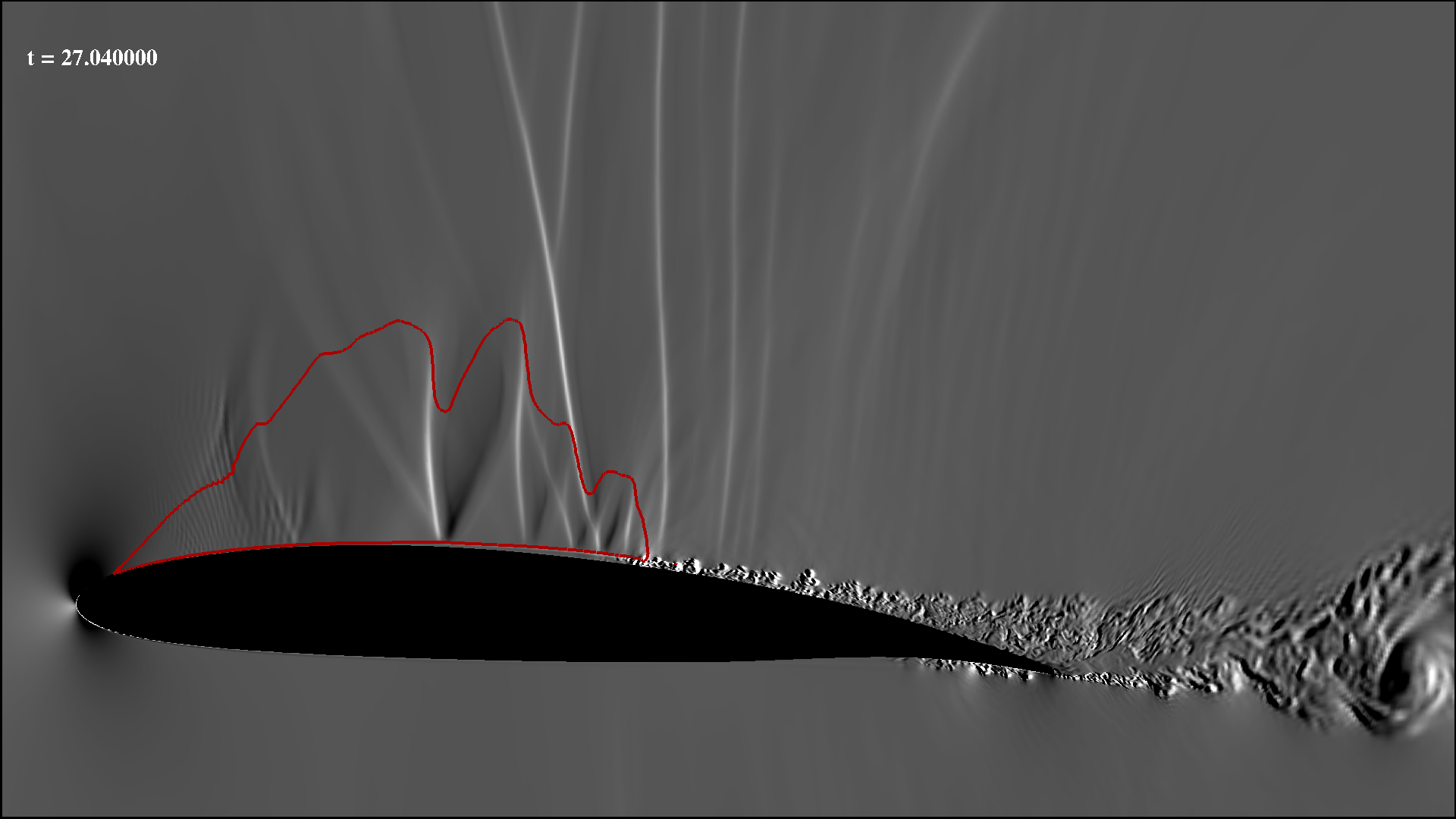} \\
  \end{tabular}
\caption{Numerical quasi-schlieren snapshots showing streamwise density gradients after approximately (a) 21 CTUs (Convective time units), (b) 24 CTUs, (c) 26 CTUs, and (d) 27 CTUs, corresponding to vertical red lines in figure \ref{fig:ref}. Black and white contours correspond to non-dimensional values of $\partial \rho / \partial x = -5$ and $10$, respectively. Sonic lines are shown in red. \label{fig:snap}}
\end{figure}
\begin{table}
\centering
\begin{tabular}{cccccc}
% \hline
Snapshot & time instant & $C_L$ & $C_{D,p}$ & $C_{D,f}$ & $C_L/C_D$ \\
\hline
a) & 21.25 & high       & increased & medium    & increased \\
b) & 23.55 & increased  & high      & low       & low  \\
c) & 25.73 & low        & low       & increased & reduced \\
d) & 27.04 & reduced    & low       & high      & high \\
%
% a) & 21.25 & $100\%$    & $67\%$    & $61\%$    & $67\%$ \\
% b) & 23.55 & $71\%$     & $100\%$   & $0\%$     & $0\%$ \\
% c) & 25.73 & $0\%$      & $13\%$    & $57\%$    & $30\%$ \\
% d) & 27.04 & $24\%$     & $0\%$     & $100\%$   & $100\%$ \\
\hline \\
\end{tabular}
\caption{Qualitative overview of aerodynamic properties for snapshots in figure \ref{fig:snap}.}
\label{tab:qual_inst}
\end{table}
Figure \ref{fig:snap} shows numerical quasi-Schlieren snapshots (streamwise density gradient contours) at representative time instants indicated by vertical black dashed lines in figure \ref{fig:ref}(b). %A corresponding movie is provided in the supplementary material. 
We will briefly outline the flow behaviour throughout one buffet cycle, while table \ref{tab:qual_inst} provides a summary of qualitative levels of aerodynamic coefficients with respect to their extrema at given time instances. Histories of corresponding coefficients are provided in figure \ref{fig:laminar_vs_tripped}(a) in appendix \S \ref{sec:appendix_lam_turb} and will not be discussed here in detail. For a more in-depth analysis of transonic buffet at moderate Reynolds numbers and free-transitional boundary layers, we refer the reader to \cite{Moise2021,Zauner2023b}. 
Looking at figure \ref{fig:snap}(a), boundary layers are mostly attached on both sides of the airfoil, when the lift reaches its maximum. Towards the trailing edge, we observe flow structures originating from Kelvin-Helmholtz instabilities in the separating shear layer \cite{Zauner2017a} and breaking down to turbulent vortices when passing through the shock wave \cite{Zauner2019c}. 
The supersonic region approaches its maximum extent and is terminated by a normal shock wave. We observe multiple upstream-propagating pressure waves within the supersonic region, which are introduced by acoustic waves circumventing this area. The present lambda-shock structure\cite{Zauner2023b} and its dynamics look qualitatively very different from tripped cases or free-transitional cases at higher Reynolds numbers, but are typical for present flow conditions \cite{Moise2021}. 
Moving towards the next snapshot, flow separation develops and spreads rapidly upstream, leading to significant flow separation and shear layers diverging from the upper airfoil geometry. 
At the time instant in figure \ref{fig:snap}(b), the pressure drag reaches its maximum as aerodynamic forces are high and as a significant portion of the upper-side boundary layer is separated. This also leads to reduced skin-friction drag and lift-over-drag ratios. The separation wave of the lambda-shock structure strengthens and moves towards the leading edge. Strong vortical structures induce supersonic pockets above the shear layer and we can also see a well-pronounced vortex street in the wake behind the airfoil\cite{Moise2021}.
Eventually, the supersonic region degenerates into localised triangular pockets near the leading edge, as seen in figure \ref{fig:snap}(c), and the lift reduces. The airfoil is basically stalled at this time instant. At such conditions with reduced aerodynamic loads, the boundary layer can start recovering behind the shock foot and the re-attachment point starts moving downstream rapidly. This leads to a sharp increase in skin-friction, while the pressure drag grows at a much lower speed.
We observe skin-friction reaching its maximum in figure \ref{fig:snap}(d), where the boundary layer is thin and fully attached. However, lift has barely increased at that time instant. While the flow is accelerating over the suction side recovering aerodynamic forces, we do not observe the formation of large Kelvin-Helmholtz structures.

The features seen here are the same as those seen for transonic buffet on the V2C, OALT25 and NACA0012 airfoils \cite{Moise2023, Zauner2023b, Moise2024} under conditions of free transition and low to moderate Reynolds numbers. Furthermore, the periodic oscillations involving boundary layer separation and reattachment are the same as has been reported in other studies on transonic buffet at higher Reynolds numbers and forced transition (\textit{e.g.} \citep{Iovnovich2012}). 
Thus, we conclude that the present parametric airfoil is capable of reproducing transonic buffet features adequately.

% conclusions
% Both $X_U$ and $Y_U$ seem to influence buffet intensity significantly
% horizontal variation
% -> Increasing $X_U$ leads to offset characteristics
% -> L/D less affected
% vertical variations 
% -> Decreasing $Y_U$ delays onset 
% -> Improves the lift-over-drag ratio significantly.

\subsubsection{Horizontal variation of the upper crest point}

\begin{figure}[h]
    \centering
    \begin{minipage}[t]{\textwidth}
        \centering
        \begin{tabular}{ll}
            a) & b) \\
            \includegraphics[width=0.45\textwidth,trim={10mm 10mm 20mm 20mm},clip]{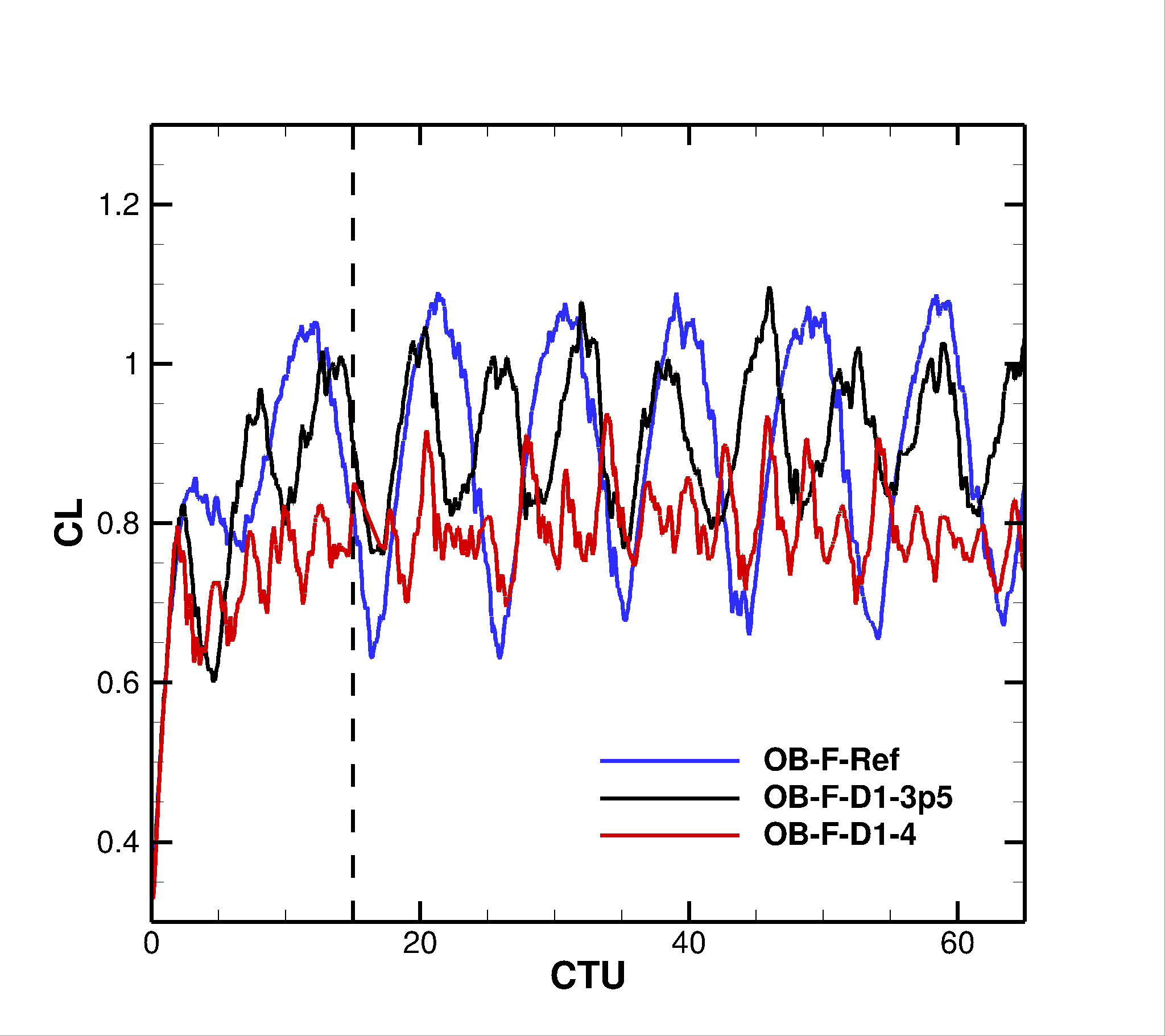} &
            \includegraphics[width=0.45\textwidth,trim={10mm 10mm 20mm 10mm},clip]{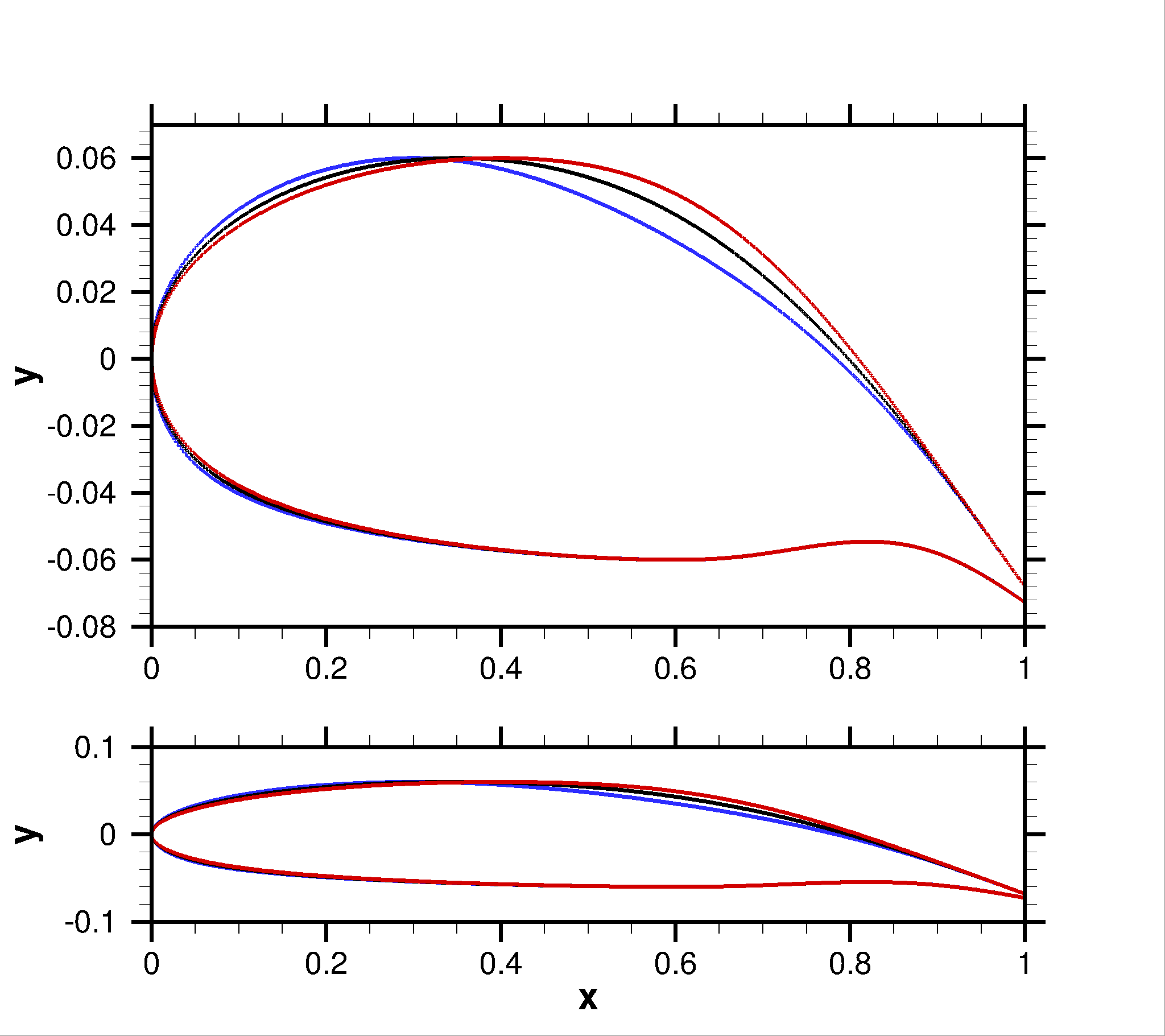} \\
        \end{tabular}
        \caption{(a) Lift coefficient ($C_L$) as a function of convective time units ($CTU$) for 4-digit airfoils at an angle of attack $\alpha=4^{\circ}$, where the first digit is varied (OB-4-X666). Corresponding airfoil contours are shown in (b), respectively, where the aspect ratio of the upper plot axes is distorted. The aspect ratio of the lower plot is x:y=1:1.  \label{fig:X666}}
    \end{minipage}
    \vspace{0.5cm} % Adjust the vertical space between the figure and table as needed
    \begin{minipage}[t]{\textwidth}
    \centering
        \begin{tabular}{lccccccccc}
        Case & $D_1$ & $St_B$ & $St_I$ & $\Delta C_{L,max}$ & $C_{L,max}$ & $C_{L,min}$ & $C_{D,max}$ & $C_{D,min}$ & $\overline{L}/\overline{D}$\\
        \hline
        OB-F-Ref    & 3 & 0.115 & - & 0.46 & 1.09 & 0.63 & 0.102 & 0.039 & 12.4 \\ %Secondary intermediate-frequency peak at St_I=0.211
        OB-F-D1-3p5  & 3.5 & 0.153 & 0.306 & 0.34 & 1.1 & 0.76 & 0.102 & 0.036 & 14.3 \\ %second freq & 0.306
        % OB-F-D1-4    & - & 0.560 & 0.24 & 0.94 & 0.70 & 0.104 & 0.048 & 11.1 \\
        %OB-F-D1-4    & - & 0.538 & 0.24 & 0.94 & 0.70 & 0.104 & 0.048 & 11.1 \\
        OB-F-D1-4    & 4 & - & 0.338 & 0.25 & 0.94 & 0.69 & 0.104 & 0.048 & 11.1 \\ % longer statistics
        \hline \\
        \end{tabular}
    \captionof{table}{Summary of buffet characteristics for free-tansitional cases varying the first digit of the reference case OB-F-Ref considering the OB-4-3666 profile.}\label{tab:X666}
    \end{minipage}
\end{figure}
Figure \ref{fig:X666} shows simulation results for OpenBuffet geometries where the streamwise location of the upper crest point is varied (\textit{i.e.} the first digit) between $0.3 \le X_U \le 0.4$ ($D_1=3.0,3.5,4.0$). While $C_L$ histories are shown in figure \ref{fig:X666}(a), corresponding airfoil geometries are illustrated in (b). Note that the upper plot is distorted to make geometric differences clearer. Parameters selected to characterise aerodynamic properties and buffet characteristics are summarised in table \ref{tab:X666}. While buffet is clearly present in cases OB-4-3666 and OB-4-3p5666, the buffet amplitude decreases when delaying the crest point in the streamwise direction. Buffet frequencies, on the other hand, increase and additional peak arises at an intermediate frequency of $St_I\approx0.3$, which appears to be similar to the separation-bubble mode reported in \cite{Zauner2023b}. The reduction of buffet amplitudes, which is mainly due to increased levels of minimum lift, leads to an overall increase of the lift-over-drag ratio ($\overline{L}/\overline{D}$). 
For $X_U=0.4$, the initial transient appears longer and we have to exclude the first $20$ instead of $15$ CTUs for spectral analysis.
When moving from $X_U=0.35$ to $X_U=0.40$, we have to acknowledge a change in trends. 
As we can see in figure \ref{fig:X666}(b), the rear part of the red geometry becomes very steep, which promotes flow separation and some sort of post-buffet or stall conditions.
While buffet seems to die out, oscillations at intermediate frequencies dominate the flow dynamics, while associated frequencies slightly increases from $St_I=0.306$ to $0.338$.
We find mean-flow separation at $x \approx 0.55$, with no re-attachment up to the trailing edge and a significant divergence of the trailing-edge pressure. Mean $C_p$ and $C_f$ distributions can be found in appendix \ref{sec:Add_F_D1}.
The mean $C_L$ seems to settle near $C_{L,min}$-levels of the OB-4-3p5666 case. We also observe frequencies around $St \approx 2$ associated with the wake mode reported by \cite{Moise2021}, which is typical for significant flow separation of the rear part.
All these characteristics are typical for buffet-offset or stall conditions.

\subsubsection{Vertical variation of the upper crest point}
\begin{figure}[hbt!]
\centering
\begin{minipage}[t]{\textwidth}
  \begin{tabular}{ll}
    a) & b) \\
    \includegraphics[width=0.45\textwidth,trim={10mm 10mm 20mm 20mm},clip]{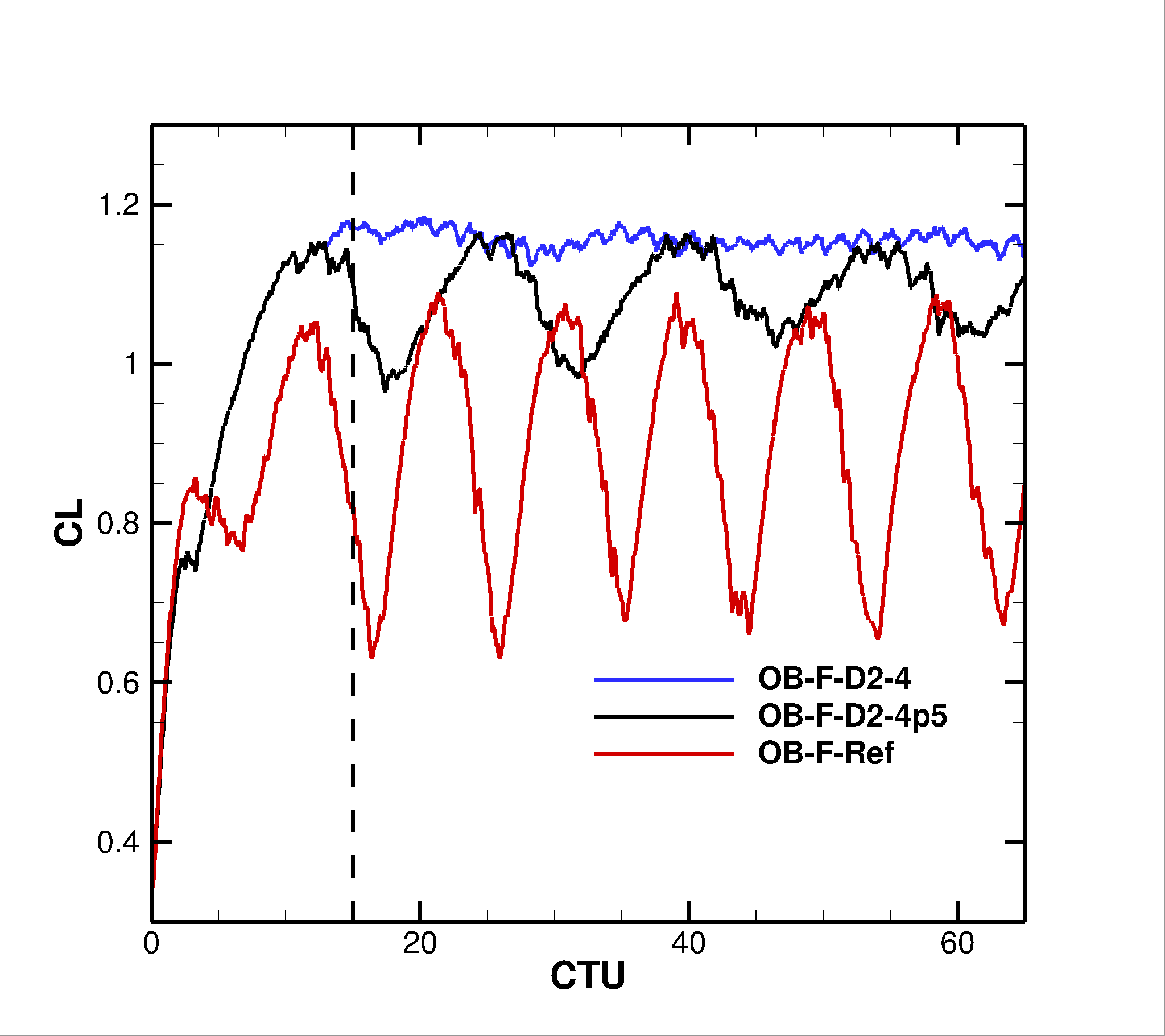} &
    \includegraphics[width=0.45\textwidth,trim={10mm 10mm 20mm 10mm},clip]{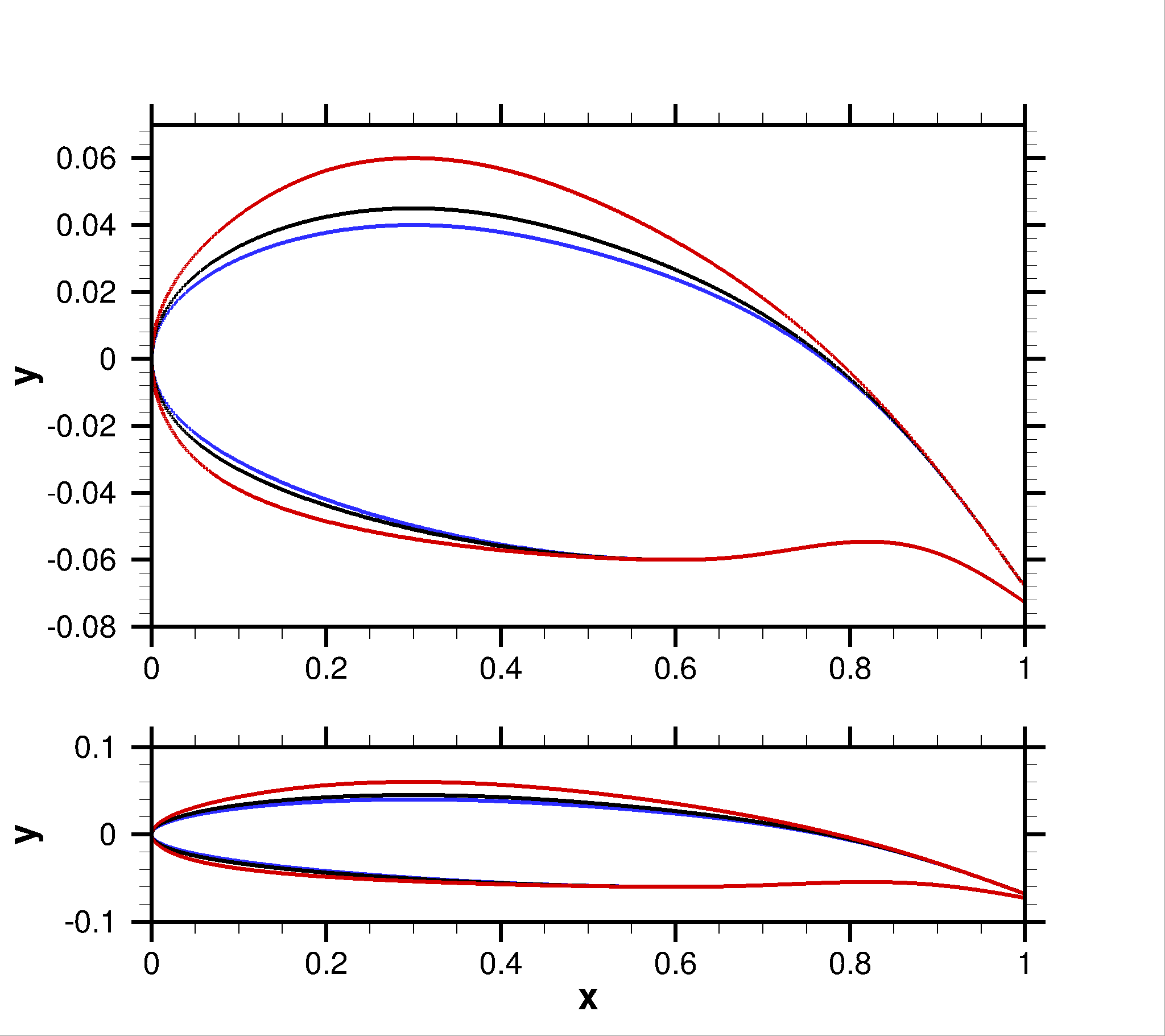} \\
  \end{tabular}
\caption{(a) Lift coefficient ($C_L$) as a function of convective time units ($CTU$) for 4-digit airfoils at an angle of attack $\alpha=4^{\circ}$, where the second digit is varied (OB-4-3X66). Corresponding airfoil contours are shown in (b), respectively, where the aspect ratio of the upper plot axes is distorted. The aspect ratio of the lower plot is x:y=1:1.  \label{fig:3X66}}
\end{minipage}
\vspace{0.5cm} % Adjust the vertical space between the figure and table as needed
\begin{minipage}[t]{\textwidth}
% \begin{table}
\centering
\begin{tabular}{lccccccccc}
Case & $D_2 $ & $St_B$ & $St_I$ & $\Delta C_{L,max}$ & $C_{L,max}$ & $C_{L,min}$ & $C_{D,max}$ & $C_{D,min}$ & $\overline{L}/\overline{D}$\\
\hline
OB-F-Ref    & 6 & 0.115 & - & 0.46 & 1.09 & 0.63 & 0.102 & 0.039 & 12.4 \\ %Secondary intermediate-frequency peak at St_I=0.211
OB-F-D2-4p5  & 4.5 & 0.077 & 0.573 & 0.20 & 1.17 & 0.96 & 0.059 & 0.033 & 22.3 \\ %second freq & 0.268
OB-F-D2-4    & 4 & 0.050 & 0.547 & 0.06 & 1.19 & 1.12 & 0.051 & 0.040 & 25.0 \\ %second freq & 0.12
\hline \\
\end{tabular}
\captionof{table}{Summary of buffet characteristics for free-tansitional cases varying the second digit of the reference case OB-F-Ref considering the OB-4-3666 profile.}\label{tab:3X66}
% \end{table}
\end{minipage}
\end{figure}
Figure \ref{fig:3X66} shows simulation results for airfoil geometries where the vertical location of the upper crest point is varied (\textit{i.e.} second digit) between $0.04 \le Y_U \le 0.06$ ($D_4=$ X $=4.0,4.5,6.0$), mainly affecting the airfoil thickness towards the fore part as well as the rounding of the leading edge. 
Considering also table \ref{tab:3X66}, we can observe buffet frequencies as well as corresponding amplitudes clearly increasing with $Y_U$. For cases with $Y_U<0.06$, we see a spectral peak arising at intermediate frequencies in the range of $St_I\approx0.5-0.6$. Decreasing $Y_U$ delays buffet onset and improves the lift-over-drag ratio significantly. The observed increase of buffet frequencies with associated amplitudes agrees with the Mach-number study of \cite{Zauner2023b} considering ONERA's OALT25 profile.

For test case OB-F-D2-4p5, we also applied a scaling for intermediate-frequency phenomena associated with an unsteady separation bubble proposed by \cite{Zauner2023b}. We multiply the associated Strouhal number of $St_I=0.573$ with the normalised mean length of the separation bubble $L_{sep}/c=0.356$ and the maximum mean reverse velocity within the separation bubble $U_R/U_{\infty}=0.102$. The obtained scaled Strouhal number of $St_{R}=0.021$ agrees well with those reported in \cite{Zauner2023b}, ranging for different airfoils from $St_R=0.20$ to $0.26$. Details about the meanflow data used for this scaling is provided in appendix \ref{sec:App_scaling}.

\subsection{Tripped boundary layers}\label{sec:OpenBuffet_Upper_Tripped}

We will investigate in this sub-section parametric OpenBuffet profiles at tripped boundary-layer conditions to see, whether we can confirm certain observations and trends of free-transitional cases of the previous sub-section. 
While we aim for flow conditions similar to experiments of \citet{JMDMS2009} corresponding to $\alpha=3.5^{\circ}$ and $M=0.73$, the Reynolds number is reduced to $Re=500,\!000$, which also lowers computational costs and aligns with free-transitional computations of the previous section.
Table \ref{tab:parameters_tripped} provides an overview of four-digit OpenBuffet geometries considered for this study. 
This airfoil family is representative of conventional supercritical airfoils, where case OB-T-D2-4p8 (OB-3p5-2p8'4p8'3p8'7p8) is similar to ONERA's OAT15A profile. %, even though the axial crest position is located further upstream for the OAT15A profile. 
Labels are composed of the profile name (`OB' for OpenBuffet), the transition mode (`T' for tripped), the digit (`D1' or `D2') which differs from the reference case (OB-T-Ref), and the value of this digit.
\begin{table}[h]
\centering
\begin{tabular}{lcccccccccc}
Case & Angle of Attack & $X_U$ & $Y_U$ & $X_L$ & $Y_L$ & $a_{LE}$ & $b_{LE}$ & $\alpha_{TE}$ & $\beta_{TE}$ & $\Delta_{TE}$\\
\hline
OB-T-Ref     & $3.5^{\circ}$ & 0.28   & 0.057 & 0.38 & -0.078 & $X_U$ & $Y_U$ & $17^{\circ}$ & $10^{\circ}$ & 0.000\\
\hline
OB-T-D1-3p0    & $3.5^{\circ}$ & \textbf{0.30}   & 0.057 & 0.38 & -0.078 & $X_U$ & $Y_U$ & $17^{\circ}$ & $10^{\circ}$ & 0.000\\
OB-T-D1-3p2    & $3.5^{\circ}$ & \textbf{0.32}   & 0.057 & 0.38 & -0.078 & $X_U$ & $Y_U$ & $17^{\circ}$ & $10^{\circ}$ & 0.000\\
\hline
OB-T-D2-4p8     & $3.5^{\circ}$ & 0.28   & \textbf{0.048} & 0.38 & -0.078 & $X_U$ & $Y_U$ & $17^{\circ}$ & $10^{\circ}$ & 0.000\\
OB-T-D2-5p3     & $3.5^{\circ}$ & 0.28   & \textbf{0.053} & 0.38 & -0.078 & $X_U$ & $Y_U$ & $17^{\circ}$ & $10^{\circ}$ & 0.000\\
OB-T-D2-6p1     & $3.5^{\circ}$ & 0.28   & \textbf{0.061} & 0.38 & -0.078 & $X_U$ & $Y_U$ & $17^{\circ}$ & $10^{\circ}$ & 0.000\\
\hline \\
\end{tabular}
\caption{Summary and labels for test cases considering tripped boundary layers.}
\label{tab:parameters_tripped}
\end{table}

\subsubsection{Reference case OB-3p5-2p8'5p7'3p8'7p8}\label{sec:reference_tripped}

\begin{figure}[hbt!]
\centering
  \begin{tabular}{ll}
    a) & b) \\
    \includegraphics[width=0.45\textwidth,trim={30mm 10mm 80mm 40mm},clip]{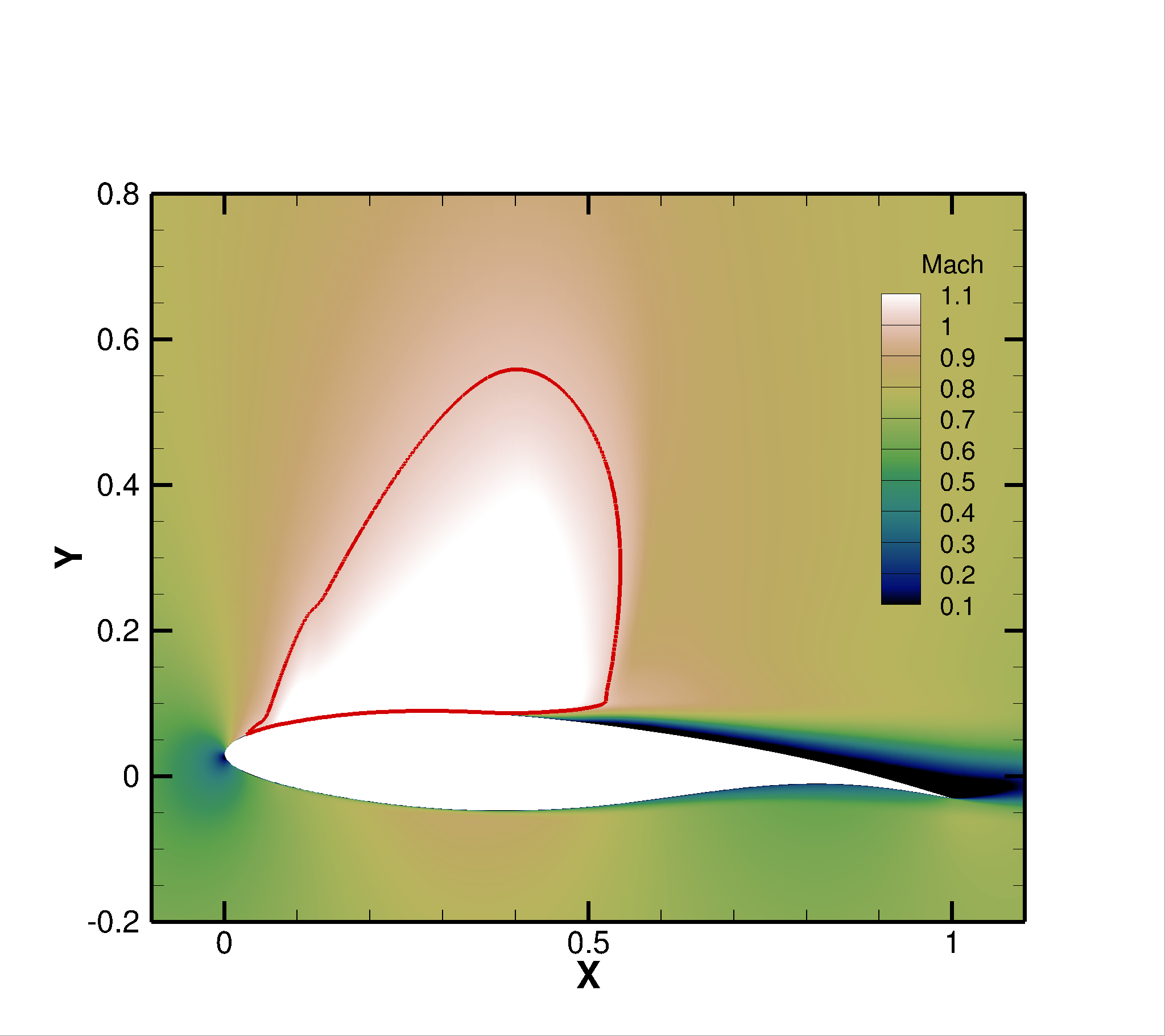} &
    \includegraphics[width=0.45\textwidth,trim={10mm 10mm 10mm 10mm},clip]{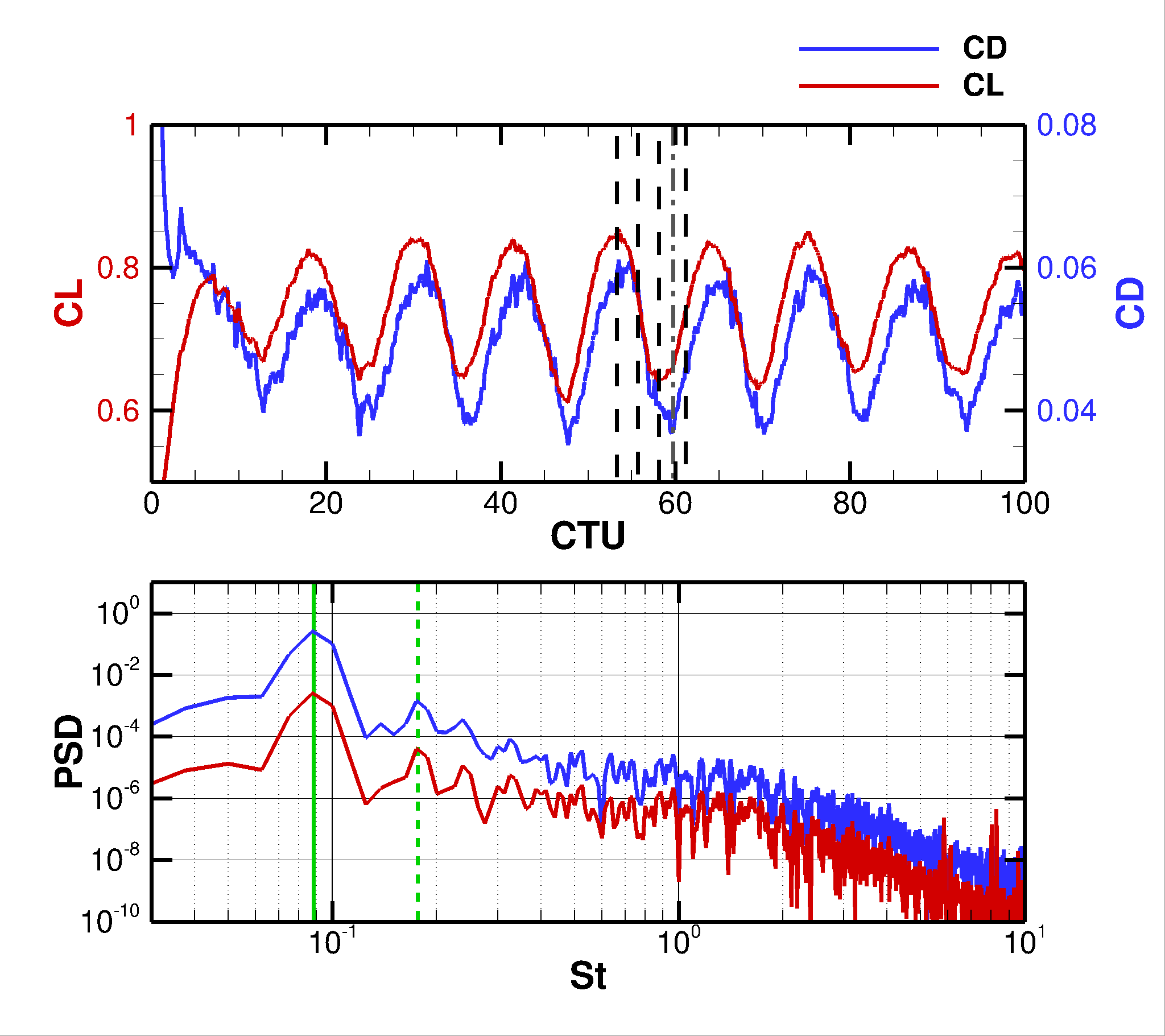} \\
  \end{tabular}
\caption{(a) Time- and span-averaged flow field showing Mach number contours for the OB-3p5-2p8'5p7'3p8'7p8 airfoil at $\alpha=3.5^{\circ}$, considering tripped boundary layers, $M=0.73$, and $Re=500,\!000$. (b) Histories and spectra of lift (red) and drag coefficients (blue). Dashed lines mark time instances of snapshots in figure \ref{fig:snap}\label{fig:ref_tripped}}
\end{figure}
As a reference test case for tripped conditions, we select the four-digit OB-3p5-2p8'5p7'3p8'7p8 profile denoted as case OB-T-Ref. Figure \ref{fig:ref_tripped}(a) shows time- and span-averaged Mach-number contours at $\alpha=3.5^{\circ}$, $M=0.73$, and $Re=500,\!000$. Compared to the free-transitional case, the supersonic region of the turbulent mean flow (delineated by the red curve) extends further into the free stream for the tripped case, where the shock foot is also much steeper. A weak kink in the sonic line at $x \approx 0.1$ is caused by an impinging wave originating from the tripping location.
Figure \ref{fig:ref_tripped}(b) shows histories (upper plot) and spectra (lower plot) of lift (red curves) and drag coefficients (blue curves). 
Spectra are computed using FFT considering a single Hanning window and ignoring the initial transient of the first 20 convective time units (CTU). The spectral resolution for the present time series corresponds to $\Delta St \approx 0.013$.
Similar to the free-transitional test case, we can observe a clear peak marked by the green solid line at $St_B\approx0.09$, which is associated with transonic buffet. The harmonic peak is also visible and marked by the green dashed line at $St\approx 0.18$. Two additional peaks around $St\approx6$ and $8$ are caused by the forcing used to trip the boundary-layers.

\begin{figure}[hbt!]
\centering
  \begin{tabular}{ll}
    a) & b) \\
    \includegraphics[width=0.45\textwidth,trim={10mm 10mm 10mm 10mm},clip]{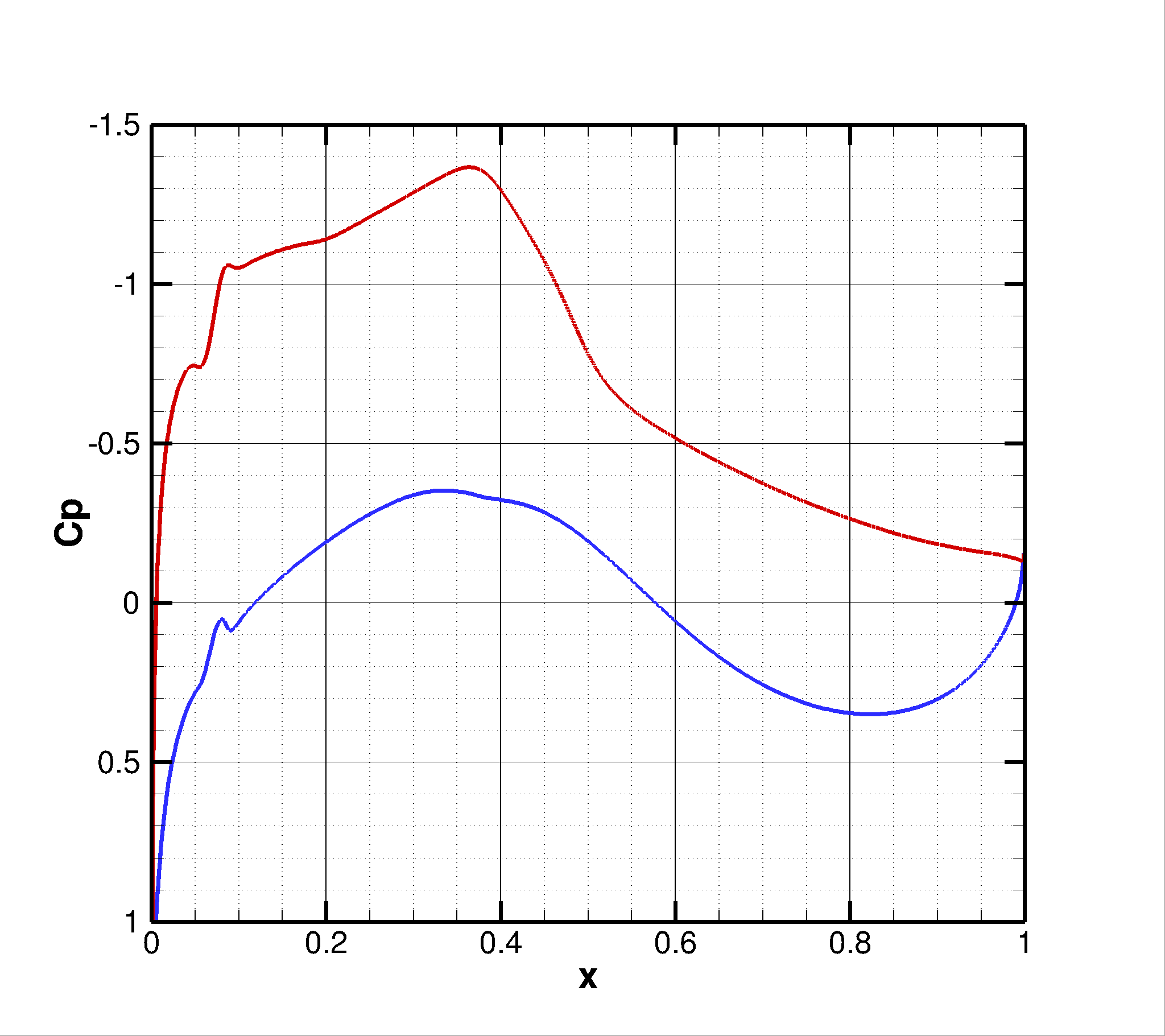} &
    \includegraphics[width=0.45\textwidth,trim={10mm 10mm 10mm 10mm},clip]{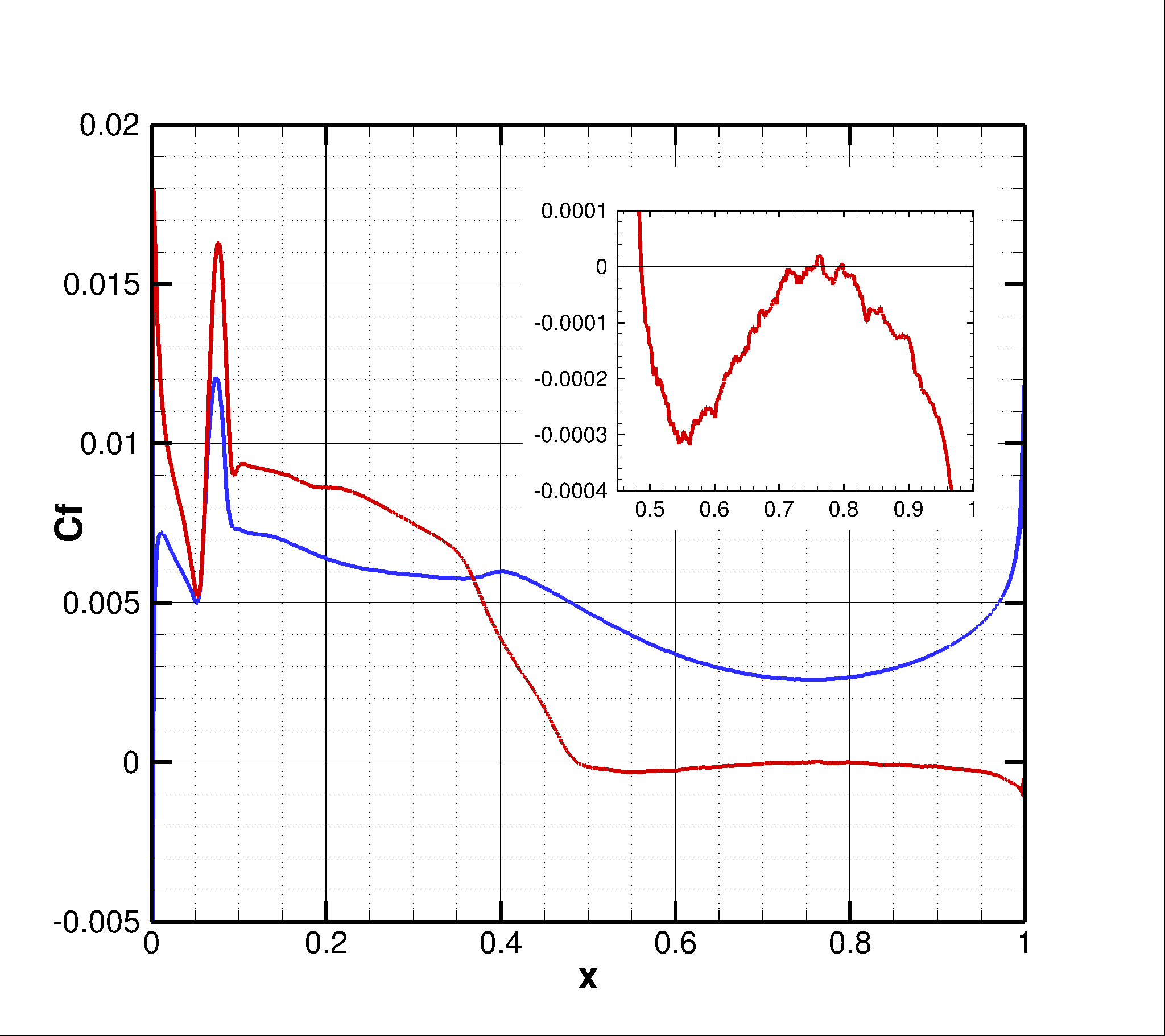} \\
  \end{tabular}
\caption{Time-averaged (a) wall-pressure coefficient $C_p$ and (b) skin-friction coefficient $C_f$ distributions for the OB-3p5-2p8'5p7'3p8'7p8 profile with boundary-layer tripping at $X_{tr}=0.07$. Blue and red curves correspond to pressure and suction side distributions, respectively.\label{fig:CpCf_tripped}}
\end{figure}
Figure \ref{fig:CpCf_tripped} shows time-averaged distributions of (a) wall-pressure coefficient $C_p$ and (b) skin-friction coefficient $C_f$ for the baseline profile. 
The boundary-layer tripping location is evident in both plots at $x=x_{tr}=0.07$, but it is more pronounced in the skin-friction distribution, showing a spike on both sides. The $C_p$ reaches its maximum around $x\approx0.35$. Due to the back and forth motion of the shock wave, the sharp pressure gradient across the shock wave appears as a smooth slope within the range of $0.35<x<0.55$ in the time-averaged flow field. The same applies to the $C_f$ distribution in figure \ref{fig:CpCf_tripped}(b), where the slope terminates with $C_f<0$ indicating mean flow separation. While the mean flow re-attaches at $x\approx0.75$ (see inset), we observe another mean-flow recirculation region near the trailing edge.

\begin{figure}[hbt!]
\centering
  \begin{tabular}{ll}
    a) & b) \\
    \includegraphics[width=0.45\textwidth,trim={0mm 0mm 0mm 0mm},clip]{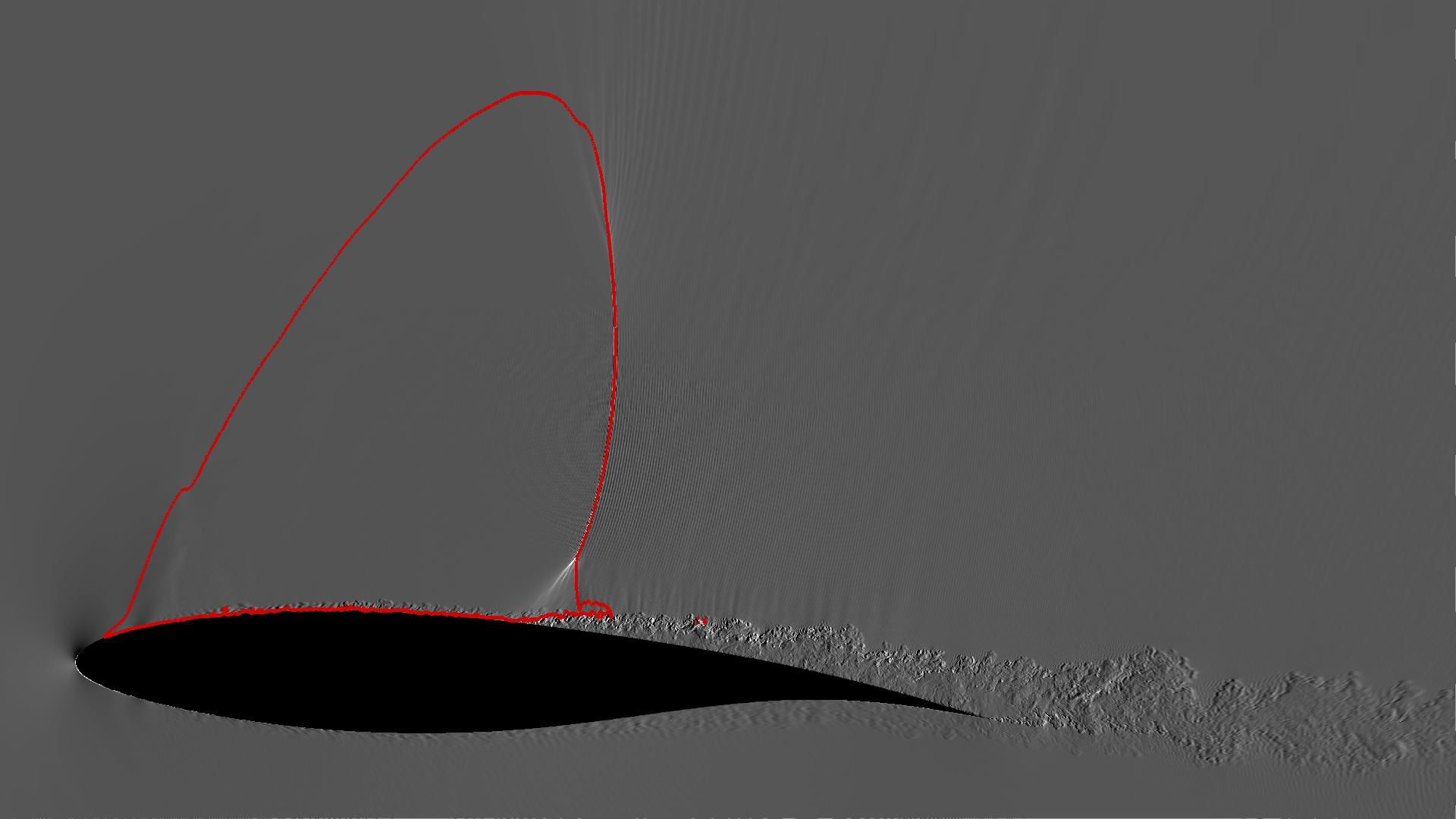} &
    \includegraphics[width=0.45\textwidth,trim={0mm 0mm 0mm 0mm},clip]{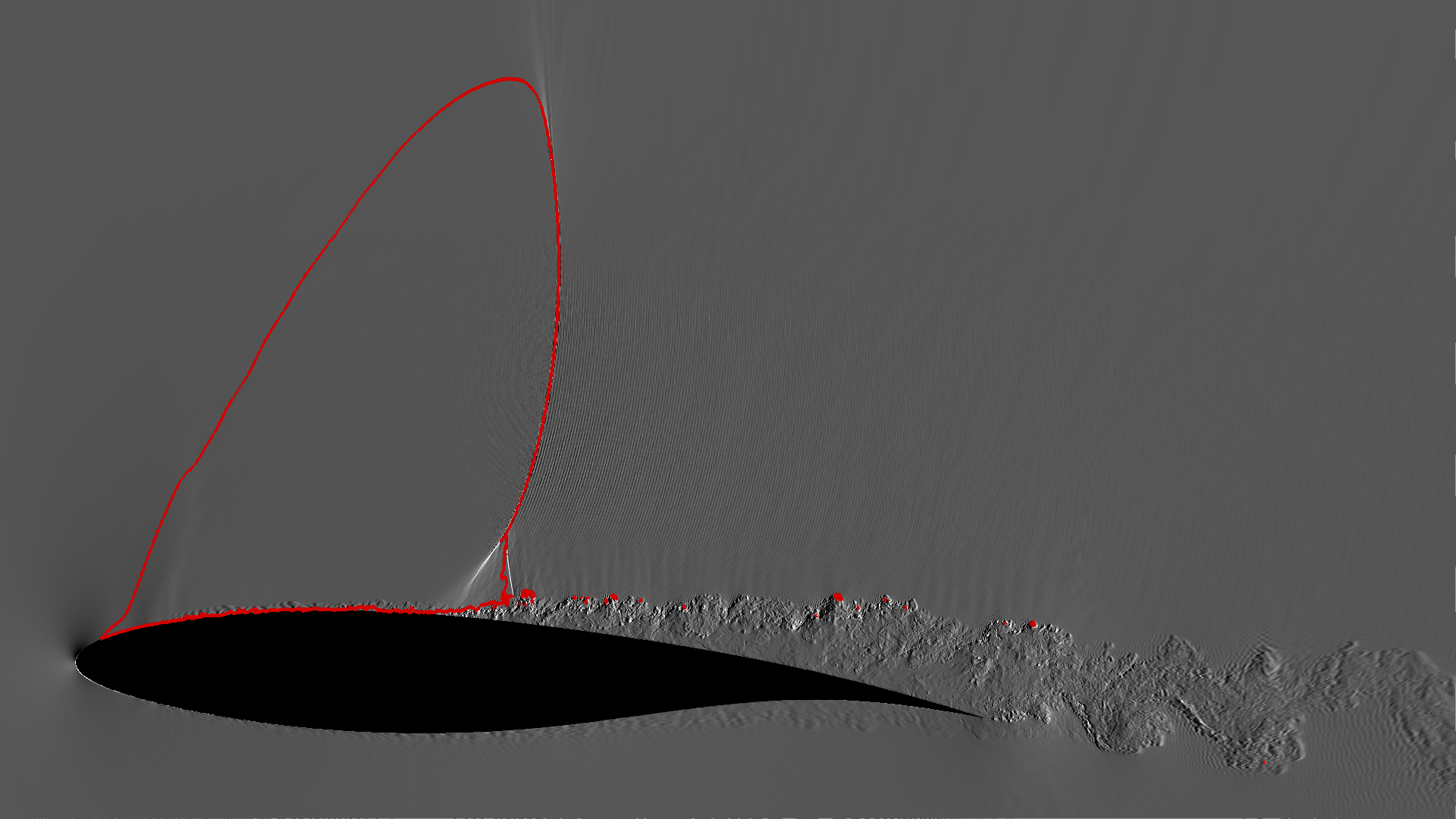} \\
    c) & d) \\
    \includegraphics[width=0.45\textwidth,trim={0mm 0mm 0mm 0mm},clip]{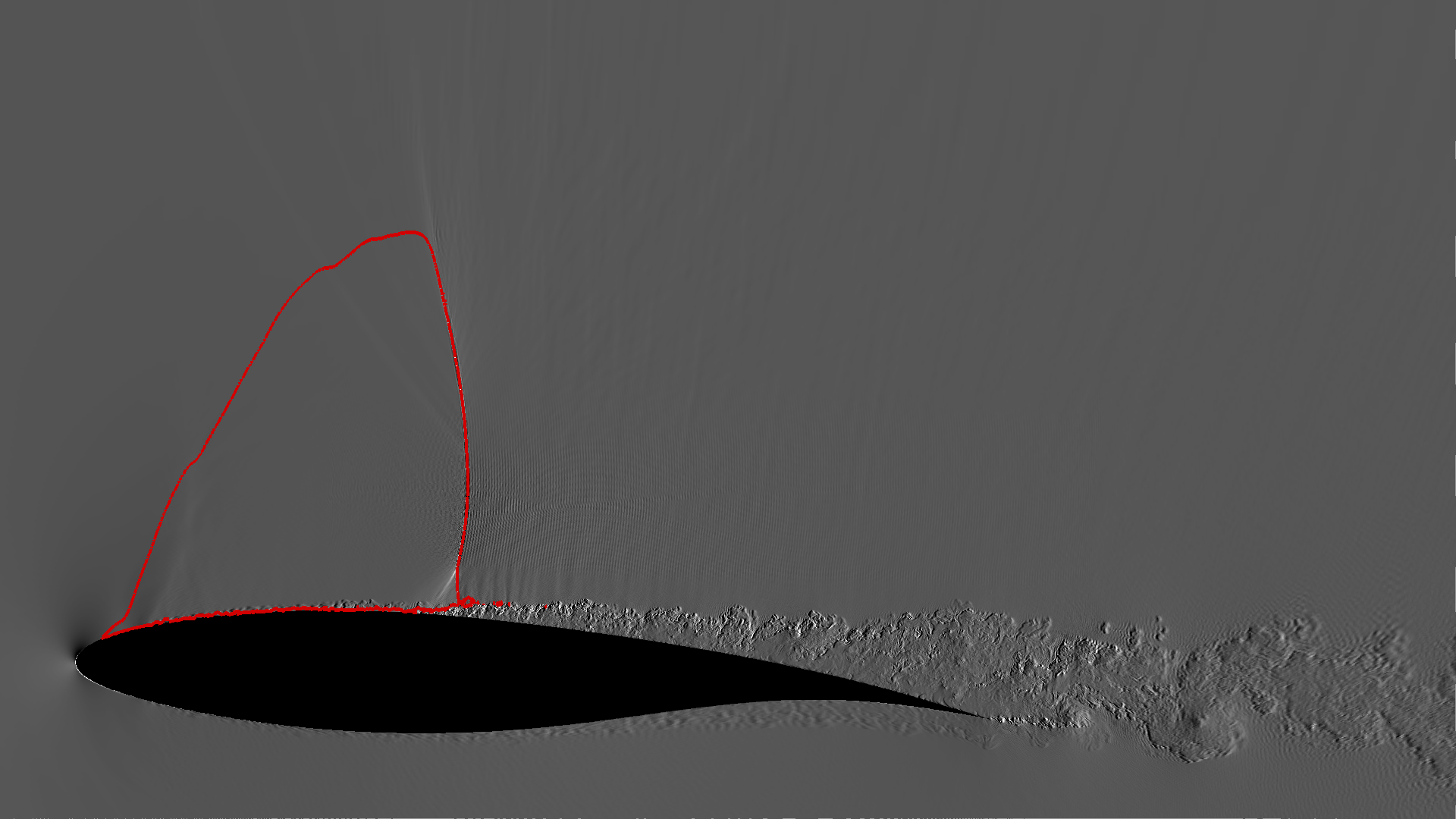} &
    \includegraphics[width=0.45\textwidth,trim={0mm 0mm 0mm 0mm},clip]{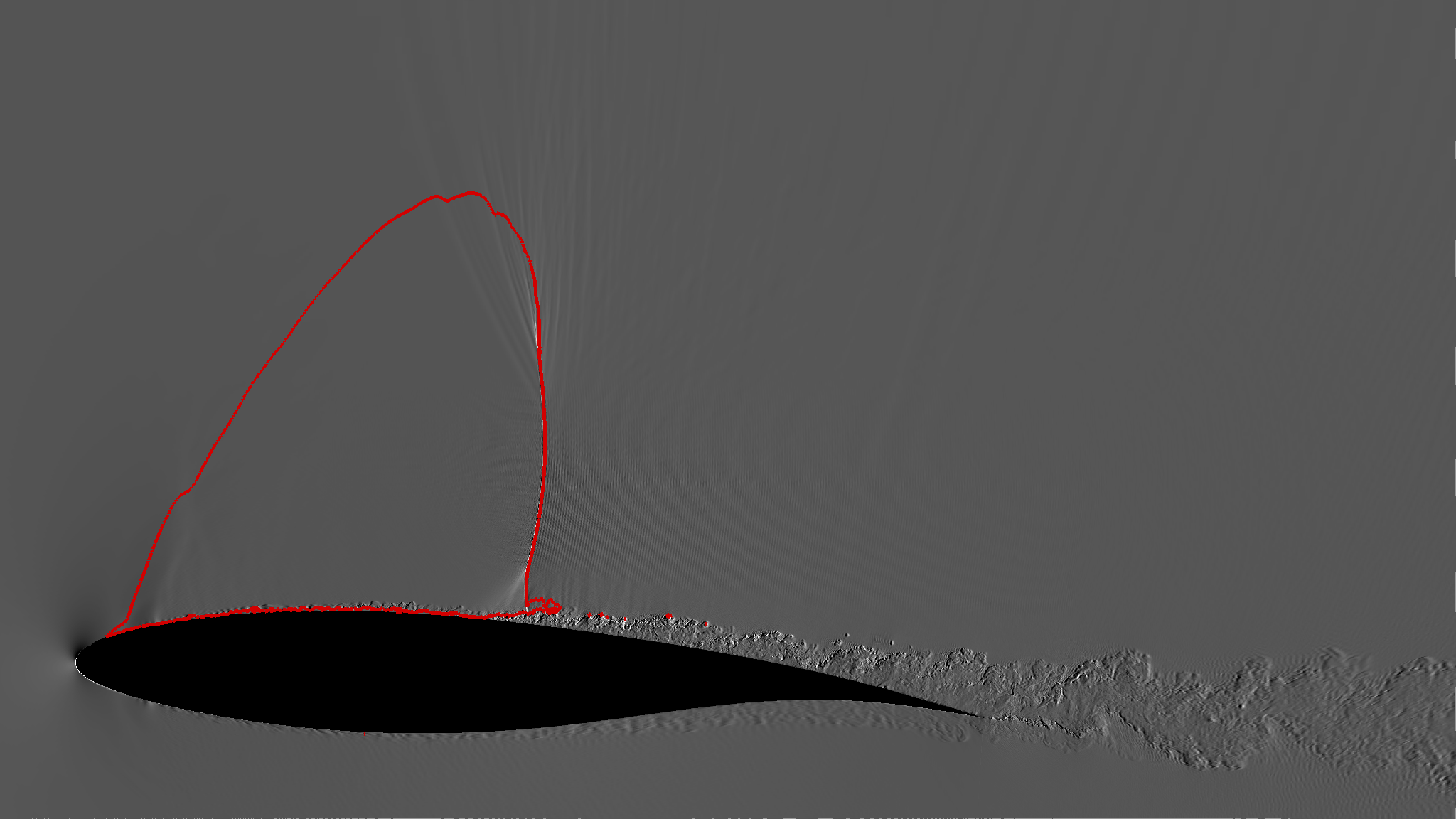} \\
  \end{tabular}
\caption{Numerical quasi-schlieren snapshots showing streamwise density gradients after approximately (a) 53 CTUs (Convective time units), (b) 56 CTUs, (c) 59 CTUs, and (d) 61 CTUs, corresponding to vertical red lines in figure \ref{fig:ref_tripped}. Black and white contours correspond to non-dimensional values of $\partial \rho / \partial x = -20$ and $40$, respectively. Sonic lines are shown in red. \label{fig:snap_tripped}}
\end{figure}
\begin{table}[h]
\centering
\begin{tabular}{cccccc}
% \hline
Snapshot & time instant & $C_L$ & $C_{D,p}$ & $C_{D,f}$ & $C_L/C_D$ \\
\hline
a) & 53.3 & high        & high      & increased & reduced \\
b) & 55.7 & increased   & high      & low       & low  \\
c) & 58.7 & low         & reduced   & reduced   & increased \\
-  & 59.7 & reduced     & low       & increased & high \\
d) & 61.15& medium      & reduced   & high      & increased \\
\hline \\
\end{tabular}
\caption{Qualitative overview of aerodynamic properties for snapshots in figure \ref{fig:snap_tripped}.}
\label{tab:qual_inst_tripped}
\end{table}
Figure \ref{fig:snap_tripped} shows numerical quasi-schlieren snapshots (streamwise density gradient contours) at representative time instants indicated by vertical black dashed lines in figure \ref{fig:ref_tripped}(b). The vertical dash-dotted grey line marks an additional time-instant, which is qualitatively similar to figure \ref{fig:snap_tripped}(c) and will be included in the discussion.
We will briefly outline the flow behaviour throughout one buffet cycle, while table \ref{tab:qual_inst_tripped} provides a summary of qualitative levels of aerodynamic coefficients with respect to their extrema. Histories of corresponding coefficients are provided in figure \ref{fig:laminar_vs_tripped}(b) in the appendix and will not be discussed in detail here. While we will outline main differences between tripped and free-transitional baseline cases, we refer the reader to \cite{Brion2019, Moise2023} for a more in-depth analysis of differences between transonic buffet subjected to tripped and free-transitional boundary layers. 
Starting at a high-lift phase, where $C_L$ reaches its maximum in figure \ref{fig:snap_tripped}(a), we can observe a large supersonic region with its shock foot reaching the most down-stream location. Similar to the free-transitional case, we can observe a lambda-shock structure. However, due to the significantly smaller shock-induced separation bubble, the lambda-shock structure is much more localised for turbulent boundary layers and is reminiscent of experimental observations in \cite{JMDMS2009}. The shock wave is slightly bent and reaches its maximum strength at a wall-distance of about $\Delta y \approx 0.2$. In the instantaneous contour plots, we can observe weak non-physical oscillations downstream of the part of the sonic line, which is subjected to the strongest density gradient. However, as they mainly affect the downstream part of the free-stream and not the boundary layers, we did not apply more frequent filtering when solving the Navier/Stokes equations, which could adversely impact large-scale phenomena \cite{Zauner2018e}.
At this time instance, we observe already weak flow separation, which grows continuously until reaching its maximum extent in figure \ref{fig:snap_tripped}(b), where skin-friction drag as well as lift-over-drag ratio hit their minimum. However, as we still observe a large supersonic region, also aerodynamic forces are still increasing. 
The lift eventually reaches its minimum in snapshot \ref{fig:snap_tripped}(c), where the turbulent boundary layer is already recovering ($C_{D,f}$ is increasing). At the same time the pressure drag keeps decreasing and reaches its minimum slightly delayed at the time instant marked by the grey dash-dotted line in figure \ref{fig:ref_tripped}(b) (no additional snapshot is shown for brevity). %At this point, also the lift-over-drag ratio reaches its maximum. 
In snapshot \ref{fig:snap_tripped}(d), the boundary layer is fully recovered leading to maximum skin-friction drag. When aerodynamic forces are strengthening, boundary layers again start to separate and the cycle starts over again.

Comparing tables \ref{tab:qual_inst} and \ref{tab:qual_inst_tripped}, we can see very clear parallels between transonic buffet at free-transitional and tripped conditions across one buffet cycle. 
Without showing details (histories are provided in the appendix in figure \ref{fig:laminar_vs_tripped}), for both baseline cases, $C_L$ histories appear quite sinusoidal. While this remains true for $C_{D}$ of the tripped case, we observe a saw-tooth pattern (slow increase and rapid decrease) for the free-transitional counterpart, where laminar boundary layers recover relatively fast. 
Another striking similarity between both cases is the phase shift between $C_L$ and $C_D$, where lift is leading by about $90^{\circ}$.
Given the present observations and detailed discussion of \citet{Moise2023}, we are confident that we look at the same underlying mechanism for both boundary-layer transition modes, leading to large-scale oscillations of aerodynamic forces associated with transonic buffet.

As for the free-transition cases, we will characterise buffet features for tripped cases by measuring frequencies ($St_B$) and extreme vales of the lift coefficient ($C_{L,max}$ and $C_{L,min}$, respectively) as well as a maximum amplitude ($\Delta C_{L,max}$). For this baseline case we obtain $\Delta C_{L,max}=0.24$, $C_{L,max}=0.85$, $C_{L,min}=0.61$. 

% CL for both cases quite sinusoidal
% CDp for turbulent quite sinusoidal, while sawtooth shape for laminar (sharp drop)
% CDf skewed but fairly sinusoidal for turbulent case, but sawtooth shape for laminar (sharp increase)
% after maximum lift, CDp spikes for laminar case
% min CDf leads min CDp for both cases
% min CDf and min CDp phase shifted by about 90deg for both cases
% CDf recovers very fast for laminar case, so it reaches max CDf when CDp is still low 
% for both, minimum of CL leads minimum of CDp
% CL/CD plateaus for laminar case, but very spikey for turbulent case 

\subsubsection{Horizontal variation of the upper crest point}

\begin{figure}[hbt!]
\centering
\begin{minipage}[t]{\textwidth}
  \begin{tabular}{ll}
    a) & b) \\
    \includegraphics[width=0.45\textwidth,trim={10mm 10mm 20mm 20mm},clip]{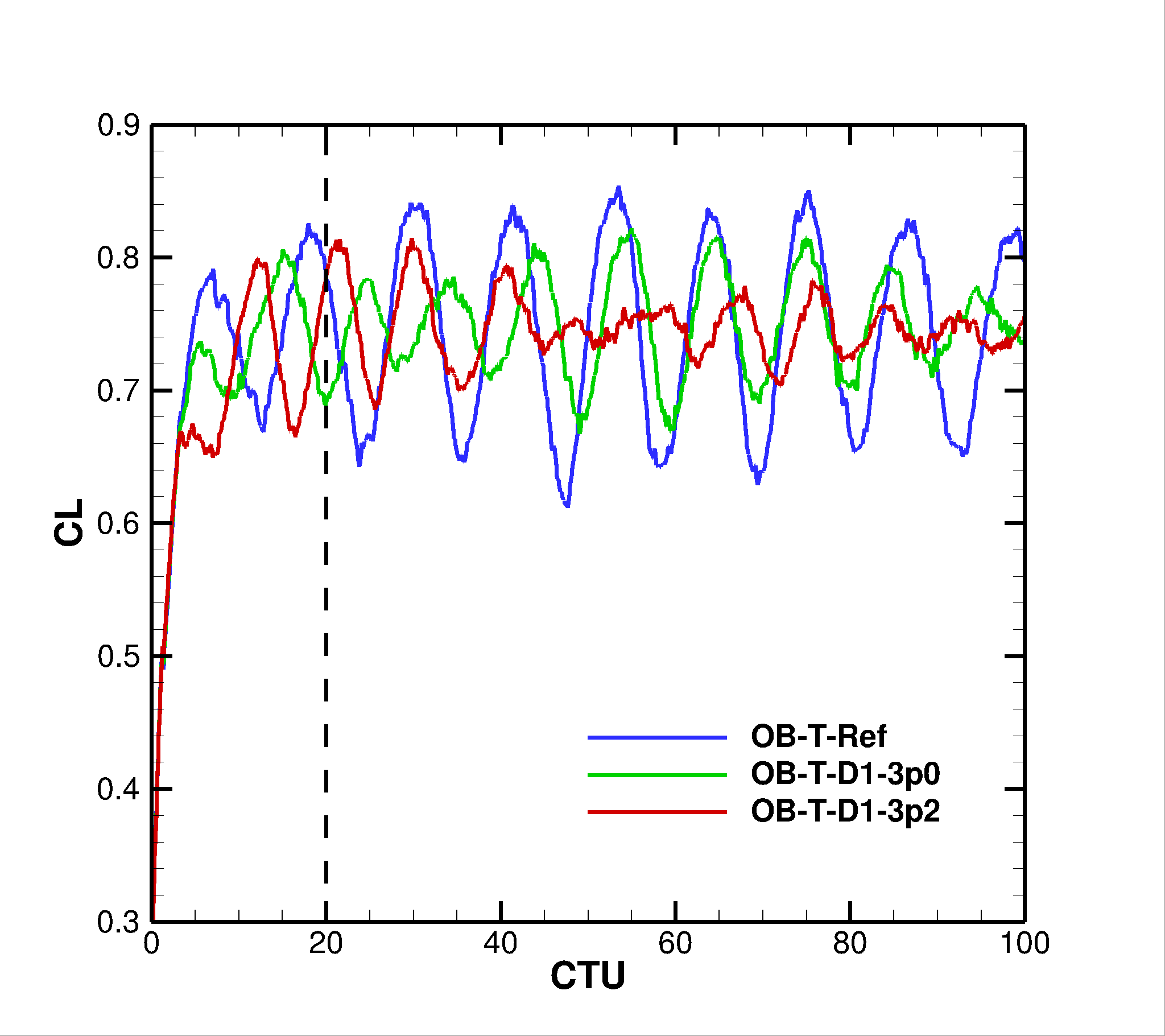} &
    \includegraphics[width=0.45\textwidth,trim={10mm 10mm 20mm 10mm},clip]{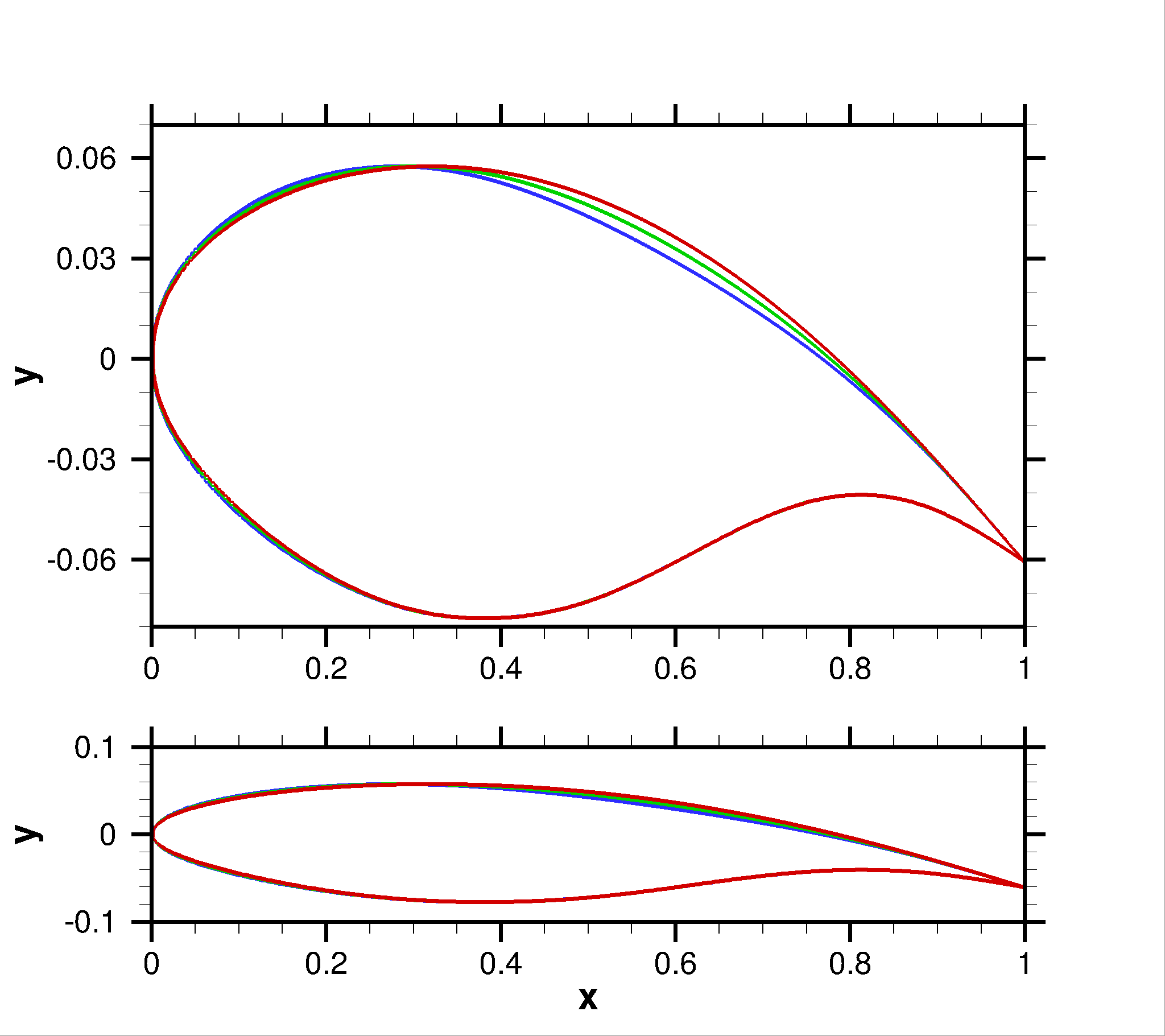} \\
  \end{tabular}
\caption{(a) Lift coefficient ($C_L$) as a function of convective time units ($CTU$) for 4-digit OB airfoils at an angle of attack $\alpha=3.5^{\circ}$, where the first digit is varied for tripped simulations. Corresponding airfoil contours are shown in (b), respectively, where the aspect ratio of the upper plot axes is distorted. The aspect ratio of the lower plot is x:y=1:1.  \label{fig:D1_tripped}}
% \end{figure}
\end{minipage}
\vspace{0.5cm} % Adjust the vertical space between the figure and table as needed
\begin{minipage}[t]{\textwidth}
% \begin{table}
\centering
\begin{tabular}{lcccccccc}
Case &  $D_1$ & $St_B$ & $\Delta C_{L,max}$ & $C_{L,max}$ & $C_{L,min}$ & $C_{D,max}$ & $C_{D,min}$ & $\overline{L/D}$\\
\hline
OB-T-Ref    & 2.8 & 0.088 & 0.24 & 0.85 & 0.61 & 0.061 & 0.035 & 15.1 \\ %second freq & 0.088
OB-T-D1-3p0 & 3.0 & 0.100 & 0.15 & 0.82 & 0.67 & 0.055 & 0.037 & 16.4 \\ %second freq & 0.1
OB-T-D1-3p2 & 3.2 & 0.113 & 0.13 & 0.81 & 0.69 & 0.053 & 0.036 & 17.3 \\ %second freq & 0.113
\hline \\
\end{tabular}
\captionof{table}{Summary of buffet characteristics for tripped cases varying the first digit of the reference case OB-T-Ref considering the OB-3p5-2p8'5p7'3p8'7p8 profile.}
\label{tab:D1_tripped}
%\end{table}
\end{minipage}
\end{figure}
Figure \ref{fig:D1_tripped} shows simulation results for OpenBuffet geometries where the streamwise location of the upper crest point is varied (\textit{i.e.} first digit) between $0.28 \le X_U \le 0.32$ ($D_1=2.8,3.0,3.2$). While $C_L$ histories are shown in figure \ref{fig:X666}(a), corresponding airfoil geometries are illustrated in (b) in the same manner as in the previous section. Table \ref{tab:D1_tripped} summarises the parameters used to characterise buffet. 
Similar to free-transitional cases, we observe decreasing buffet amplitudes when moving the crest point in the downstream direction, while associated frequencies and lift-over-drag ratios increase continuously. 
For the red curve ($D1 = 3.2$) in figure \ref{fig:D1_tripped}(a) buffet becomes very weak after an initial transient and we observe a rather irregular behaviour. We do not observe the appearance of the intermediate frequency peak that was seen in free-transitional cases. This is most probably due to significantly smaller separation bubbles for turbulent shock-wave/boundary-layer interactions.
It is interesting to note that the mean values (after initial transients) are similar for all curves in figure \ref{fig:D1_tripped}(a). 

\subsubsection{Vertical variation of the upper crest point}

\begin{figure}[hbt!]
\centering
\begin{minipage}[t]{\textwidth}
  \begin{tabular}{ll}
    a) & b) \\
    \includegraphics[width=0.45\textwidth,trim={10mm 10mm 20mm 20mm},clip]{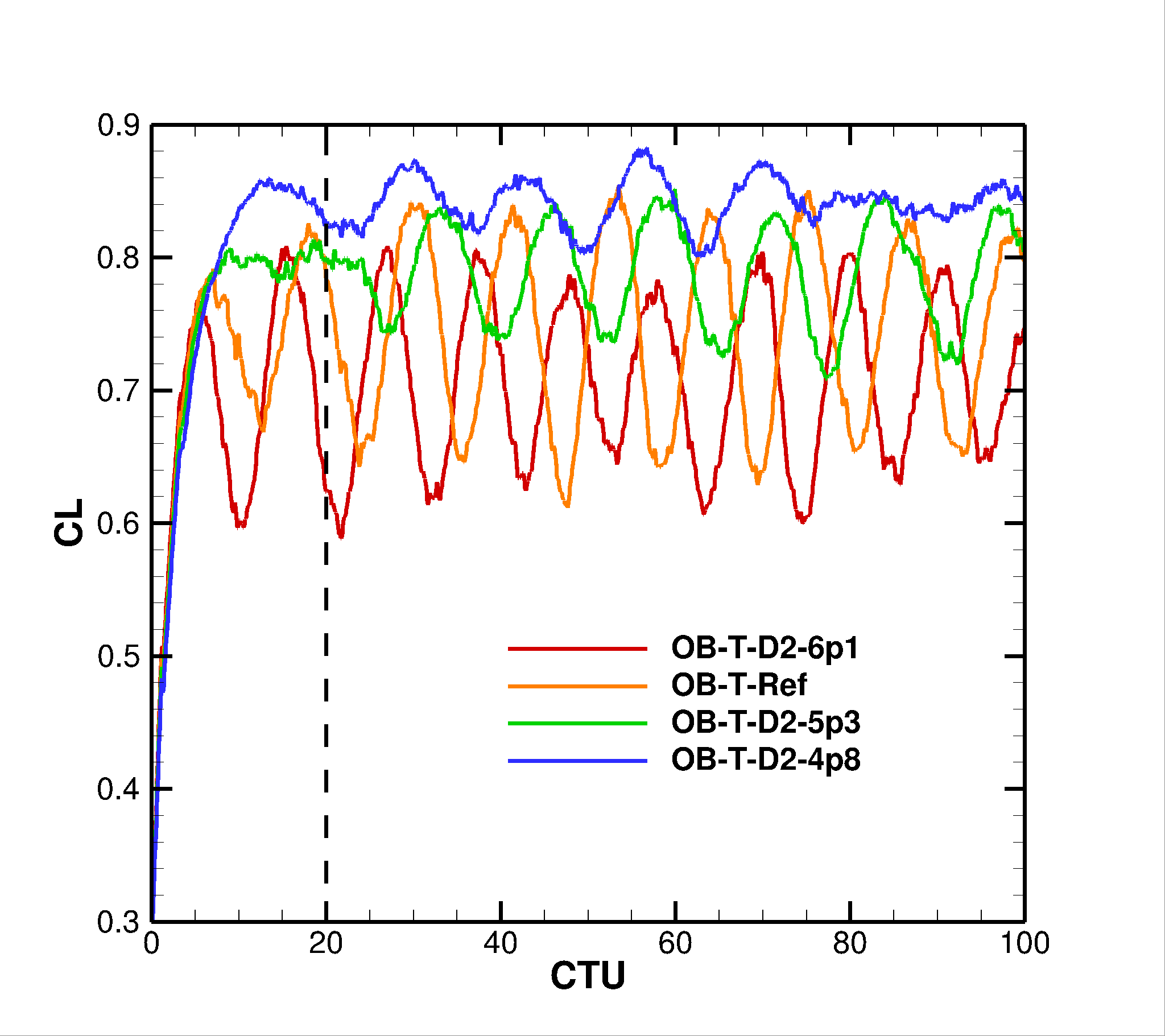} &
    \includegraphics[width=0.45\textwidth,trim={10mm 10mm 20mm 10mm},clip]{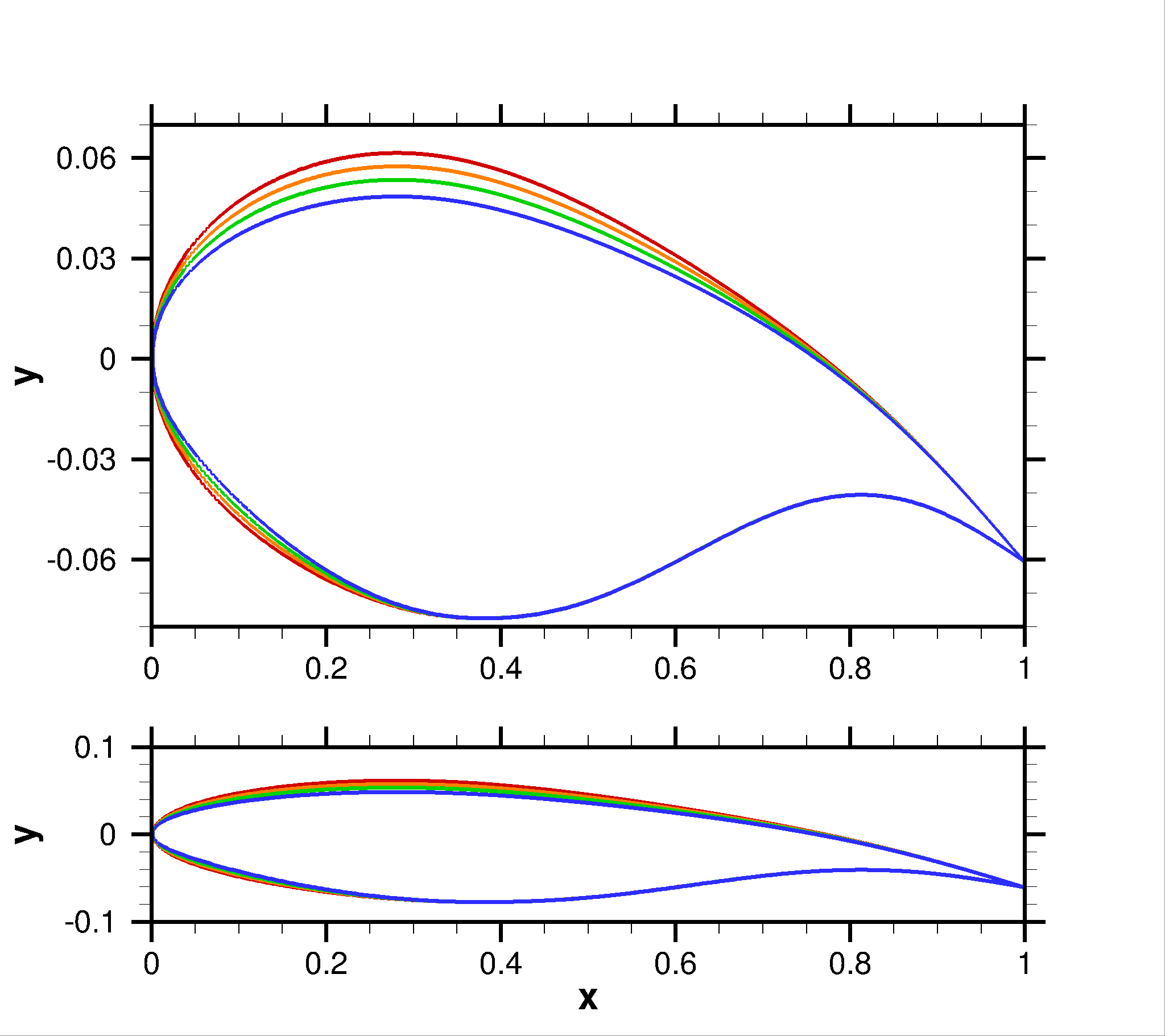} \\
  \end{tabular}
\caption{(a) Lift coefficient ($C_L$) as a function of convective time units ($CTU$) for 4-digit OB airfoils at an angle of attack $\alpha=4^{\circ}$, where the second digit is varied for tripped simulations. Corresponding airfoil contours are shown in (b), respectively, where the aspect ratio of the upper plot axes is distorted. The aspect ratio of the lower plot is x:y=1:1.  \label{fig:D2_tripped}}
% \end{figure}
\end{minipage}
\vspace{0.5cm} % Adjust the vertical space between the figure and table as needed
\begin{minipage}[t]{\textwidth}
% \begin{table}
\centering
\begin{tabular}{lcccccccc}
Case & $D_2$ & $St_B$ & $\Delta C_{L,max}$ & $C_{L,max}$ & $C_{L,min}$ & $C_{D,max}$ & $C_{D,min}$ & $\overline{L/D}$\\
\hline
OB-T-D2-6p1  & 6.1 & 0.100 & 0.22 & 0.81 & 0.59 & 0.065 & 0.038 & 13.6 \\ %second freq & 0.1
OB-T-Ref     & 5.7 & 0.088 & 0.24 & 0.85 & 0.61 & 0.061 & 0.035 & 15.1 \\ %second freq & 0.088
OB-T-D2-5p3  & 5.3 & 0.075 & 0.14 & 0.85 & 0.71 & 0.055 & 0.037 & 17.2 \\ %second freq & 0.163
OB-T-D2-4p8  & 4.8 & 0.075 & 0.08 & 0.88 & 0.8 & 0.047 & 0.038 & 19.9 \\ %second freq & 0.163
\hline \\
\end{tabular}
\captionof{table}{Summary of buffet characteristics for tripped cases, varying the second digit of the reference case OB-T-Ref considering the OB-3p5-2p8'5p7'3p8'7p8 profile.}\label{tab:D2_tripped}
% \end{table}
\end{minipage}
\end{figure}
Figure \ref{fig:D2_tripped} shows tripped simulation results for OpenBuffet airfoil geometries where the vertical location of the upper crest point is varied (\textit{i.e.} second digit) between $0.048 \le Y_U \le 0.061$ ($D_4=4.8,5.3,5.7,6.1$), mainly affecting the airfoil thickness towards the fore part as well as the rounding of the leading edge. Parameters summarised in table \ref{tab:D2_tripped} confirm these trends. Again similar to free-transitional cases, we observe very clear trends. Reducing the thickness via $Y_U$, buffet frequencies and associated amplitudes decrease. While buffet appears to be intermittent for case OB-T-D2-4p8, the mean lift settles around maximum $C_L$-levels of the baseline test case. Also similar to the free-transitional cases, the lift-over-drag ratios increase with decreasing $Y_U$. For the considered profiles at tripped conditions, we do not observe intermediate frequencies arising when buffet becomes weak.

\section{Buffet sensitivities for flattened versions of the OpenBuffet geometries}\label{sec:modified_profiles}

\begin{figure}[hbt!]
\centering
	\includegraphics[width=0.45\textwidth,trim={10mm 10mm 10mm 10mm},clip]{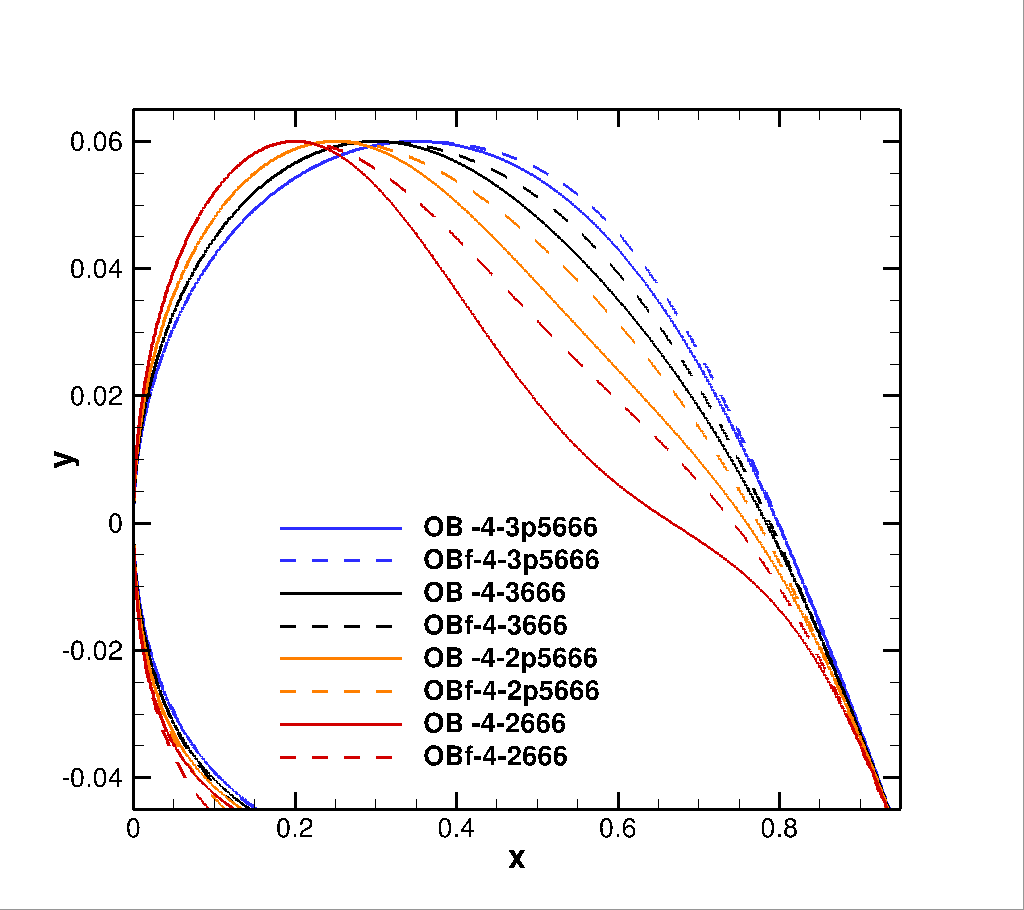}
\caption{Comparison of original OpenBuffet geometries (solid curves) and modified versions (dashed curves). \label{fig:airfoil_comp_mod}}
\end{figure}
In this section, we will consider a slightly modified version of the OpenBuffet parametric airfoil geometries, denoted by `OBf'. For the same airfoil designation, this is used to check the sensitivity to the upper surface curvature distribution and simulataneously allows a wider parametric study. To obtain a slightly flatter airfoil shape downstream of the crest point, the second derivatives at the crest points have been set to zero ($Y_U''(x=X_U)=Y_L''(x=X_L)=0$). This avoids an inflection point of the polynomials connecting crest points and trailing edge for combinations of parameters leading to a relatively small radius near the crest (\textit{e.g.} low $X_U$ values). Figure \ref{fig:airfoil_comp_mod} shows examples with distorted coordinates for clarity to illustrate the problem. Starting from $X_U=0.35$, the streamwise position of the upper crest point is incrementally reduced for 4-digit profiles. Original and modified OpenBuffet geometries are denoted by solid and dashed curves, respectively. While the differences are relatively small for $X_U \ge 0.3$, we see the appearance of an inflection point in the original OB-4-2p5666 profile (orange curves), which is absent in the OBf-4-2p5666 profile. Even though the modified profile also shows an deflection point for the very extreme case of $X_U=0.2$ (red curves), it is significantly less pronounced compared to the original version. 

This modification allows us to cover an extended parameter range and is also used to assess buffet sensitivities to the shape of the polynomial connecting crest points and trailing edge. We will limit this study to free-transitional test cases, as we expect similar trends for the (computationally more expensive) tripped cases. For free-transitional test cases, we can also study sensitivities of intermediate-frequency phenomena, which have not been observed for the tripped cases within the considered parameter space. 
In this section, test cases labels are composed of the type of geometry (``OBf'' or ``OB'' for flattened or original OpenBuffet profiles), the design angle $\alpha_D$, and the four digits $D1-D4$, where the decimal point is replaced by `p'.

\subsection{Comparison between flattened and original OpenBuffet profiles}\label{sec:flat_vs_original}

\begin{figure}[hbt!]
\centering
  \begin{tabular}{ll}
    a) & b) \\
    \includegraphics[width=0.45\textwidth,trim={10mm 10mm 20mm 20mm},clip]{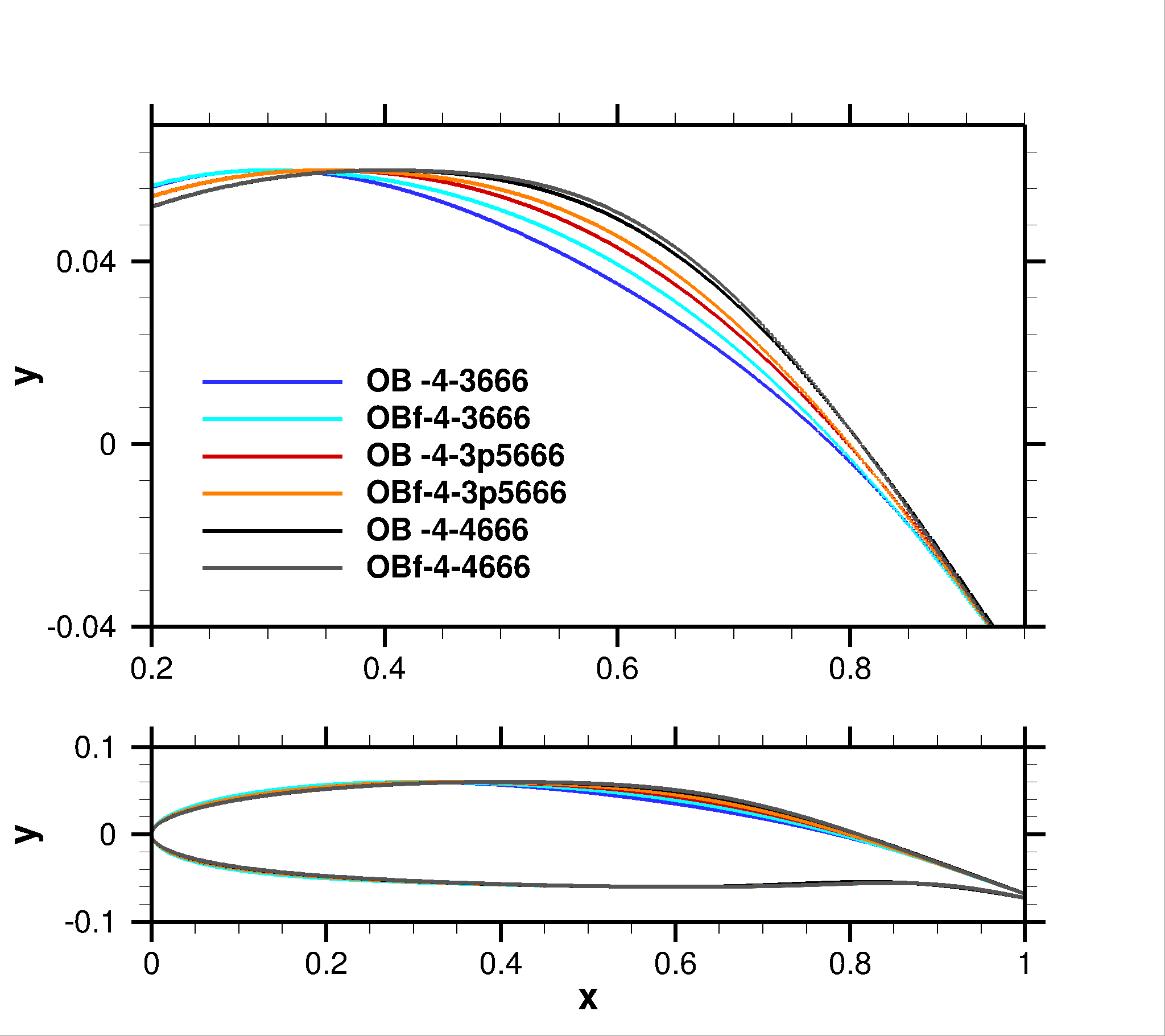} &
    \includegraphics[width=0.45\textwidth,trim={10mm 10mm 20mm 10mm},clip]{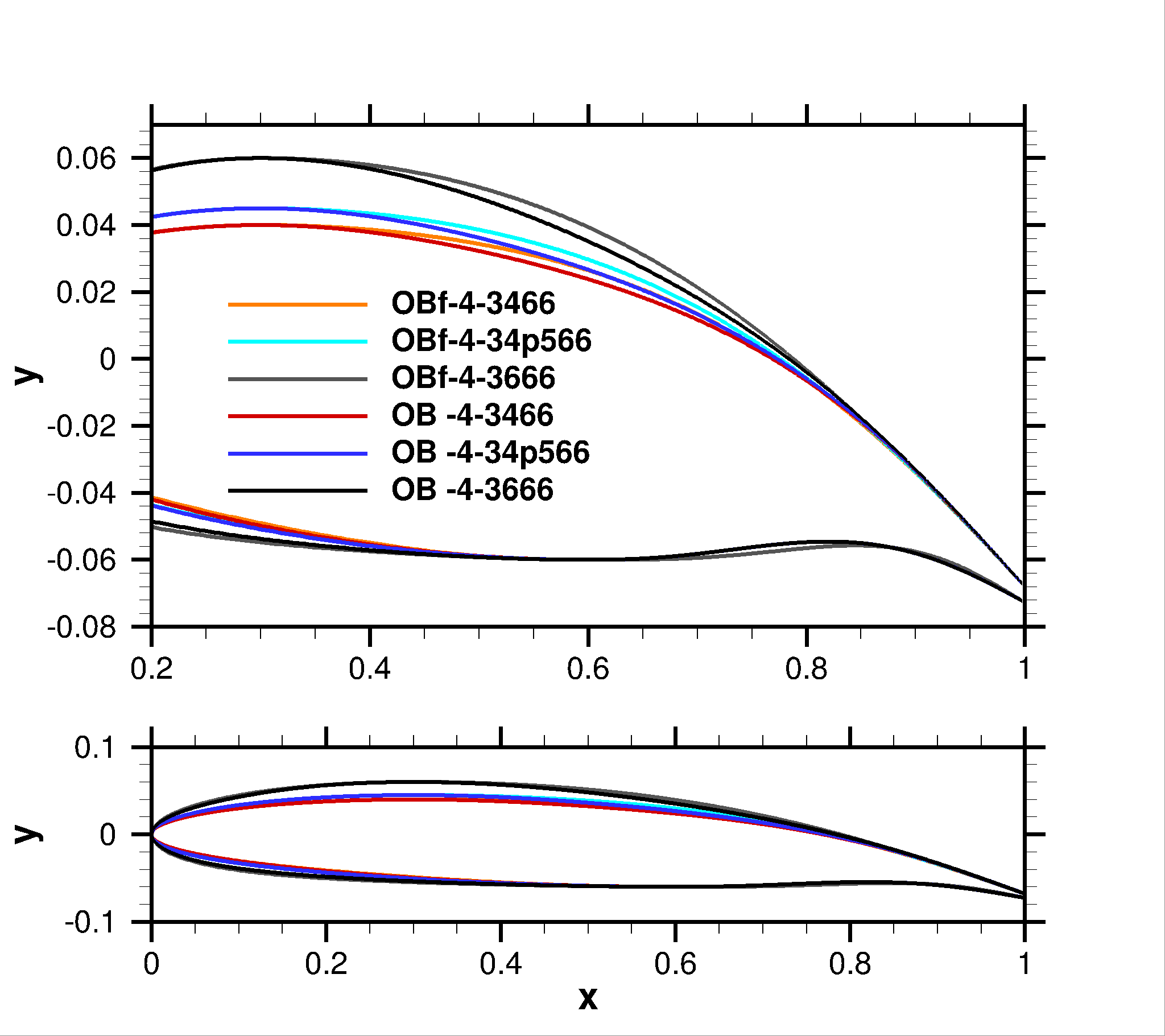} \\
  \end{tabular}
\caption{Selected flattened and original OpenBuffet airfoil profiles for variations of (a) first ($D1$) and (b) second digits ($D2$). While the aspect ratio of the upper plot axes is distorted, the aspect ratio of the lower plot is x:y=1:1.  \label{fig:mod_geom}}
\end{figure}
To compare flattened and original OpenBuffet geometries, figure \ref{fig:mod_geom} shows a selection of profiles, where (a) the axial and (b) the vertical location of the upper crest point is altered. We observe more pronounced differences for profiles with decreased $X_U$ and/or increased $Y_U$.
For brevity, we will show only tabulated results and provide corresponding lift histories in the appendix \S \ref{sec:appendix_comp_flat_orig}.
\begin{table}
\centering
\begin{tabular}{lccccccccc}
Case & $D_1$ & $St_B$ & $St_I$ & $\Delta C_{L,max}$ & $C_{L,max}$ & $C_{L,min}$ & $C_{D,max}$ & $C_{D,min}$ & $\overline{L}/\overline{D}$\\
\hline
4-OBf-3666   & 3 & 0.115 & 0.520 & 0.51 & 1.15 & 0.64 & 0.100 & 0.044 & 12.6 \\ %second freq & 0.115
4-OB -3666   & 3 & 0.115 & -    & 0.46 & 1.09 & 0.63 & 0.102 & 0.039 & 12.4 \\
\hline
4-OBf-3p5666 & 3.5 & 0.165 & 0.657 & 0.28 & 1.04 & 0.76 & 0.085 & 0.040 & 15.0 \\
4-OB -3p5666 & 3.5 & 0.153 & 0.615 & 0.34 & 1.10 & 0.76 & 0.102 & 0.036 & 14.3 \\ %second freq & 0.306
\hline
% 4-OBf-4666   & -     & 0.370 & 0.32 & 1.02 & 0.70 & 0.116 & 0.051 & 10.8 \\
% 4-OB -4666   & -     & 0.560 & 0.24 & 0.94 & 0.70 & 0.104 & 0.048 & 11.1 \\
4-OBf-4666   & 4 & -     & 0.367 & 0.32 & 1.02 & 0.7 & 0.116 & 0.051 & 10.9 \\
4-OB -4666   & 4 & -     & 0.338 & 0.25 & 0.94 & 0.69 & 0.104 & 0.048 & 11.1 \\ % longer statistics
%0.326 & 0.24 & 0.94 & 0.70 & 0.104 & 0.048 & 11.1 \\
\hline \\
\end{tabular}
\caption{Summary of buffet characteristics comparing original and modified OpenBuffet geometries, varying the first digit.}
\label{tab:D1_orig_flat}
\end{table}
Table \ref{tab:D1_orig_flat} summarises the usual quantities for flattened OBf-profiles and original OB-profiles considering variations of the axial location of the upper crest point ($D1$). While we observe good overall agreement of $C_L$ and aerodynamic performance, we observe moderate variations of $C_D$. Buffet frequencies seem to be hardly affected by the geometric differences. 
% At intermediate frequencies, however, we see significant differences. While geometries appear very similar for cases with $X_U=4.0$ in figure \ref{fig:mod_geom}(a), intermediate frequencies are significantly reduced for the flatter profile. However, we have to bear in mind that the flow for these cases is quite irregular and simulation run times have been limited. Cutting of the first $30$ CTUs instead of $15$ CTUs, both signals delivers $St_I \approx 0.35$.
% Also for the cases with $X_U=3.0$, we observe difference in that frequency range. 
While spectra for the original OpenBuffet geometry do not exhibit a clear peak for $X_U=0.3$, we observe one at intermediate frequencies of $St_I=0.52$ for the modified profile. Based on \cite{Zauner2023b}, intermediate frequencies associated with separation-bubble phenomena scale with the length of the separation bubble and maximum reverse velocities, which are highly sensitive to flow conditions.

\begin{table}
\centering
\begin{tabular}{lccccccccc}
Case & $D_2$ & $St_B$ & $St_I$ & $\Delta C_{L,max}$ & $C_{L,max}$ & $C_{L,min}$ & $C_{D,max}$ & $C_{D,min}$ & $\overline{L}/\overline{D}$\\
\hline
4-OBf-3466   & 4 & 0.057 & 0.50 & 0.09 & 1.23 & 1.14 & 0.057 & 0.041 & 24.8 \\
4-OB -3466   & 4 & 0.050 & 0.55 & 0.06 & 1.19 & 1.12 & 0.051 & 0.040 & 25.0 \\ %second freq & 0.12
\hline
4-OBf-34p566 & 4.5 & 0.077 & 0.54 & 0.15 & 1.18 & 1.03 & 0.062 & 0.035 & 22.9 \\
4-OB -34p566 & 4.5 & 0.077 & 0.57 & 0.20 & 1.17 & 0.96 & 0.059 & 0.033 & 22.3 \\ %second freq & 0.268
\hline
4-OBf-3666   & 6 & 0.115 & 0.52 & 0.51 & 1.15 & 0.64 & 0.100 & 0.044 & 12.6 \\ %second freq & 0.115
4-OB -3666   & 6 & 0.115 & -    & 0.46 & 1.09 & 0.63 & 0.102 & 0.039 & 12.4 \\ % mean CL 8.914391e-01 CD 2*3.418608e-02
\hline \\
\end{tabular}
\caption{Summary of buffet characteristics comparing original and modified OpenBuffet geometries, varying the second digit.}
\label{tab:D2_orig_flat}
\end{table}
Table \ref{tab:D2_orig_flat} summarises the usual quantities for flattened OBf-profiles and original OB-profiles considering variations of the vertical location of the upper crest point ($D2$). Again, we observe very good agreement in terms of mean properties. Extreme values for $C_L$ and $C_D$ do not show significant sensitivities to the geometric modifications. While buffet frequencies show excellent agreement for cases with $Y_U=0.06$ and $0.045$, we observe some variations for $Y_U=0.04$. Also for these cases, intermediate frequencies show slightly increased sensitivities to the geometric variations.

Overall, we can confirm the same trends for modified as well as original OpenBuffet geometries, when varying the upper crest position. We do observe difference of extreme values for $C_L$ and $C_D$ (which may be also due to limited run times), but mean flow quantities (\textit{e.g.} $\overline{L}/\overline{D}$) show very good agreement. Also, while buffet frequencies show minor sensitivities, reduced curvature (flatter profiles) seems to promote separation-bubble instabilities at intermediate frequencies as well as shifting $St_I$ for most cases to lower frequencies. According to \citet{Zauner2023b}, this may be due to reduced separation bubble length or decreased reverse velocities. 

\subsection{Analysis of extended parameter space using flattened OpenBuffet profiles}
As mentioned at the beginning of this section, the flattened OpenBuffet profiles allow us to study an extended parameter space without additional inflection points appearing within the rear part of the suction side.
Table \ref{tab:results_flat} shows a summary of our usual quantities to characterise aerodynamic properties and buffet phenomena for all selected cases. For brevity, we will focus here on quantitative results of this table and provide full histories for all cases in appendix \S \ref{sec:appendix_hist_flat}. 

\begin{table}
\centering
\begin{tabular}{llcccccccc}
Case & $D_x$ & $St_B$ & $St_I$ & $\Delta C_{L,max}$ & $C_{L,max}$ & $C_{L,min}$ & $C_{D,max}$ & $C_{D,min}$ & $\overline{L}/\overline{D}$\\
\hline
4-OBf-2p5666 & $D_1=2.5$ & 0.128 & 0.43 & 0.41 & 0.99 & 0.59 & 0.122 & 0.032 & 11.6 \\
4-OBf-3666   & $D_1=3$ & 0.115 & 0.52 & 0.51 & 1.15 & 0.64 & 0.100 & 0.044 & 12.6 \\ 
4-OBf-3p5666 & $D_1=3.5$ & 0.165 & 0.66 & 0.28 & 1.04 & 0.76 & 0.085 & 0.040 & 15.0 \\
4-OBf-4666   & $D_1=4$ & -     & 0.37 & 0.32 & 1.02 & 0.70 & 0.116 & 0.051 & 10.8 \\
\hline
4-OBf-3466   & $D_2=4$ & 0.057 & 0.50 & 0.09 & 1.23 & 1.14 & 0.057 & 0.041 & 24.8 \\
4-OBf-34p566 & $D_2=4.5$ & 0.077 & 0.53 & 0.15 & 1.18 & 1.03 & 0.062 & 0.035 & 22.9 \\
4-OBf-3566   & $D_2=5$ & 0.096 & 0.52-0.61 & 0.40 & 1.19 & 0.79 & 0.079 & 0.031 & 18.3 \\
4-OBf-3666   & $D_2=6$ & 0.115 & 0.52 & 0.51 & 1.15 & 0.64 & 0.100 & 0.044 & 12.6 \\ 
4-OBf-3766   & $D_2=7$ & 0.134 & 0.63 & 0.48 & 1.13 & 0.65 & 0.130 & 0.051 & 10.2 \\
\hline
\hline
4-OBf-3646   & $D_3=4$ & 0.115 & 0.40 & 0.49 & 1.20 & 0.71 & 0.111 & 0.044 & 12.7 \\
4-OBf-3656   & $D_3=5$ & 0.115 & 0.52 & 0.48 & 1.17 & 0.68 & 0.117 & 0.041 & 12.5 \\
4-OBf-3666   & $D_3=6$ & 0.115 & 0.52 & 0.51 & 1.15 & 0.64 & 0.100 & 0.044 & 12.6 \\ 
\hline
4-OBf-3663   & $D_4=3$ & 0.115 & 0.46 & 0.43 & 1.30 & 0.87 & 0.120 & 0.050 & 13.2 \\
4-OBf-3666   & $D_4=6$ & 0.115 & 0.52 & 0.51 & 1.15 & 0.64 & 0.100 & 0.044 & 12.6 \\ 
4-OBf-3669   & $D_4=9$ & 0.096 & 0.54 & 0.45 & 0.89 & 0.44 & 0.074 & 0.034 & 12.6 \\
\hline \\
\end{tabular}
\caption{Summary of buffet characteristics for flattened OpenBuffet geometries considering variations of $D_1-D_4$. Each section of the table shows variations of a single digit.}
\label{tab:results_flat}
\end{table}
The first (upper) section of table \ref{tab:results_flat} contains simulation results for variations of the axial position on the upper crest point. 
Given the additional simulation with $X_U=0.25$, it indeed seems as the axial position may impact buffet on-set and off-set characteristics. Buffet amplitudes increase from $X_U=0.25$ to $X_U=0.30$ before incrementally decreasing again for $X_U>0.30$. While the shock position moves downstream when increasing $X_U$, flow separation on the rear part of the airfoil is promoted by the steep slope of the upper surface contour. Eventually we observe off-set conditions at $X_U=0.40$. Tendencies of $C_L$, $C_D$, and lift-over-drag ratio align with this explanation as well. 
Clear trends are also observed in the second section of table \ref{tab:results_flat}, where buffet frequencies as well as amplitudes continuously increase with increasing $Y_U$. Separation-bubble instabilities at intermediate frequencies do not seem to be significantly affected, but they become less pronounced for $Y_U=0.05$, where drag is low.

\begin{figure}[hbt!]
\centering
  \begin{tabular}{ll}
    a) & b) \\
    \includegraphics[width=0.45\textwidth,trim={10mm 10mm 20mm 100mm},clip]{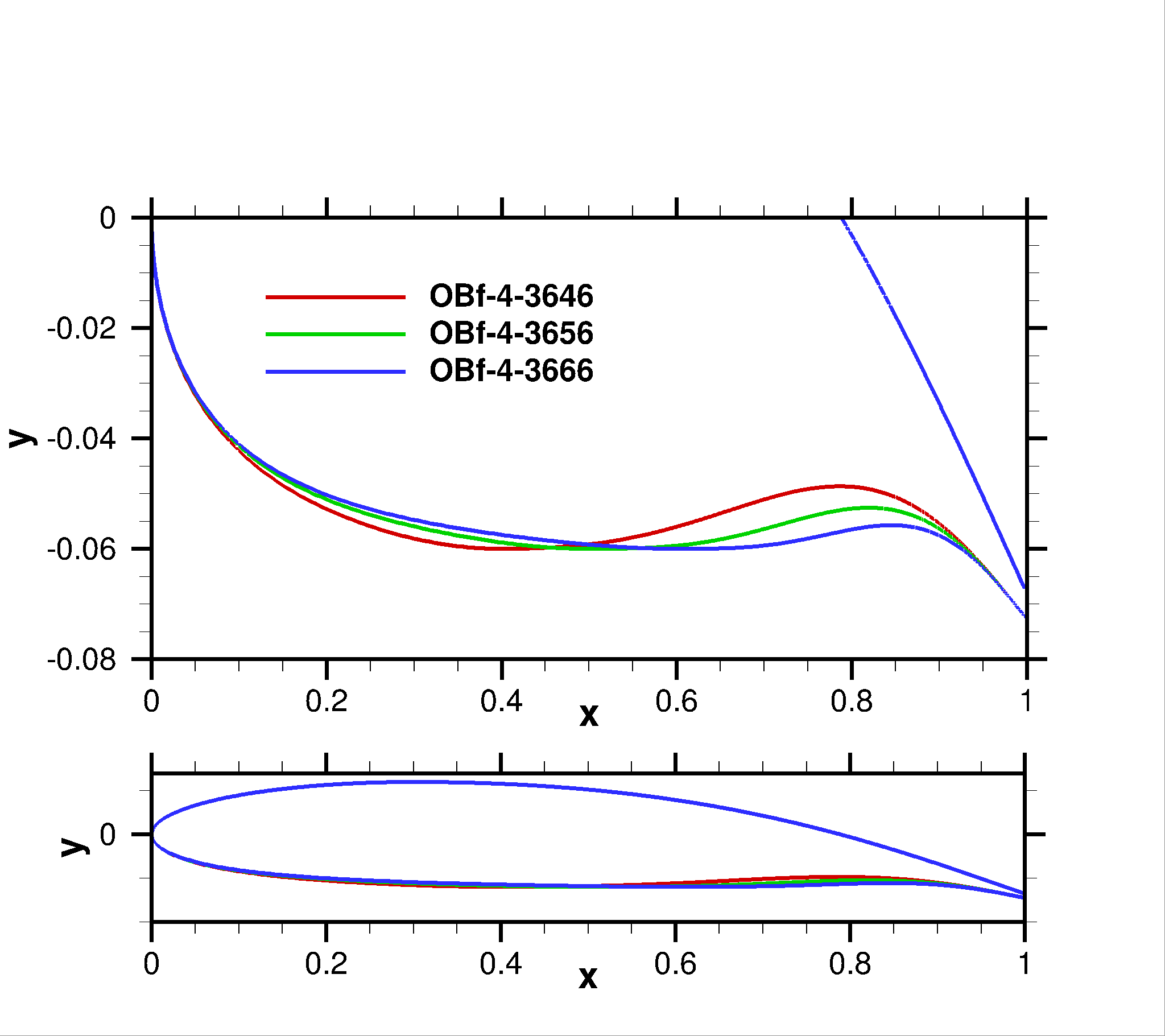} &
    \includegraphics[width=0.45\textwidth,trim={10mm 10mm 20mm 100mm},clip]{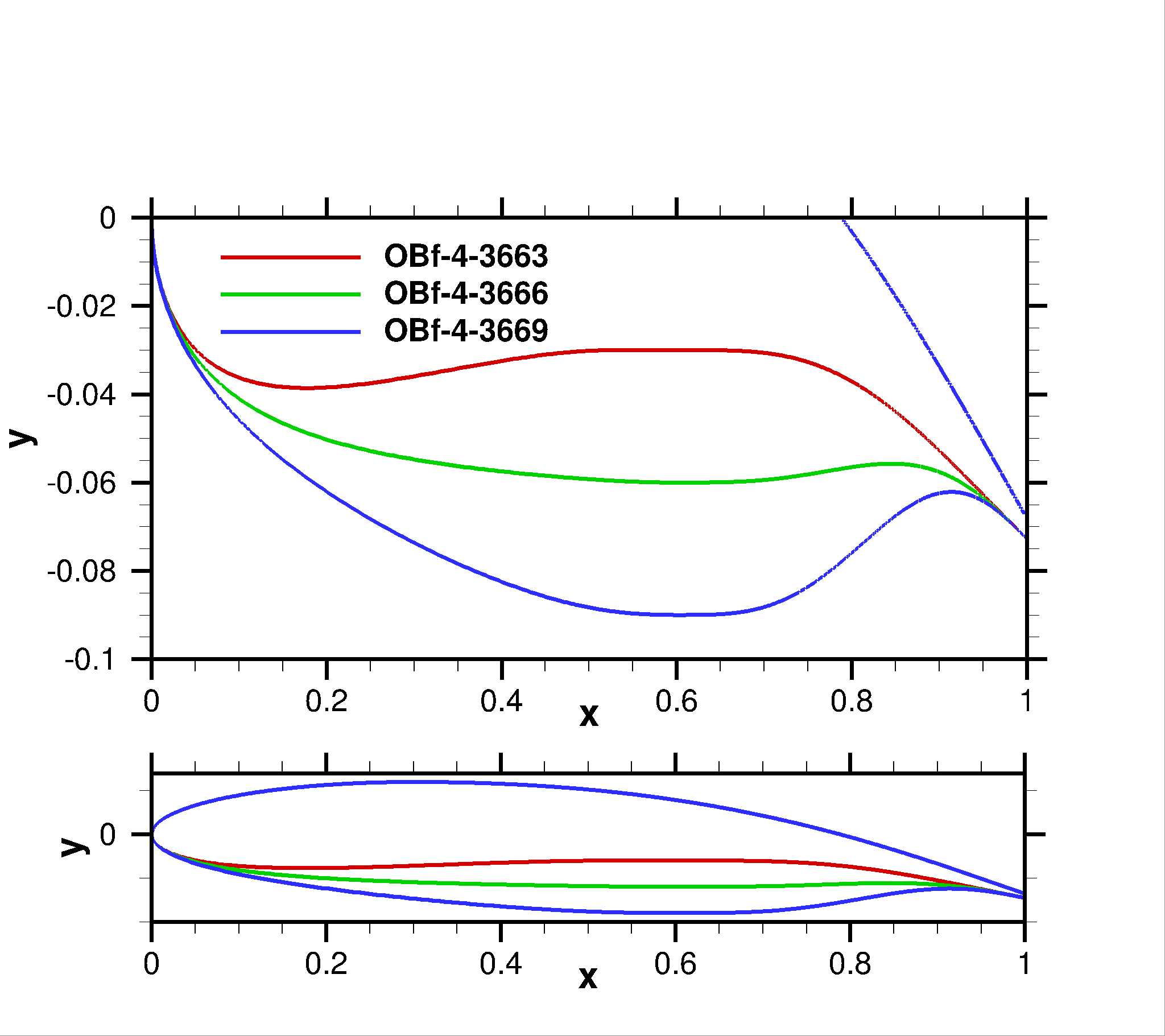} \\
  \end{tabular}
\caption{Flattened OpenBuffet airfoil profiles for variations of (a) third ($D3$) and (b) fourth digits ($D4$). While the aspect ratio of the upper plot axes is distorted, the aspect ratio of the lower plot is x:y=1:1\label{fig:lower_geometries}}
\end{figure}
While the upper crest points have already been analysed for original OpenBuffet geometries as well, the lower (pressure side) crest-point locations have only been studied for the flattened OpenBuffet profiles of this section. Geometries are shown in figure \ref{fig:lower_geometries} and corresponding simulation results are summarised in the bottom two sections of table \ref{tab:results_flat}. Although the axial location of the lower crest point ($X_L$) shows hardly any influence on the aerodynamic properties or buffet characteristics at present conditions, we do observe influence of the vertical location $Y_L$. We should emphasise that the lower crest point becomes concave for $D_4=3$. While the inflection point of the lower-side geometry is usually located between the lower crest and the trailing edge, it shifts in the upstream direction when increasing $Y_L$. For the red profile in figure \ref{fig:lower_geometries}(b), the axial location of the inflection point is eventually located upstream of $X_L$. Nonetheless, this effect shows no significant effect on the buffet characteristics, even though aerodynamic forces increase with increasing $Y_L$. Intermediate frequencies, however, tend to decrease with increasing $Y_L$. When moving from $D_4=6$ to $D_4=9$, however, we observe a decrease of the buffet frequency, but no significant changes in the lift-over-drag ratio.

\begin{figure}[hbt!]
\centering
\includegraphics[width=0.70\textwidth,trim={10mm 10mm 20mm 20mm},clip]{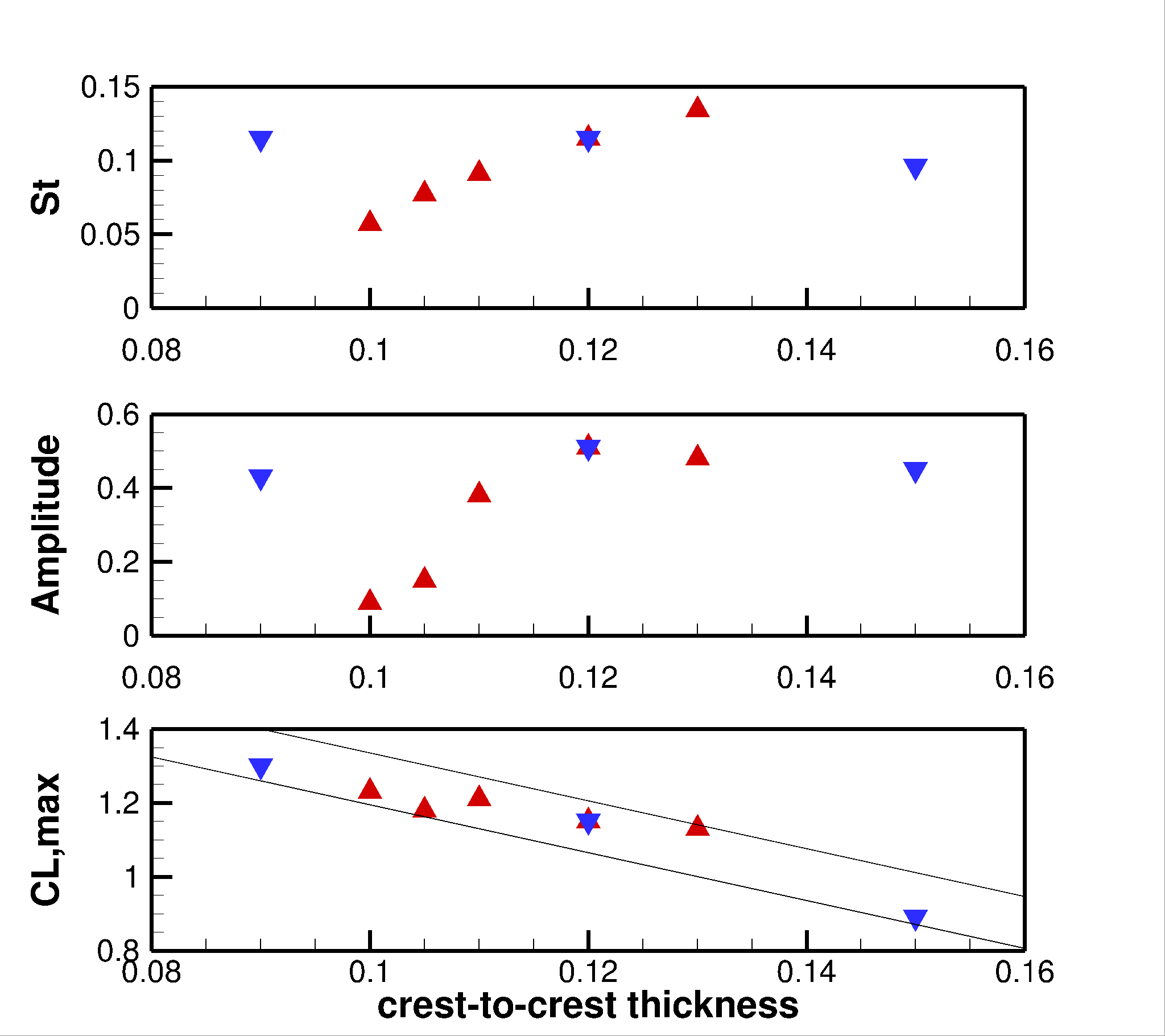} 
  % \begin{tabular}{ll}
  %   a) & b) \\
  %   \includegraphics[width=0.48\textwidth,trim={10mm 10mm 20mm 20mm},clip]{Figures_4digit/Scaling.png}  &
  %   \includegraphics[width=0.48\textwidth,trim={10mm 10mm 20mm 20mm},clip]{Figures_4digit/Scaling_2.png}  \\
  % \end{tabular}
\caption{Buffet frequency (top), amplitude (center), and maximum lift (bottom) as a function of the crest-to-crest thickness ($Y_U-Y_L$). Red and blue symbols denote variations of $Y_U$ and $Y_L$, respectively.
%(b) Oscillation amplitudes ($\Delta C_L$) are shown for all cases as a function of mean lift-over-drag ratio. 
\label{fig:trends_thickness}}
\end{figure}
Figure \ref{fig:trends_thickness} shows buffet frequency (top plot), buffet amplitude (center plot), and maximum lift (bottom plot) as a function of the crest-to-crest thickness ($Y_U-Y_L$). Simulation results for variations of the vertical crest position of the upper and lower sides are denoted by red and blue symbols, respectively.
Adjusting the suction side, the buffet frequency increases almost linearly with the crest-to-crest thickness. Variations of the crest-to-crest thickness via the lower crest position show only little effect. 
Looking at the $\Delta C_{L,max}$ associated with the buffet amplitude, we observe an initial increase, peaking at a crest-to-crest thickness of $0.12$. When buffet is established, amplitudes show only minor sensitivity to changes of the crest-to-crest thickness.
The maximum lift, however, shows a common trend across all simulations modifying $Y_U$ or $Y_L$. Despite some variations, it steadily decreases with increasing crest-to-crest thickness. Black lines indicate a corridor corresponding to a $\Delta C_L=0.14$ (corresponding to approximately $10\%$ of $C_{L,max}$) for visual aid with respect to the spread of results. Due to the complex nature of present simulations and limited run-time such variations are not surprising. For longer run times and statistical convergence, the corridor containing simulation results may become narrower. 
% Figure \ref{fig:trends_thickness}(b) shows oscillation amplitudes ($\Delta C_L$) are shown for all cases as a function of the mean lift-over-drag ratio. We observe a general increase of 

% $St_B$ Differences due to upper crest location 
%-> -0.028 v2c vs OALT25
%-> +0.010 oalt25 vs OAT15A
% \begin{table}[htb]    
%     \caption{Comparison of buffet characteristics of OALT25 and OAT15A airfoils. Cases at $Re=3\!,000\!,000$ correspond to experimental data of \cite{Brion2019} and \cite{JMDMS2009}. The power-spectral density (PSD) measurements for OALT25 and OAT15A profiles are respectively based on pressure-sensor data at $x/c=0.73$ and $0.45$.}
%     \begin{center}
%     \def~{\hphantom{0}}
%         \begin{tabular}{l|c|c|c|c|c|c|c}
%         Airfoil  & $\alpha$ & $St_B$ & $X_U$ & $Y_U$ & $X_L$ & $Y_L$ & crest-to-crest thickness                         \\
%         \hline
%         V2C      & $4.0^{\circ}$ & 0.12-0.15 & 0.33 & 0.055 & 0.59 & -0.095 & 0.150 \\
%         OALT25   & $4.0^{\circ}$ & 0.05-0.07 & 0.26 & 0.044 & 0.54 & -0.087 & 0.131  \\
%         % OAT15A   & $3.5^{\circ}$ & 0.07      & 0.25 & 0.048 & 0.37 & -0.077 & 0.125 \\
%         \hline
%         \end{tabular}
%   \label{tab:airfoils_tripped}
%   \end{center}
% \end{table}

\section{Conclusions} \label{sec:conclusion}
An airfoil construction method has been devised to explore the geometric sensitivities of large-scale oscillations at typical buffet conditions in an intuitive way. 
Using Large-Eddy Simulations, an extensive parametric study has been carried out for four-digit OpenBuffet profiles considering free-transitional as well as tripped boundary layers at moderate Reynolds numbers of $Re=500,\!000$. 

General trends for variations of the suction-side geometry agree very well for both free-transitional and tripped boundary-layer conditions.
We observe significant influence of the upper-side crest location to buffet on- and off-set conditions. Shifting the upper crest towards the trailing edge promotes buffet off-set, while promoting separation effects and intermediate-frequency phenomena for free-transitional cases. 
Within the considered parameter space, no intermediate-frequency phenomena associated with separation-bubble instabilities have been observed for tripped cases. 
Reducing the airfoil thickness via the vertical location of the upper crest ($Y_U$) tends to stabilise the flow and increase the aerodynamic efficiency. Also the Strouhal number of the buffet frequency increases with $Y_U$. The maximum lift coefficient, however, does not change significantly. A similar trend was observed in \cite{Zauner2023b} when increasing the Mach number for ONERA's OALT25 profile at a fixed angle of attack. Variations of the axial location of the pressure-side crest show minor impact and can be neglected for future studies. The vertical displacement of the crest, however, can affect levels of maximum as well as buffet frequencies.

Scaling of intermediate frequencies based on mean-flow properties of the separation bubble was applied to the OB-F-D2-4p5 test case (OB-4-34p566 profile) and agrees well with reference data of different airfoil geometries in \cite{Zauner2023b}.

Modified versions of the OpenBuffet geometries have been studied as well, where the upper-side geometry downstream of the crest is slightly flattened to increase the parameter space. Flatter profiles show small effects on mean properties, but promote intermediate-frequency phenomena.

For present airfoil geometries, the maximum lift levels scale well with the crest-to-crest thickness. For cases with established buffet, the crest-to-crest thickness has a minor effect on buffet amplitudes. Buffet frequencies scale almost linearly with variations of the vertical position of the upper crest point. Therefore, the airfoil thickness is not sufficient to explain differences of frequencies between different airfoils. Instead, the upper crest location of the rotated airfoil shows significant influence on buffet characteristics. For a given geometry, buffet characteristics do not only depend on the shape at zero inclination, but also the angle of attack, which has significant influence on the crest locations.

The present study provides data for a wide parameter range, which can be used by the research community for comparison with lower-fidelity methods like RANS or URANS. Present OpenBuffet geometries are publicly available via \url{https://doi.org/10.5281/zenodo.12204411}.

% Mind that when we change the angle of attack, we do not only modify the inclination of the chord, but also the location of the crest points with respect to the free stream. While effect of chord-rotation on aerodynamic properties can be explained by the thin-arifoil theory (i.e. potential flow), we could show in the present study that buffet characteristics are highly sensitive to the crest-point locations. Therefore, we need to be careful when concluding form angle-of-attack studies on the nature of transonic buffet.

\section*{Appendix}

\subsection{Comparison of tripped and free-transitional OpenBuffet reference cases}\label{sec:appendix_lam_turb}
\begin{figure}[hbt!]
\centering
  \begin{tabular}{ll}
    a) & b) \\
    \includegraphics[width=0.45\textwidth,trim={10mm 10mm 10mm 10mm},clip]{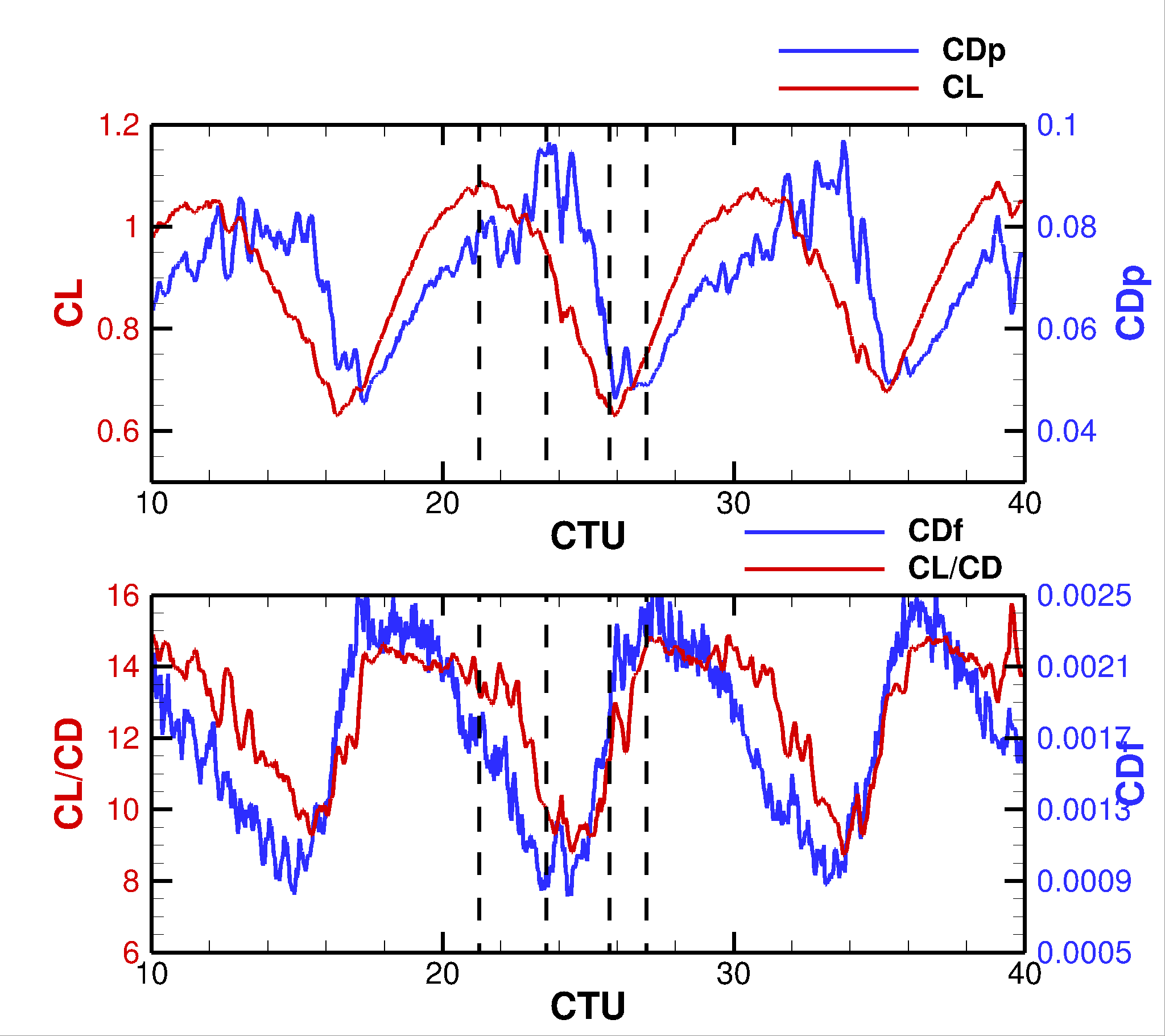} &
    \includegraphics[width=0.45\textwidth,trim={10mm 10mm 10mm 10mm},clip]{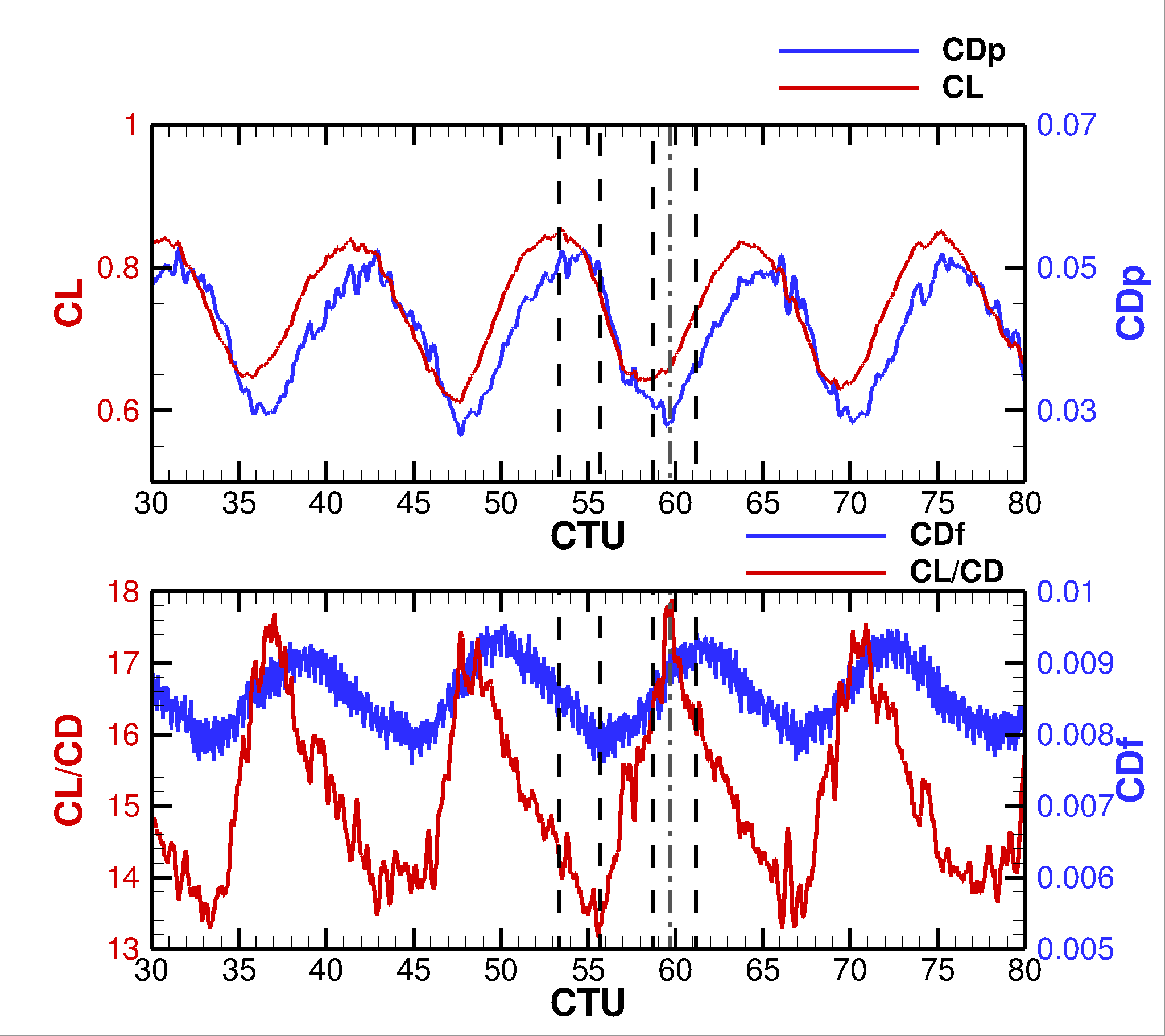} \\
  \end{tabular}
\caption{Histories of aerodynamic coefficients for OpenBuffet airfoils considering (a) free-transitional (OB-4-3666) and (b) turbulent boundary layers (OB-3p5-2p8'4p8'3p8'7p8).  \label{fig:laminar_vs_tripped}}
\end{figure}
Comparing tables \ref{tab:qual_inst} and \ref{tab:qual_inst_tripped} of the main text, we can see very clear parallels between transonic buffet at free-transitional and tripped conditions. Figure \ref{fig:laminar_vs_tripped} provides additional histories for (a) free-transitional and (b) tripped reference cases. 
Top plots show lift coefficient $C_L$ (red curves and left-hand-side scale) and pressure-drag coefficient $C_{D,p}$ (blue curves and right-hand-side scale) as a function of convective time units. 
Bottom plots show lift-over-drag ratio $C_L/C_D$ (red curves and left-hand-side scale) and skin-friction-drag coefficient $C_{D,f}$ (blue curves and right-hand-side scale) as a function of convective time units. 
For both baseline cases, $C_L$ histories are approximately sinusoidal. While this remains true for $C_{D,p}$ of the tripped case, we observe a saw-tooth pattern (slow increase and rapid decrease) for the free-transitional counterpart. 
For $C_{D,f}$ of free-transitional cases, the saw-tooth pattern is reversed, as we observe rapid increase and slower decay. For the turbulent case, the $C_{D,f}$ history appears slightly skewed. 
As boundary layers recover much faster for the free-transitional case, we observe some sort of plateau for $C_{D,f}$ as well as lift-over-drag histories. The latter appears very spiky with flat troughs for the tripped case.
Buffet amplitudes are notably higher for the free-transitional case compared to the tripped case. 
Despite some variations in signal shapes, a striking similarity between both cases is the phase shift between $C_L$ and $C_D$, where lift leads drag by about $90^{\circ}$.
% CL for both cases quite sinusoidal
% CDp for turbulent quite sinusoidal, while sawtooth shape for laminar (sharp drop)
% CDf skewed but fairly sinusoidal for turbulent case, but sawtooth shape for laminar (sharp increase)
% after maximum lift, CDp spikes for laminar case
% min CDf leads min CDp for both cases
% min CDf and min CDp phase shifted by about 90deg for both cases
% CDf recovers very fast for laminar case, so it reaches max CDf when CDp is still low 
% for both, minimum of CL leads minimum of CDp
% CL/CD plateaus for laminar case, but very spikey for turbulent case 

\subsection{Additional data for free-transitional test cases}\label{sec:Add_F_D1}
\begin{figure}
\centering
  \begin{tabular}{ll}
    a) & b) \\
    \includegraphics[width=0.45\textwidth,trim={10mm 10mm 20mm 20mm},clip]{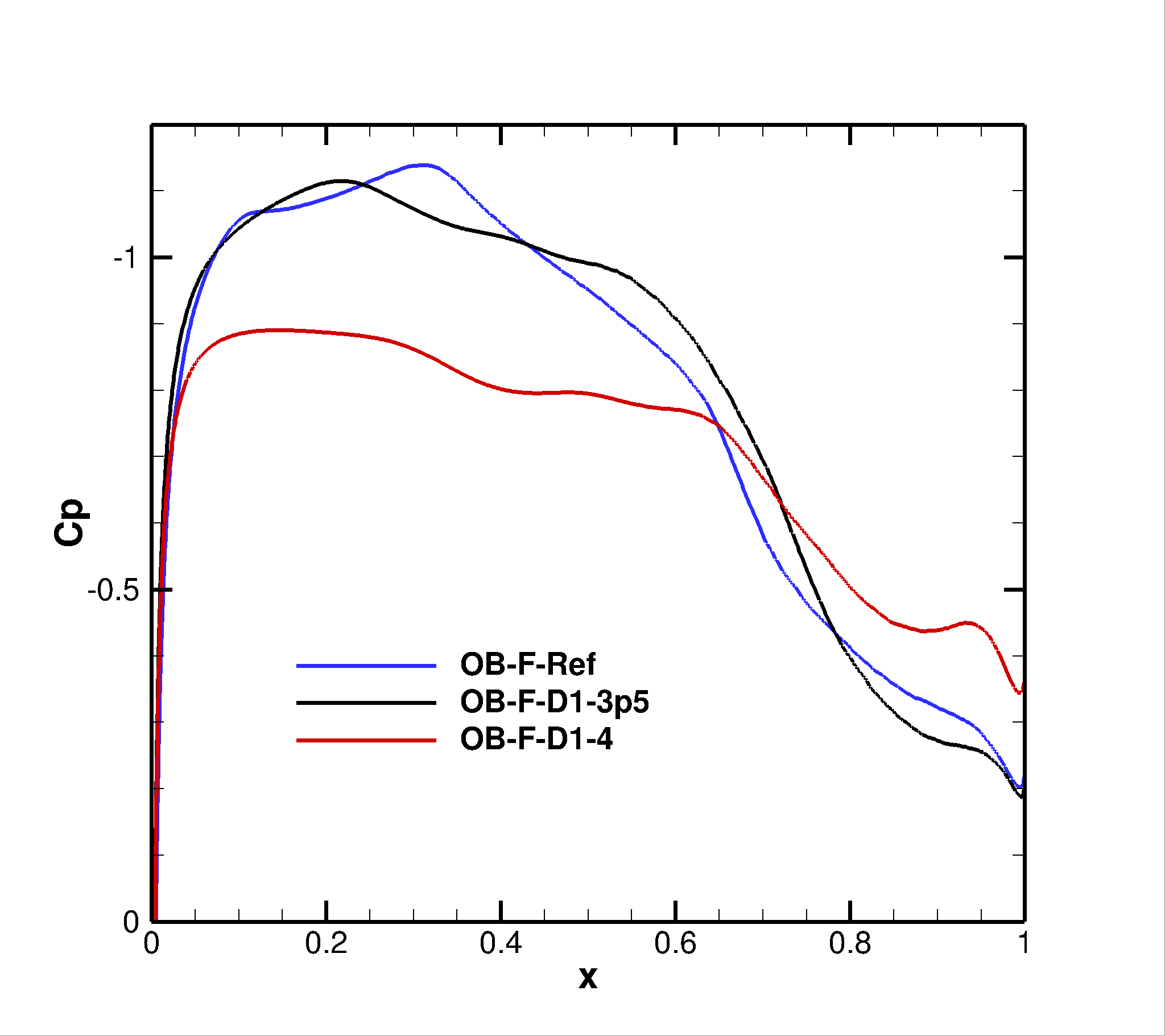} &
    \includegraphics[width=0.45\textwidth,trim={10mm 10mm 20mm 10mm},clip]{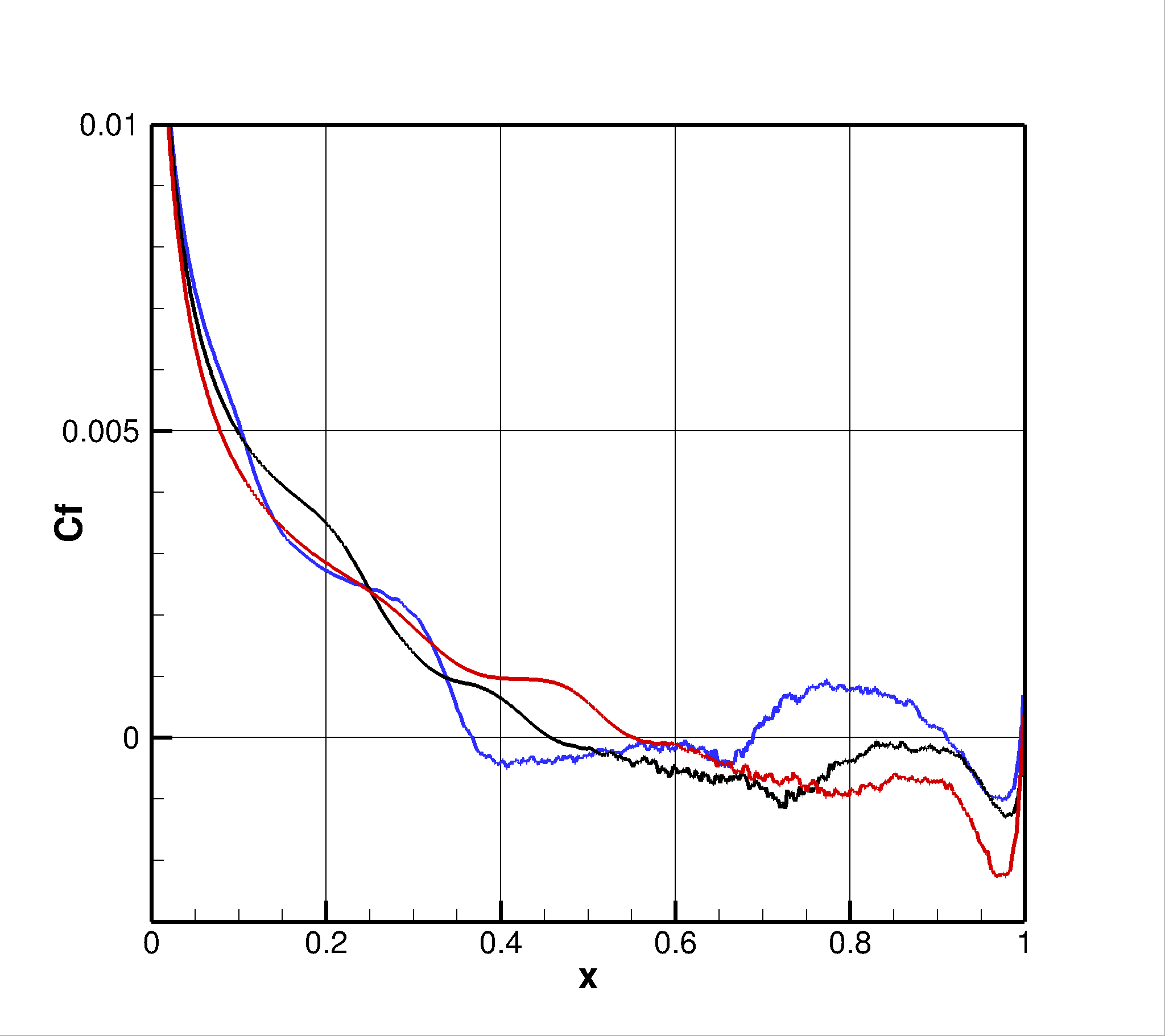} \\
  \end{tabular}
\caption{Distributions of suction-side (a) wall-pressure and (b) skin-friction coefficients for free-transitional cases varying the upper crest location.\label{fig:CpCf_OB-F-D1}}
\end{figure}

Figure \ref{fig:CpCf_OB-F-D1} shows time-averaged suction-side (a) wall-pressure and (b) skin-friction coefficients for free-transitional cases varying the upper crest location. Differences between the blue (OB-4-3666 airfoil) and the black (OB-4-3p5666 airfoil) curves are mainly due to a down-stream shift of the mean shock-wave position. Further downstream displacement of the upper crest leads to significantly reduced suction for the red curve (OB-4-4666 airfoil) in figure \ref{fig:CpCf_OB-F-D1}(a), while the trailing edge pressure diverges significantly due to increased flow separation. We can observe in figure \ref{fig:CpCf_OB-F-D1}(b) that only the blue curve shows clear re-attachment of the mean flow. For increasing $X_U$, $C_f$ levels decrease due to the dominance of separation phenomena. These tendencies are very reminiscent of buffet off-set or incipient stall.

\subsection{Comparison between flattened and original OpenBuffet profiles}\label{sec:appendix_comp_flat_orig}
Lift histories of free-transitional LES are shown for both, flattened and original OpenBuffet profiles. 
\begin{figure}[hbt!]
\centering
  \begin{tabular}{ll}
    a) & b) \\
    \includegraphics[width=0.45\textwidth,trim={10mm 10mm 20mm 20mm},clip]{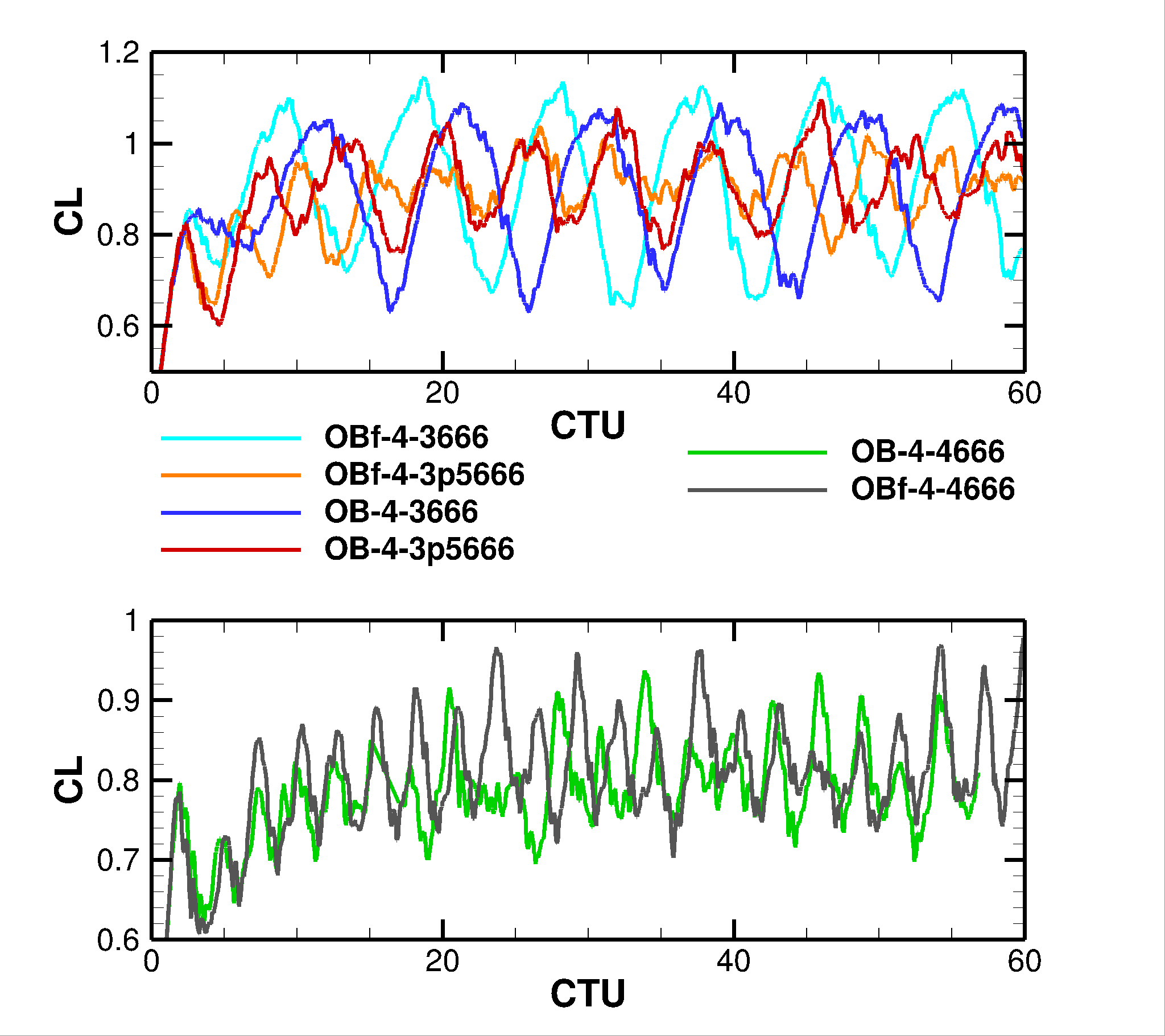} &
    \includegraphics[width=0.45\textwidth,trim={10mm 10mm 20mm 10mm},clip]{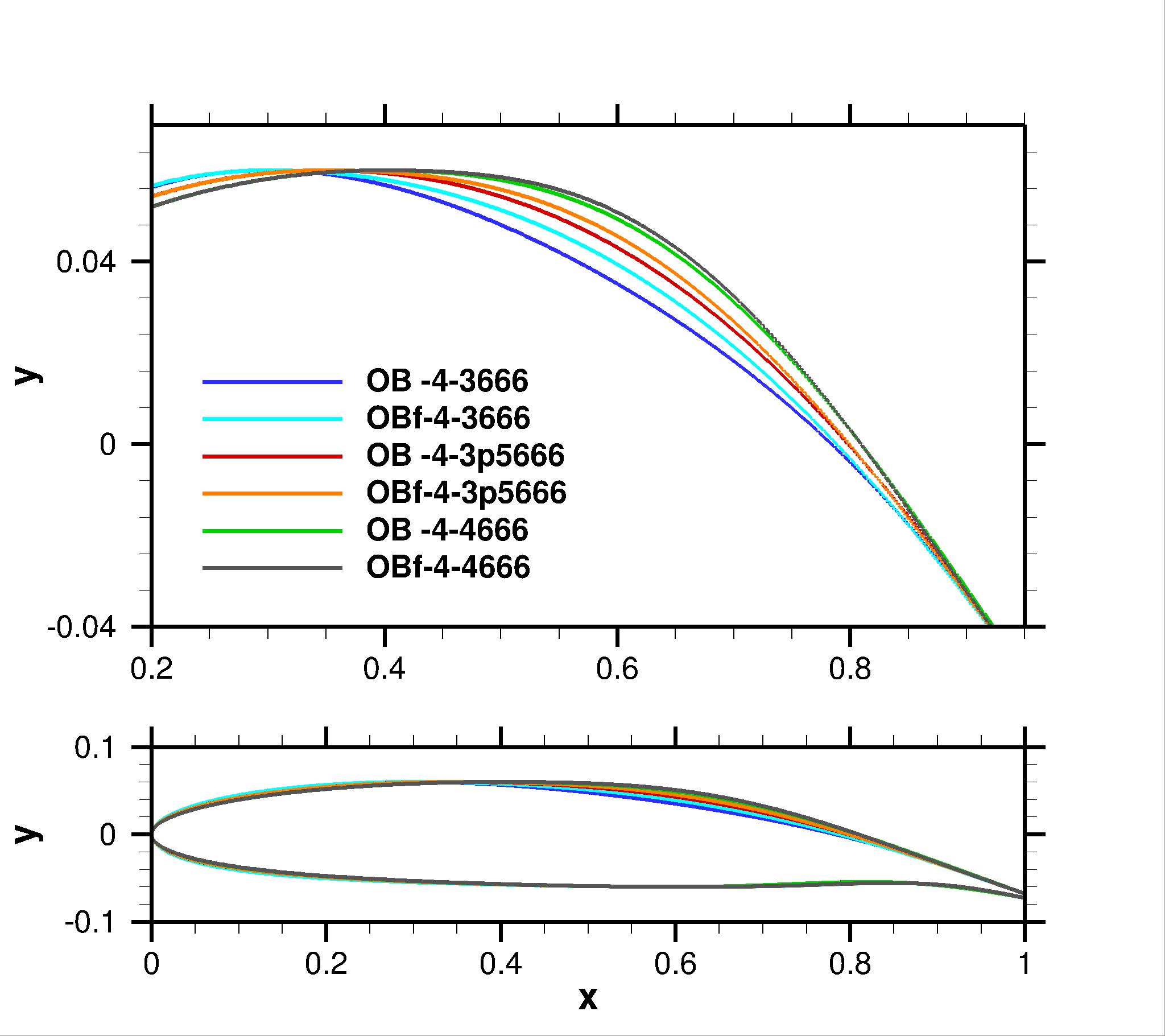} \\
  \end{tabular}
\caption{(a) Lift coefficient ($C_L$) as a function of convective time units ($CTU$) for 4-digit airfoils at an angle of attack $\alpha=4^{\circ}$, where the first digit is varied for modified (OBf-4-X666) and original OB-4-X666 profiles. Corresponding airfoil contours are shown in (b), respectively, where the aspect ratio of the upper plot axes is distorted. The aspect ratio of the lower plot is x:y=1:1.  \label{fig:D1_orig}}
\end{figure}
\begin{figure}[hbt!]
\centering
  \begin{tabular}{ll}
    a) & b) \\
    \includegraphics[width=0.45\textwidth,trim={10mm 10mm 20mm 50mm},clip]{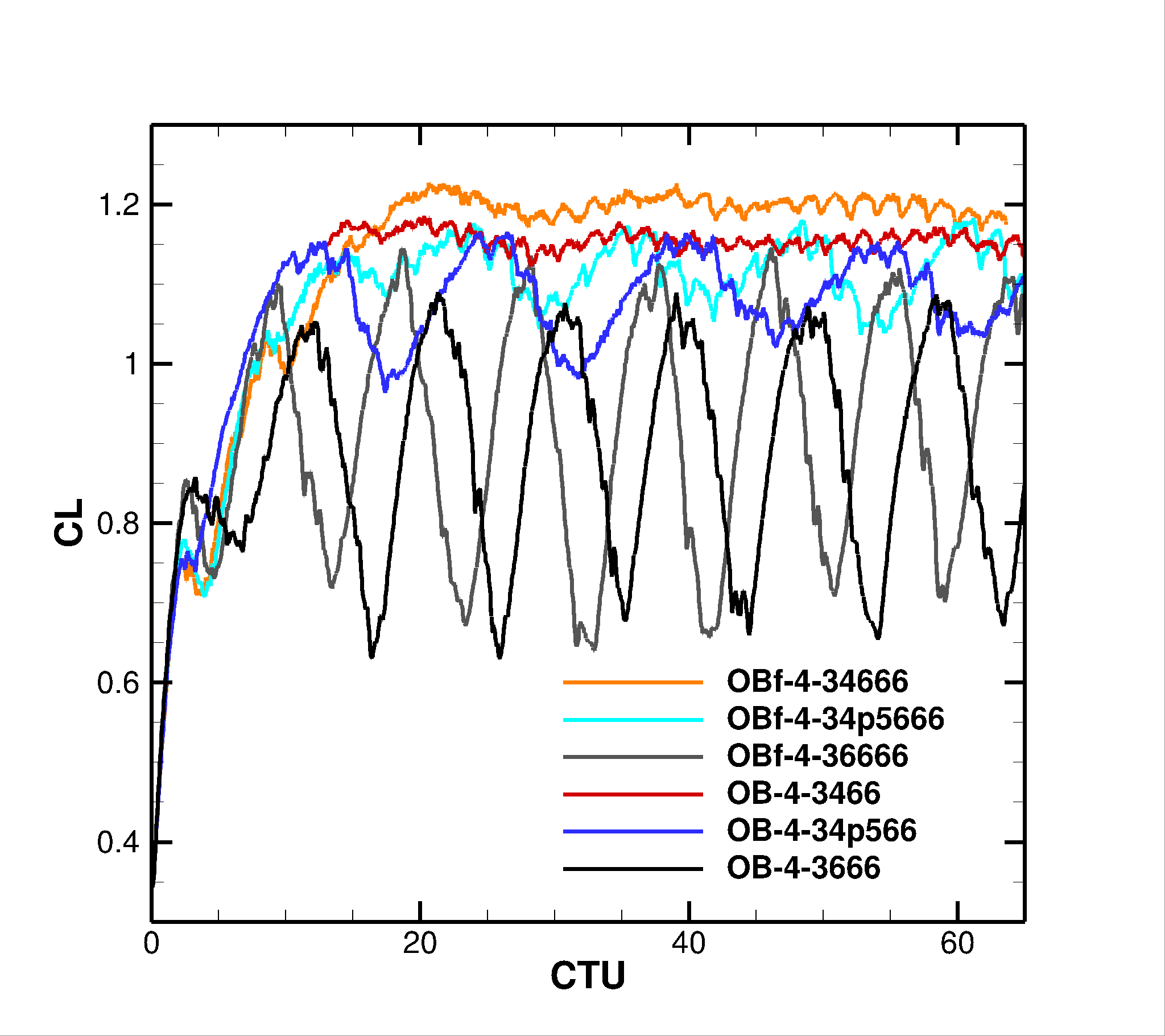} &
    \includegraphics[width=0.45\textwidth,trim={10mm 10mm 20mm 50mm},clip]{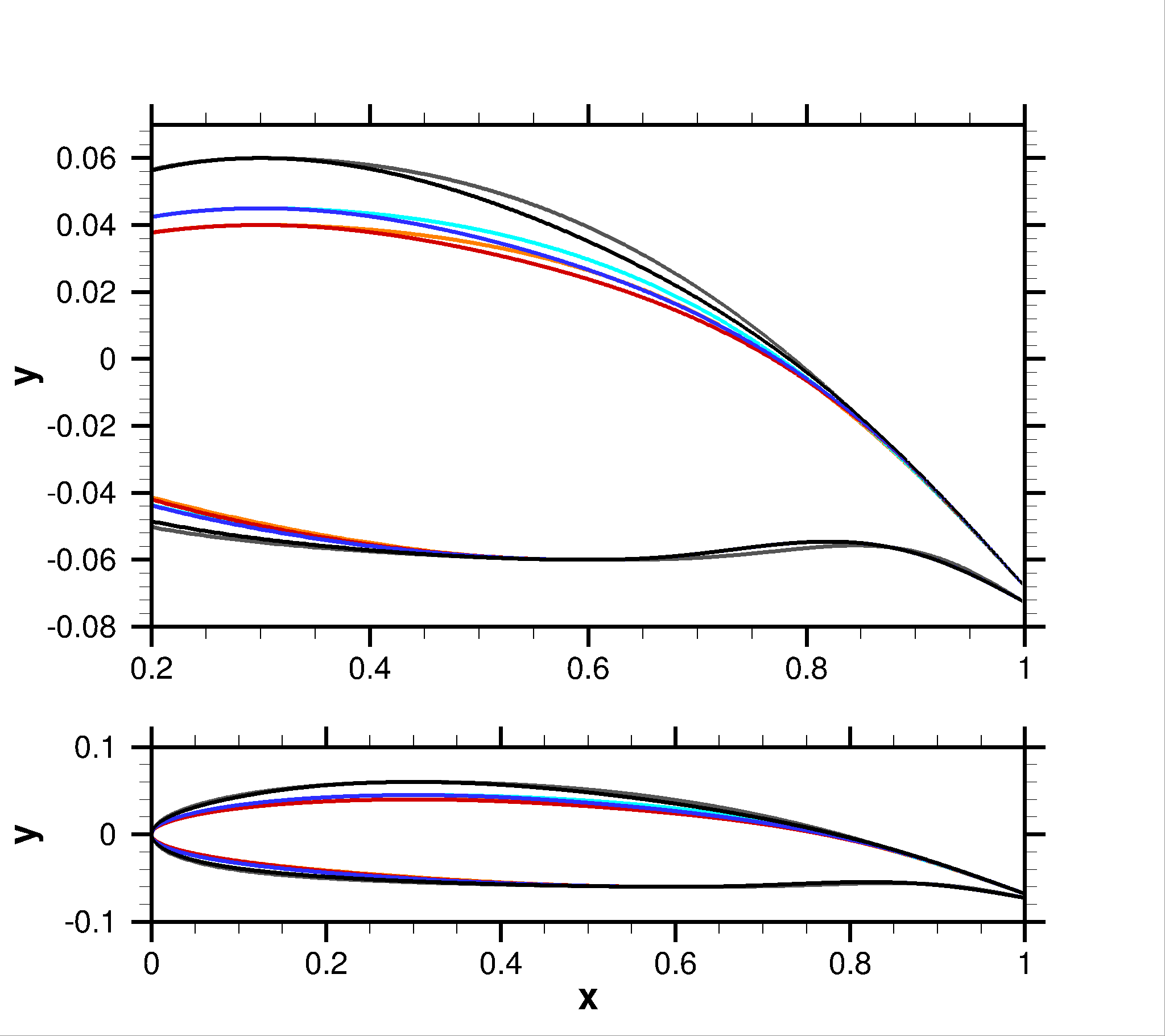} \\
  \end{tabular}
\caption{(a) Lift coefficient ($C_L$) as a function of convective time units ($CTU$) for 4-digit airfoils at an angle of attack $\alpha=4^{\circ}$, where the second digit for modified (OBf-4-X666) and original OB-4-X666 profiles. Corresponding airfoil contours are shown in (b), respectively, where the aspect ratio of the upper plot axes is distorted. The aspect ratio of the lower plot is x:y=1:1.  \label{fig:D2_orig}}
\end{figure}
Figure \ref{fig:D2_orig} shows (a) histories and (b) corresponding geometries for variations of the vertical position of the upper crest point. 
Figure \ref{fig:D1_orig} shows (a) histories and (b) corresponding geometries for variations of the axial position of the upper crest point. 
All in all, we can confirm good qualitative agreement for both airfoil families. 

\subsection{Data used for scaling intermediate frequencies of case OB-F-D2-4p5}\label{sec:App_scaling}
\begin{figure}
\centering
  \begin{tabular}{ll}
    a) & b) \\
    \includegraphics[width=0.45\textwidth,trim={10mm 10mm 20mm 10mm},clip]{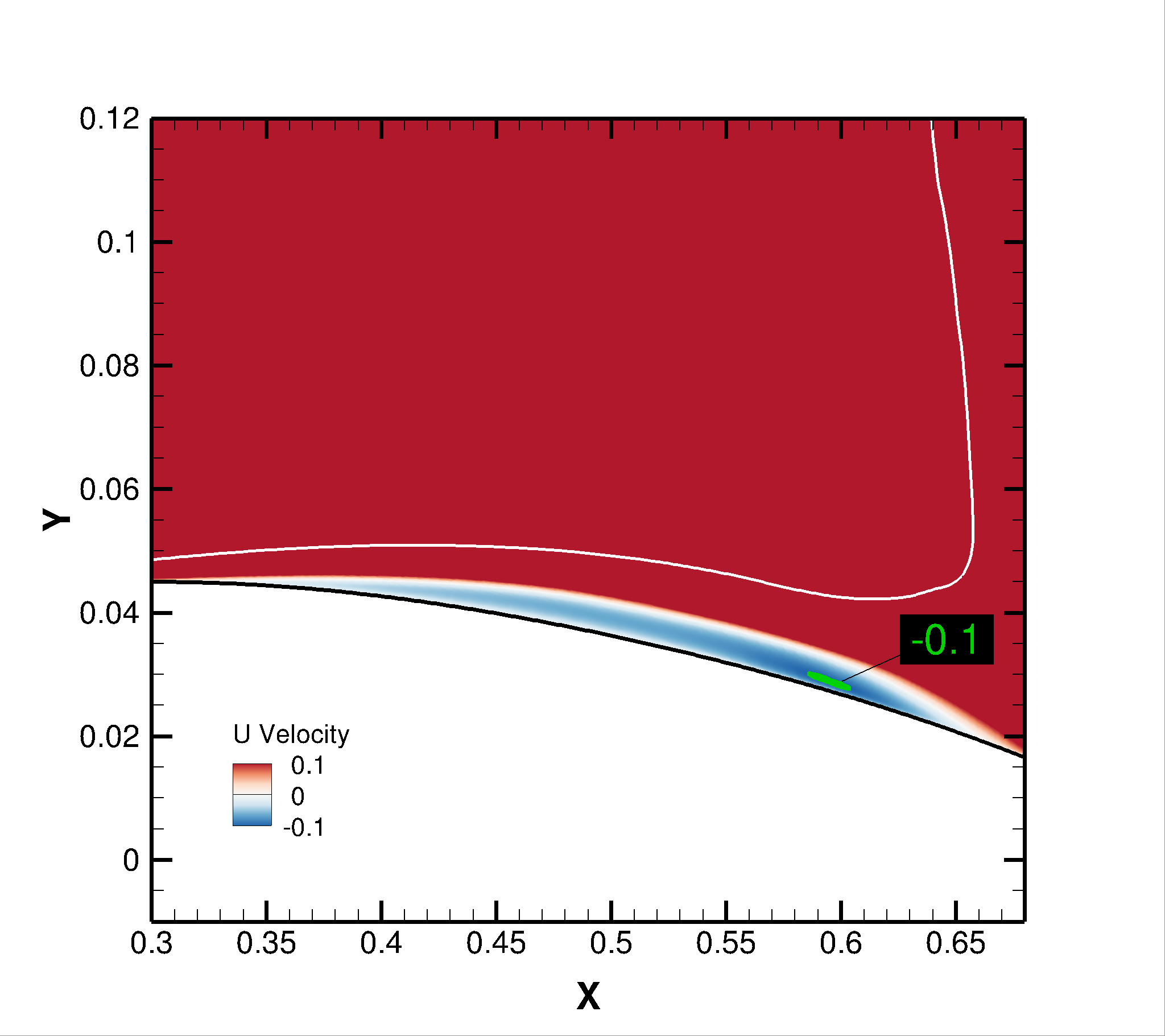} &
    \includegraphics[width=0.45\textwidth,trim={10mm 10mm 20mm 10mm},clip]{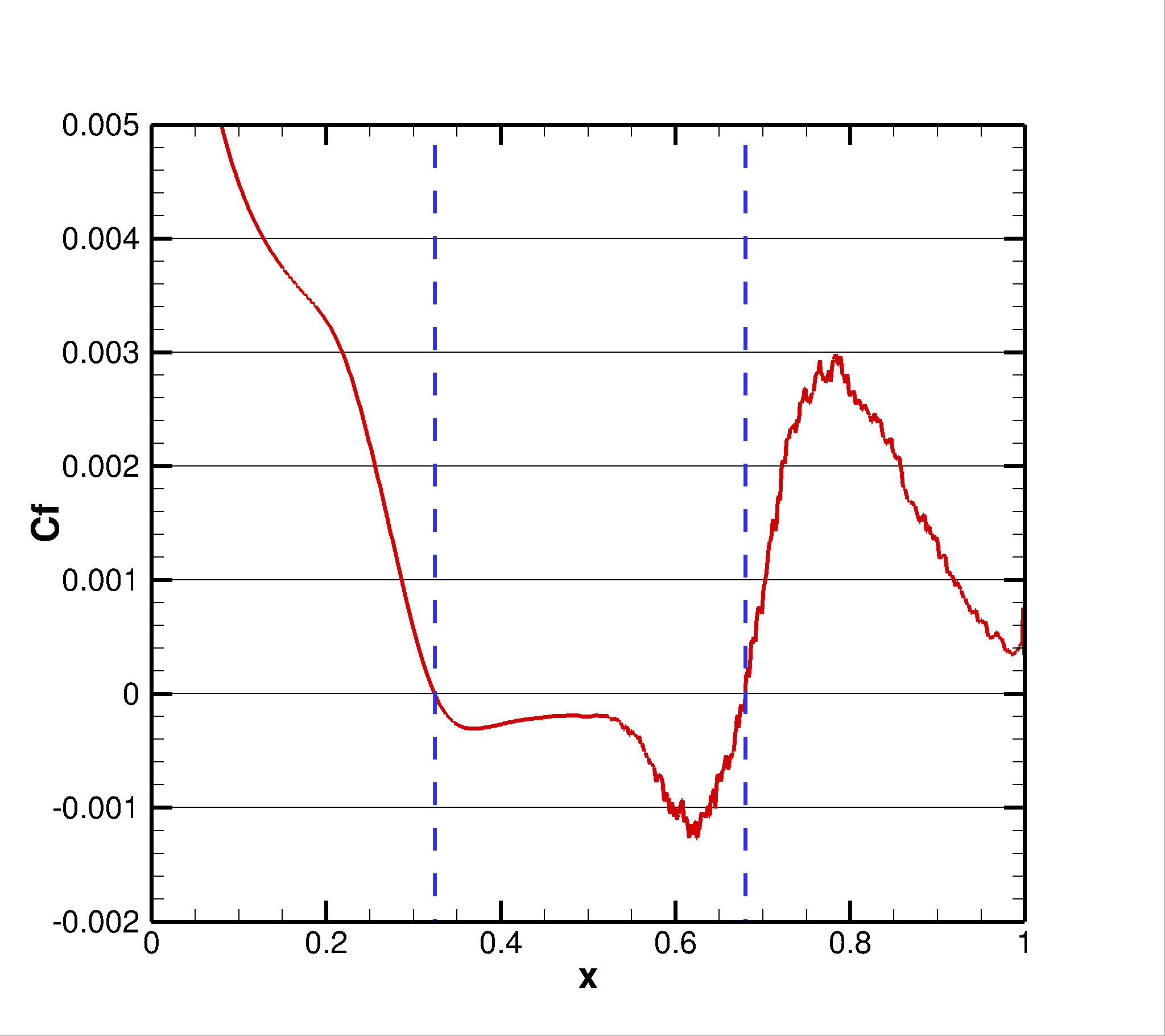} \\
  \end{tabular}
\caption{(a) Streamwise mean-velocity component and (b) suction-side skin-friction distribution for free-transitional case OB-F-D2-4p5 using the OB-4-34p566 profile.\label{fig:scaling}}
\end{figure}
For a selected test case, we attempt a scaling approach of the intermediate-frequency phenomenon proposed by \cite{Zauner2023b}. We consider test case OB-F-34p566 (OB-4-34p566 profile) of section \ref{sec:OpenBuffet_Upper_Free}. 
In a LES study of ONERA's OALT25 airfoil at similar flow conditions, \cite{Zauner2023b} scaled the Strouhal number $St_I$ associated with an unsteady recirculation bubble with mean-flow properties of the separation bubble. To obtain the scaled Strouhal number $St_R$, the chord-based Strouhal number $St_I$ needs to be multiplied by the normalised mean separation length $L_{sep}/c$ and the maximum reverse velocity $U_{R}/U_{\infty}$ observed in the mean flow within the separation bubble. The latter is approximated by the minimum streamwise mean-velocity component, which is shown in figure \ref{fig:scaling}(a). Blue and red contours show respectively negative and positive values of the streamwise mean-velocity component $U_x$, while the green iso-curve delineates a region of $U_x/U_{\infty}<-0.1$, which contains $U_{R}/U_{\infty}=|U_x/U_{\infty}|=0.102$.
Figure \ref{fig:scaling}(b) shows the distribution of the time-averaged skin-friction coefficient $C_f$ (using a reduced sampling rate of 0.035 CTUs), where we can extract the mean length of the separation bubble, bounded by the vertical dashed blue lines delivering $L_{sep}/c=0.356$. The Strouhal number $St_I=0.573$ associated with the intermediate-frequency phenomenon is provided in table \ref{tab:3X66} for the test case OB-F-D2-4p5. Applying now the scaling according to \cite{Zauner2023b}, we obtain $St_R=St_I L_{sep}/c U_{R}/U_{\infty} = 0.573 \cdot 0.356 \cdot 0.102 = 0.021$. This scaled frequency agrees very well with those of various airfoil profiles in \cite{Zauner2023b}. For ONERA's OALT25 profile they reported at the same Mach- and Reynolds number $St_R=0.022$ and for Dassault Aviations's V2C profile $St_R=0.023$. The Strouhal numbers without scaling are very different, reading $St_I=0.42$ and $0.6$, resoectively.

\subsection{Histories of flattened OpenBuffet profiles}\label{sec:appendix_hist_flat}
\begin{figure}[hbt!]
\centering
  \begin{tabular}{ll}
    a) & b) \\
    \includegraphics[width=0.45\textwidth,trim={10mm 10mm 20mm 20mm},clip]{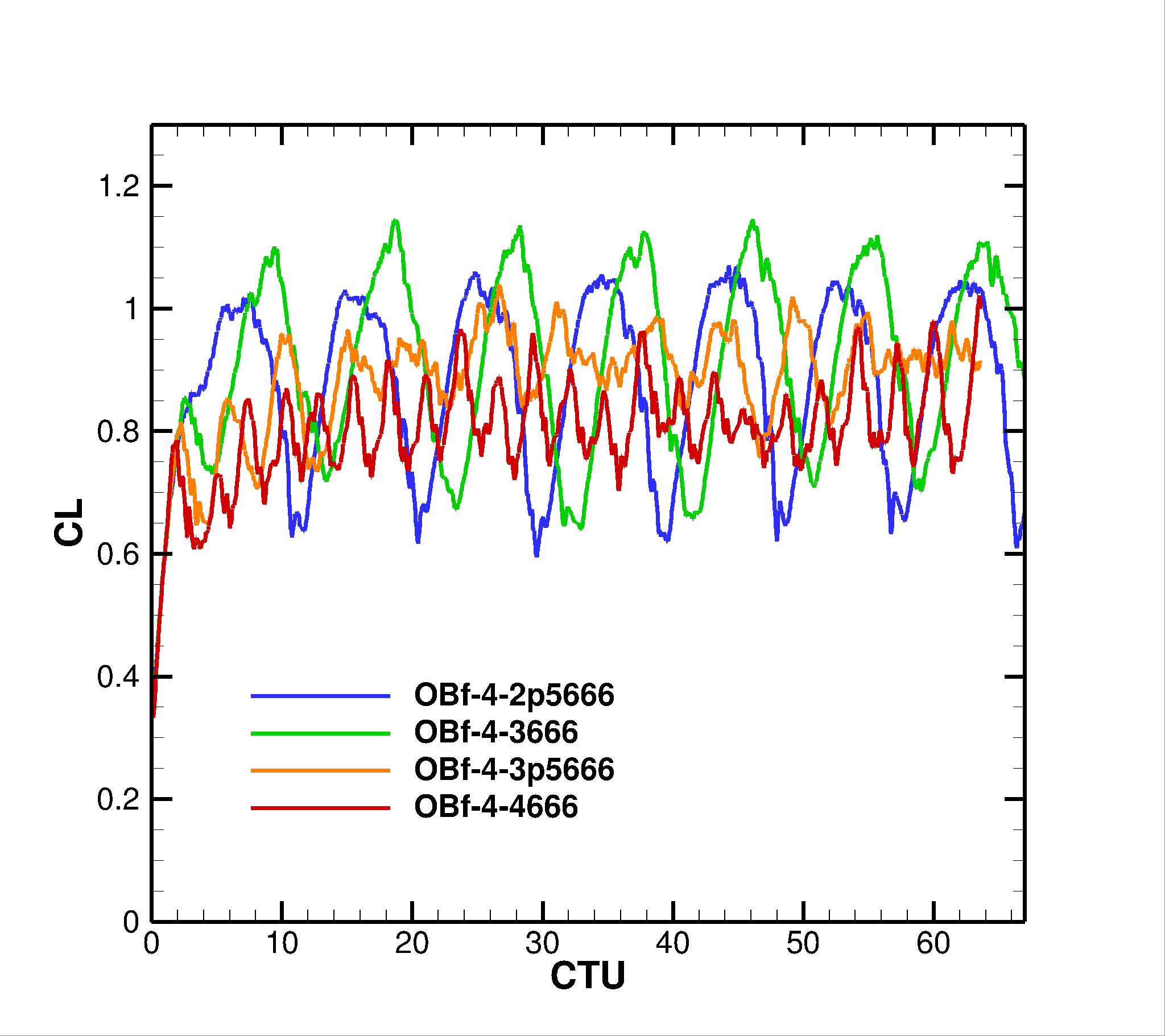} &
    \includegraphics[width=0.45\textwidth,trim={10mm 10mm 20mm 10mm},clip]{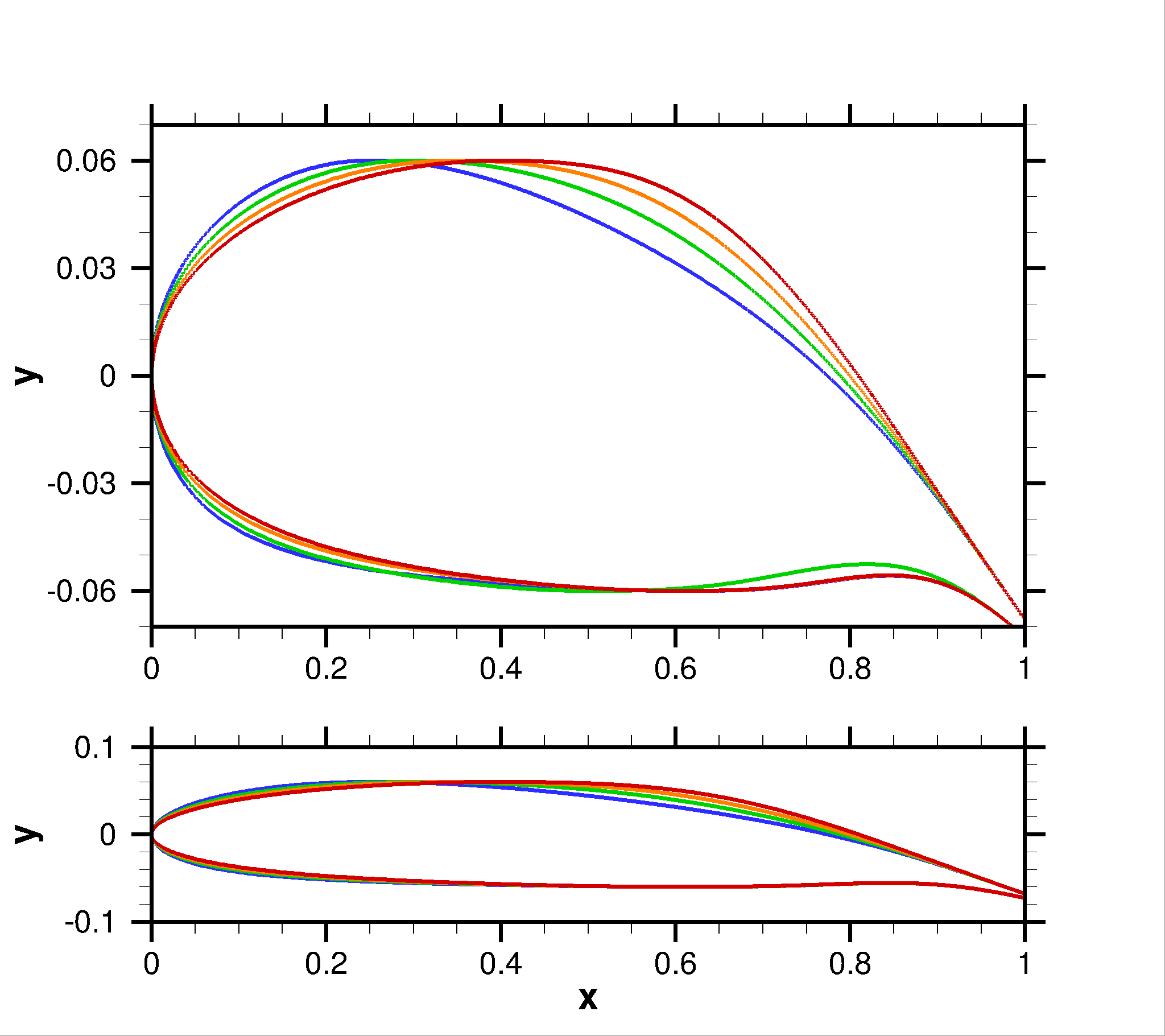} \\
  \end{tabular}
\caption{(a) Lift coefficient ($C_L$) as a function of convective time units ($CTU$) for 4-digit airfoils at an angle of attack $\alpha=4^{\circ}$, where the first digit is varied (OBf-4-X666). Corresponding airfoil contours are shown in (b), respectively, where the aspect ratio of the upper plot axes is distorted. The aspect ratio of the lower plot is x:y=1:1.  \label{fig:X666_flat}}
\end{figure}

\begin{figure}[hbt!]
\centering
  \begin{tabular}{ll}
    a) & b) \\
    \includegraphics[width=0.45\textwidth,trim={10mm 10mm 20mm 20mm},clip]{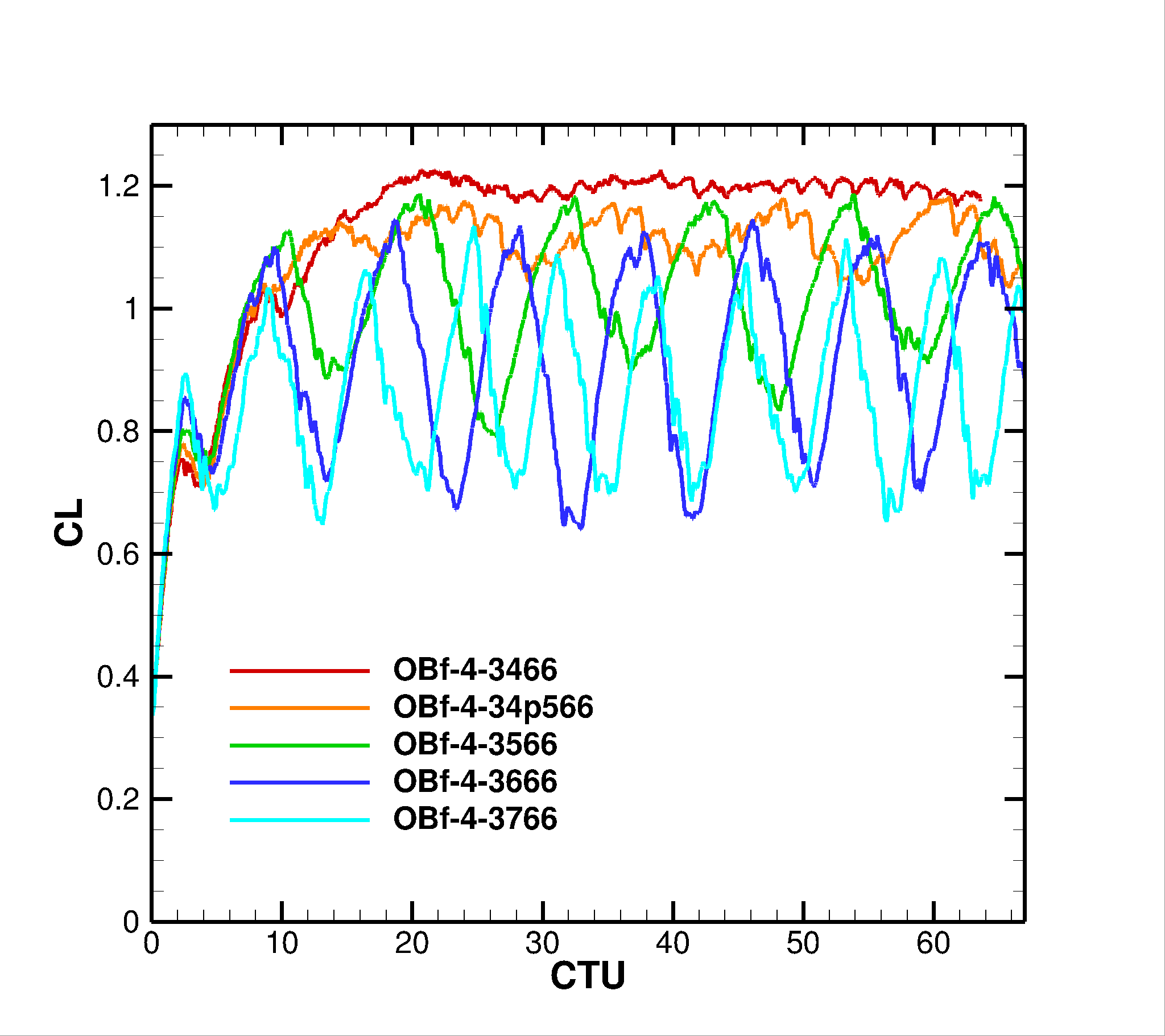} &
    \includegraphics[width=0.45\textwidth,trim={10mm 10mm 20mm 10mm},clip]{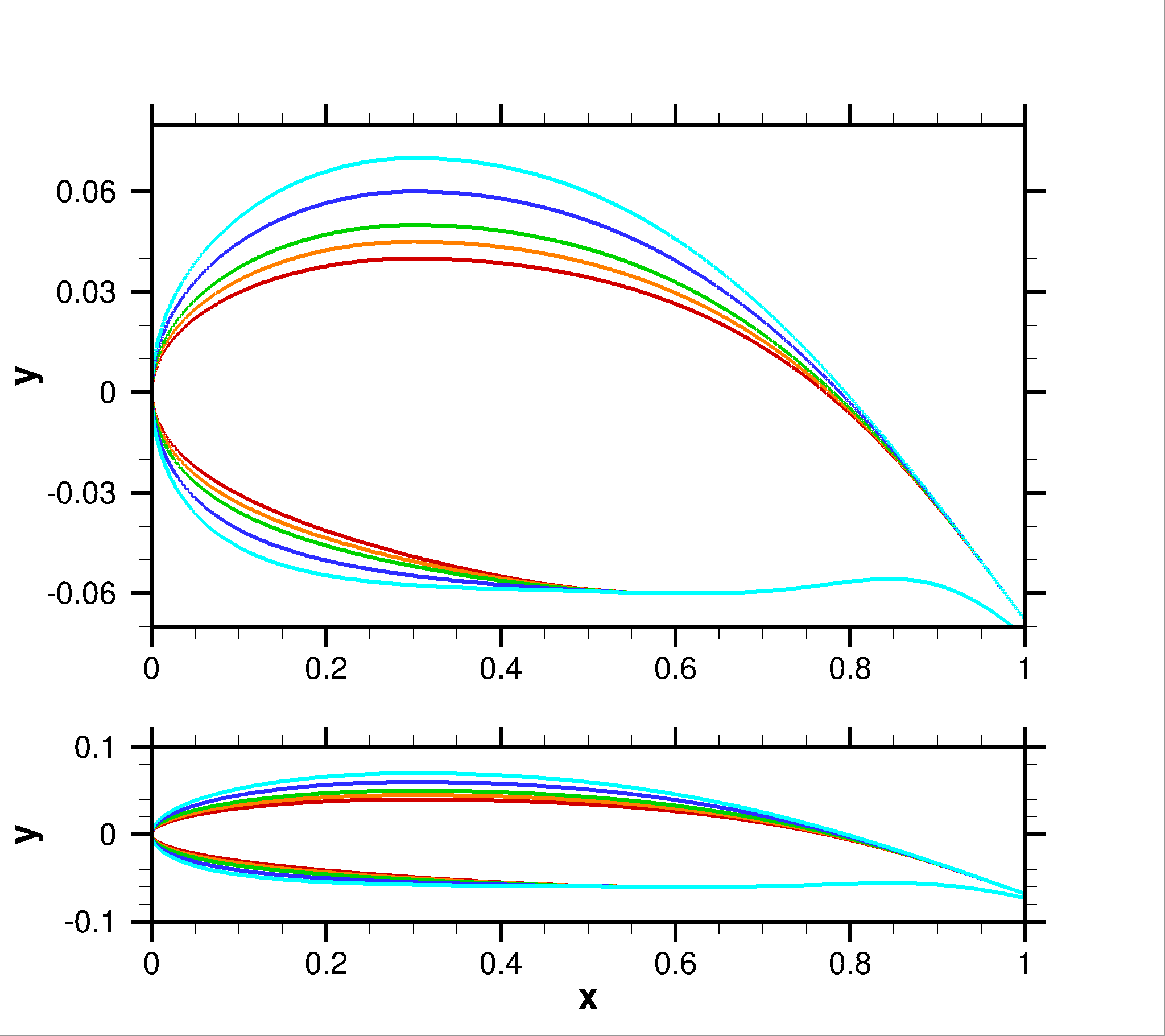} \\
  \end{tabular}
\caption{(a) Lift coefficient ($C_L$) as a function of convective time units ($CTU$) for 4-digit airfoils at an angle of attack $\alpha=4^{\circ}$, where the second digit is varied (OBf-4-3X66). Corresponding airfoil contours are shown in (b), respectively, where the aspect ratio of the upper plot axes is distorted. The aspect ratio of the lower plot is x:y=1:1.  \label{fig:3X66_flat}}
\end{figure}

\begin{figure}[hbt!]
\centering
  \begin{tabular}{ll}
    a) & b) \\
    \includegraphics[width=0.45\textwidth,trim={10mm 10mm 20mm 20mm},clip]{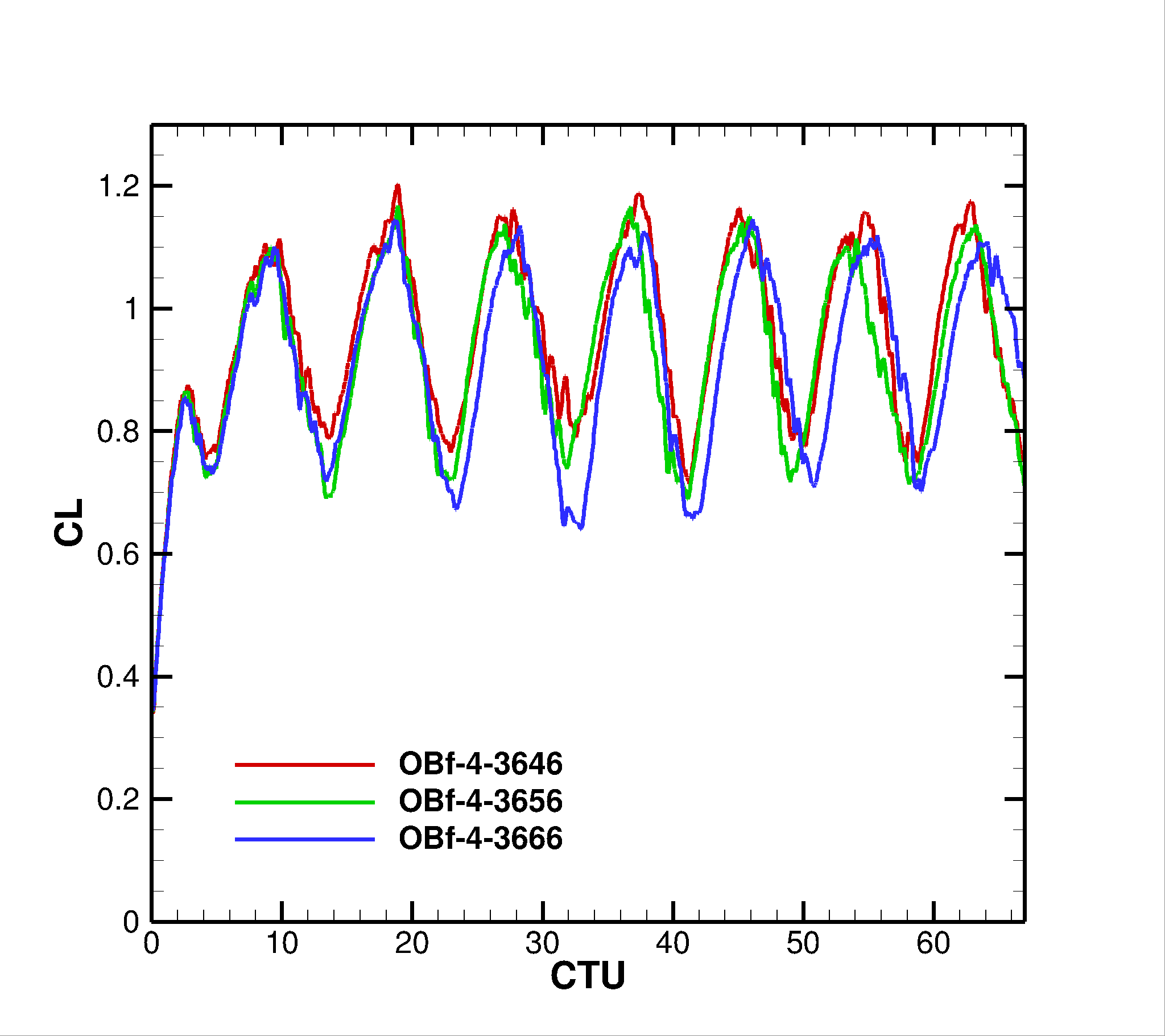} &
    \includegraphics[width=0.45\textwidth,trim={10mm 10mm 20mm 10mm},clip]{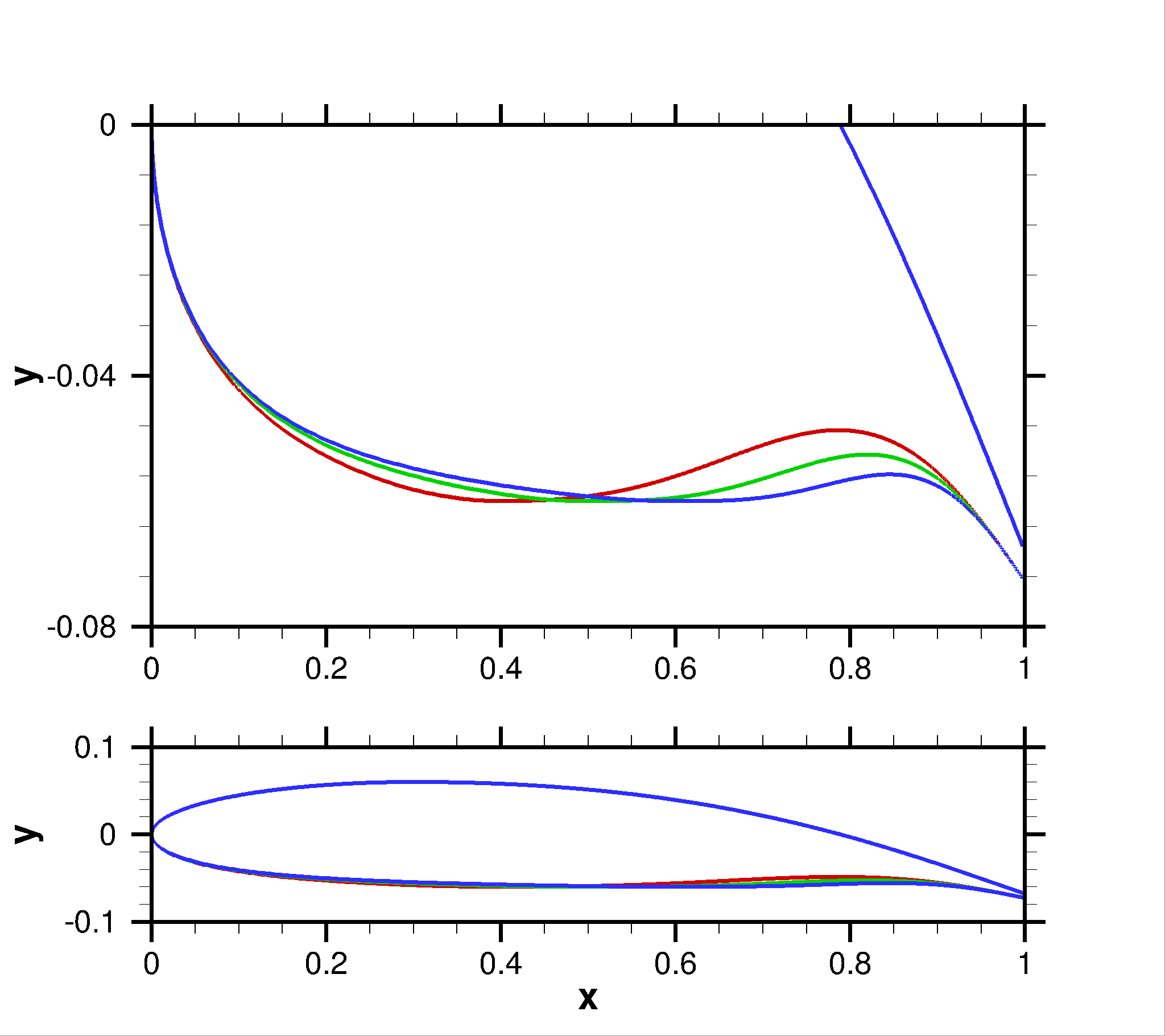} \\
  \end{tabular}
\caption{(a) Lift coefficient ($C_L$) as a function of convective time units ($CTU$) for 4-digit airfoils at an angle of attack $\alpha=4^{\circ}$, where the third digit is varied (OBf-4-36X6). Corresponding airfoil contours are shown in (b), respectively, where the aspect ratio of the upper plot axes is distorted. The aspect ratio of the lower plot is $x$:$y$=1:1. Please note that the upper side is constant for all profiles shown. \label{fig:36X6_flat}}
\end{figure}

\begin{figure}[hbt!]
\centering
  \begin{tabular}{ll}
    a) & b) \\
    \includegraphics[width=0.45\textwidth,trim={10mm 10mm 20mm 20mm},clip]{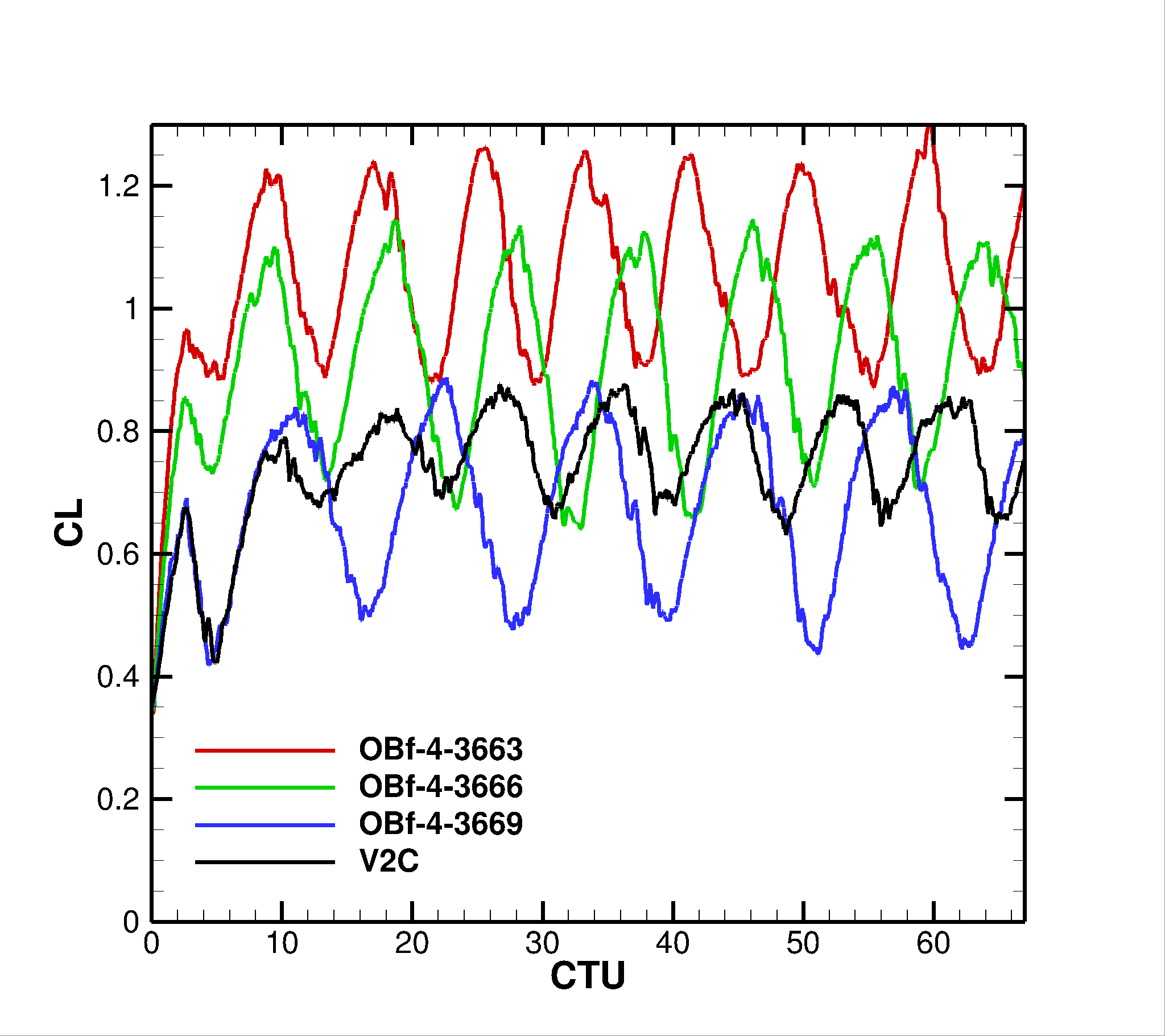} &
    \includegraphics[width=0.45\textwidth,trim={10mm 10mm 20mm 10mm},clip]{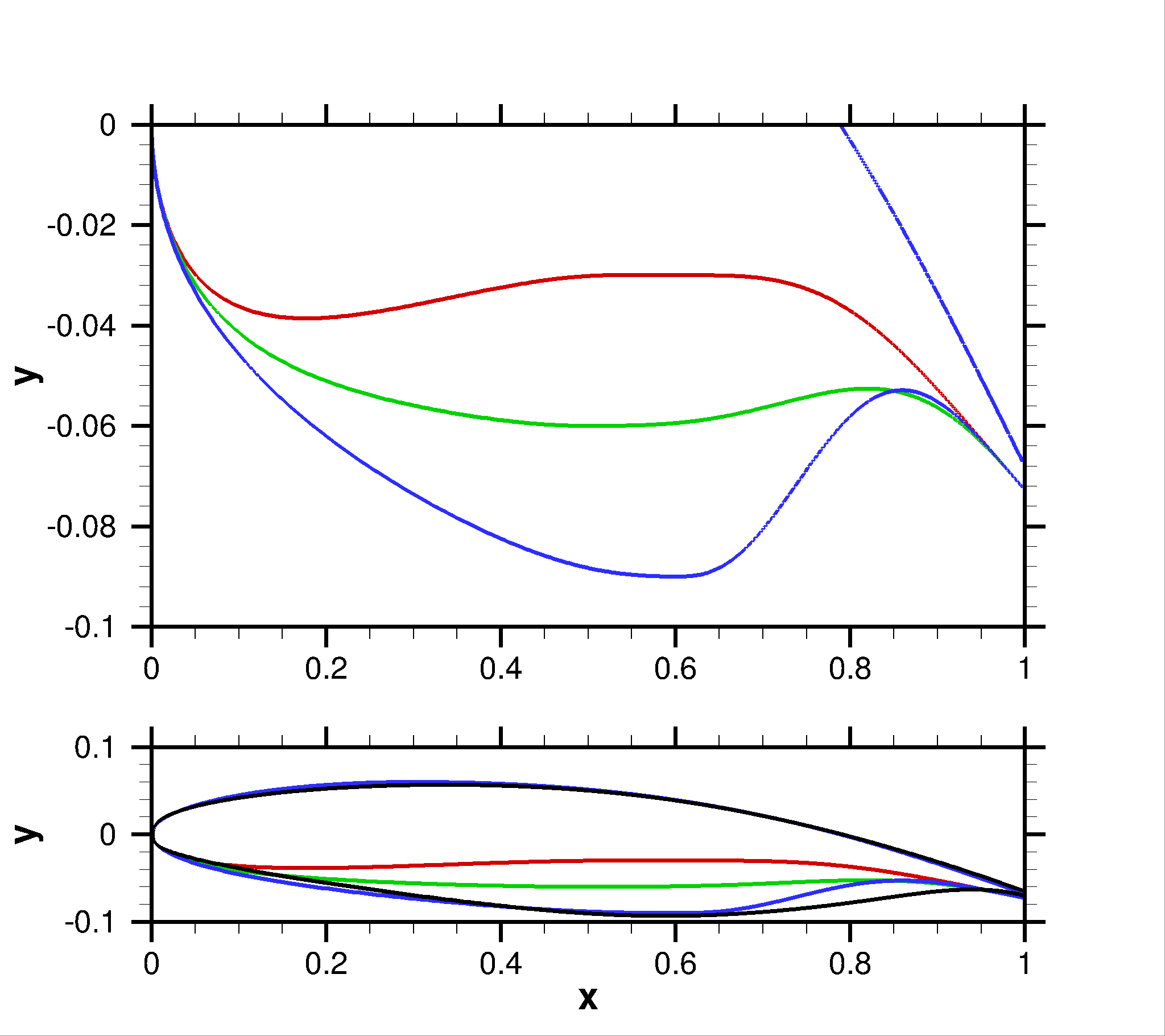} \\
  \end{tabular}
\caption{(a) Lift coefficient ($C_L$) as a function of convective time units ($CTU$) for 4-digit airfoils at an angle of attack $\alpha=4^{\circ}$, where the fourth digit is varied (OBf-4-366X). Black curves correspond to data for the V2C profile taken from \cite{Moise2021}. Corresponding airfoil contours are shown in (b), respectively, where the aspect ratio of the upper plot axes is distorted. The aspect ratio of the lower plot is x:y=1:1 and contains also Dassault Aviation's V2C profile, which it is not included in the distorted plot. Please note that the upper side is constant for all profiles shown.  \label{fig:366X_flat}}
\end{figure}

Additional histories are provided for the modified OpenBuffet geometries with flattened crest. Figures \ref{fig:3X66_flat} and \ref{fig:X666_flat} show variations of the upper crest point, while figures \ref{fig:36X6_flat} and \ref{fig:366X_flat} show variations of the lower crest point.
Plots (a) and (b) show $C_L$ histories and corresponding geometries, respectively.

% \section*{Contributions}
% This work has been done in a collaboration between the University of Southampton and the Japanese Aerospace Exploration Agency. Contributions of both institutions are detailed below.\\ \\
% \textbf{The University of Southampton:} M. Zauner (before joining JAXA), P. Moise, and N. D. Sandham \newline
% Development of the OpenBuffet airfoil geometries, generation of grids for free-transitional as well as tripped cases, simulation of free-transitional cases on IRIDIS 5, post-processing and analysis, and preparation of the manuscript.
% \textbf{Japanese Aerospace Exploration Agency:} M. Zauner (current affiliation), D. J. Lusher, A. Sansica, and A. Hashimoto \newline
% Simulation of tripped cases on JSS3, post-processing and analysis, and preparation of the manuscript.

\section*{Acknowledgments}
The authors want to acknowledge the Iridis5 (University of Southampton) and JSS3 (JAXA) supercomputing facilities and associated staff for providing computational resources and technical support. 
Neil D. Sandham acknowledges the support of EPSRC project EP/W026686/1.
David J. Lusher has been funded by the Japan Society for the Promotion of Science (JSPS), on a postdoctoral fellowship awarded to the JAXA Chofu Aerospace Center. 
Additional funding was provided by a JSPS KAKENHI grant award (22F22059).
The authors also want to thank ONERA and Dassault Aviation for providing OAT15A, OALT25, and V2C airfoil geometries.
This manuscript has been submitted for presentation at the 2024 AIAA AVIATION Forum in Las Vegas (\url{https://doi.org/10.2514/6.2024-3508}).

\clearpage

\bibliography{sample}

\end{document}